\documentclass[12pt]{article}
\usepackage{epsfig}
\usepackage{amssymb,amsmath}
\usepackage{setspace}
\pdfoutput=1
\usepackage{subfigure}
\usepackage{amssymb,amsmath}
\usepackage{graphicx}
\usepackage{color}
\usepackage{cancel}
\usepackage[colorlinks=true
,urlcolor=blue
,citecolor=blue
,linkcolor=blue
,pagecolor=blue
,linktocpage=true
,pdfproducer=medialab
]{hyperref}
\def\bi{\begin{itemize}}
\def\ei{\end{itemize}}

\def\alt{\lesssim}
\def\agt{\gtrsim}

\usepackage{amsmath}

\newcommand{\bea}{\begin{eqnarray}}
\newcommand{\eea}{\end{eqnarray}}

\newcommand{\beq}{\begin{equation}}
\newcommand{\eeq}{\end{equation}}

 \makeatletter
\def\alt{\mathrel{\mathpalette\gl@align<}}
\def\agt{\mathrel{\mathpalette\gl@align>}}
\def\gl@align#1#2{\lower.6ex\vbox{\baselineskip\z@skip\lineskip\z@
\ialign{$\m@th#1\hfil##\hfil$\crcr#2\crcr\sim\crcr}}} \makeatother

\usepackage{longtable}

\begin{document}
%
\vspace*{1.0cm}

\begin{center}
  \baselineskip 20pt {\Large\bf
    The Super-Natural Supersymmetry and Its Classic Example: 
M-Theory Inspired NMSSM
}
\vspace{1cm}

{\large
Tianjun Li$^{a,b,}$,
Shabbar Raza$^{a}$,
Xiao-Chuan Wang$^{a,c}$
} \vspace{.5cm}

{\baselineskip 20pt \it $^a$
State Key Laboratory of Theoretical Physics and Kavli Institute for Theoretical Physics China (KITPC),
Institute of Theoretical Physics, Chinese Academy of Sciences, Beijing 100190, P. R. China \\
}
{\it $^b$
School of Physical Electronics, University of Electronic Science and Technology of China,
Chengdu 610054, P. R. China \\
}
{\it $^c$
Department of Physics, Henan Normal University, Xinxiang, Henan, 453007, P.R.China
}

\vspace{.1cm}

\vspace{1.5cm} {\bf Abstract}
\end{center}

We briefly review the super-natural supersymmetry (SUSY), which provides a most promising solution to
the SUSY electroweak fine-tuning problem. In particular, we address its subtle issues as well.
Unlike the Minimal Supersymmetric Standard model (MSSM),
the Next to MSSM (NMSSM) can be scale invariant and has no mass parameter in its Lagrangian
before SUSY and gauge symmetry breakings. Therefore, the NMSSM is a perfect framework for
super-natural SUSY. To give the SUSY breaking soft mass to the singlet, we consider the
moduli and dilaton dominant SUSY breaking scenarios in M-theory on $S^1/Z_2$.
In these scenarios, SUSY is broken by one and only one $F$-term of moduli or dilaton,
and the SUSY breaking soft terms can be determined via the K\"ahler potential and
superpotential from Calabi-Yau compactification of M-theory on $S^1/Z_2$. Thus, as
predicted by super-natural SUSY, the SUSY electroweak fine-tuning measure is of unity order.
In the moduli dominant SUSY breaking scenario,  the right-handed sleptons
are relatively light around 1 TeV, stau can be even as light as 580 GeV and degenerate
with the lightest neutralino, chargino masses are larger than 1 TeV,
the light stop masses are around 2 TeV or larger, the first two-generation
squark masses are about 3 TeV or larger, and gluinos are heavier than squarks.
In the dilaton dominant SUSY breaking scenario, the qualitative picture remain the same
but we have heavier spectra as compared to moduli dominant SUSY breaking scenario.
In addition to it,  we have Higgs $H_{2}/A_{1}$-resonance solutions for dark matter (DM).
In both scenarios, the minimal value of DM relic density is about 0.2. To obtain the observed
DM relic density, we can consider the dilution effect from  supercritical string cosmology  or introduce the axino
as the lightest supersymmetric particle.

\thispagestyle{empty}

\newpage

\addtocounter{page}{-1}

\baselineskip 18pt
\section{Introduction}

Supersymmetry (SUSY) provides a natural solution to the gauge hierarchy problem in the Standard Model (SM).
In the supersymmetric SMs (SSMs) with $R$-parity, 
gauge coupling unification can be obtained, the Lightest Supersymmetric Particle (LSP) such as
neutralino can be a dark matter (DM) candidate, and the electroweak (EW) gauge symmetry can be broken
radiatively due to the large top quark Yukawa coupling, etc. Moreover,
gauge coupling unification strongly implies the Grand Unified Theories (GUTs), 
and the SUSY GUTs can be constructed from  superstring theory, which is the
most competitive candidate for quantum gravity. 
Therefore, supersymmetry is not only the most promising new physics beyond the SM,
but also a bridge between the low energy phenomenology and high-energy
fundamental physics.

It is well-known that a SM-like Higgs boson with mass $m_H$ around 125 GeV was discovered
during the first run of the LHC~\cite{ATLAS, CMS}. In the MSSM, to realize
such a Higgs boson mass, we need the multi-TeV top squarks with small mixing or TeV-scale 
top squarks with large mixing~\cite{Carena:2011aa}. 
There also exists strong constraints on the parameter space in the SSMs from the LHC SUSY searches.
For example, the gluino mass $m_{\tilde g}$ and first two-generation squark mass $m_{\tilde q}$  
should be heavier than about 1.7 TeV if they are roughly degenerate $m_{\tilde q} \sim m_{\tilde g}$, 
and the gluino mass is heavier than about 1.3 TeV
for $m_{\tilde q} \gg m_{\tilde g}$~\cite{Aad:2014wea, Chatrchyan:2013wxa}.
Naively, from the naturalness of the electroweak scale, the 
bilinear Higgs mass parameter $\mu$, which is related to the Higgs
boson mass, may need to be of the order of 100 GeV. 
Thus, the naturalness in the SSMs is challenged from both the Higgs boson mass
and the LHC SUSY searches.

To quantize the size of fine-tuning in the SSMs, we need to define the measure.
There are two kinds of definitions for fine-tuning measures:
the low energy definition~\cite{Kitano:2005wc, Papucci:2011wy, Baer:2012mv} and
high energy definition~\cite{Ellis:1986yg,Barbieri:1987fn}. 
We emphasize that the naturalness conditions from the low energy definition can still be satisfied
in principle, but the naturalness condition from the high energy definition
is indeed a big challenge.
However, because SUSY is the connection between the low and high energy physics, we do need to
consider seriously the fine-tuning problem via the high
energy definition. To solve this problem, we proposed the super-natural SUSY,
which provides a most promising solution to
the SUSY EW fine-tuning problem. 
It was shown in Refs.~\cite{Leggett:2014mza, Leggett:2014hha, Du:2015una} that the high energy fine-tuning measure 
will automatically be at the order one ${\cal O} (1)$ in the
${\cal F}$-$SU(5)$ models~\cite{F-SU5, Jiang:2006hf, Jiang:2009zza, Li:2010ws}
and the MSSM with no-scale supergravity (SUGRA)~\cite{Cremmer:1983bf}
and Giudice-Masiero (GM) mechanism~\cite{Giudice:1988yz}.
We will briefly review the super-natural SUSY in Section 3 and
for the first time address its subtle issues publicly. Especially,
the major challenge to the previous studies is $\mu$ term, which is generated
by the GM mechanism and then is proportional to the universal gaugino mass $M_{1/2}$.
The ratio $\mu/M_{1/2}$ is of order one but cannot be determined as an exact number.
We have studied it carefully before, and did not find any loophole~\cite{Leggett:2014mza, Leggett:2014hha, Du:2015una}.

On the other hand,  the MSSM suffers from the so-called $\mu$ problem~ \cite{Kim:1983dt}.
In the Next to MSSM (NMSSM) which is 
the simplest extension of the MSSM \cite{genNMSSM1,genNMSSM2,genNMSSM3}, due to the presence of an extra singlet
superfield $\hat S$, the effective $\mu_{eff}\equiv \lambda \langle {\hat S} \rangle$ term
can be generated via the superpotential term
$\lambda {\hat S} {\hat H}_d {\hat H}_u$ after $\hat S$ acquires a Vacuum Expectation Value (VEV),
where $\lambda$ is the Yukawa coupling while ${\hat H}_d$ and ${\hat H}_u$ are one pair of Higgs doublets in the MSSM. Moreover,
the SUSY breaking scale is the only scale in the Lagrangian, since it allows for a scale invariant
superpotential~\cite{Djouadi:2008uj}.
The SM-like Higgs,  due to the above superpotentional term, gets additional contributions at tree level.
Furthermore, the SM-like Higgs mass can be pushed up by the mixing effects in diagonalizing the
mass matrix of CP-even Higgs fields~\cite{Ellwanger:2011aa, Kang:2012sy, Cao:2012fz}. This results
in a SM-like Higgs boson with mass around 125 GeV without large loop contributions, and then the SUSY EW fine-tuning problem
can be ameliorated~\cite{tuning2}. Another consequence of extra singlet field is that there are three
CP-even Higgs $H_{1}$, $H_{2}$ and $H_{3}$, two CP-odd Higgs $A_{1}$ and $A_{2}$, a pair of charged Higgs
$H^{\pm}$, and an additional neutralino (singlino), as compared to the MSSM where $H_3$, $A_2$ and
singlino are absent. Similar to the constrained MSSM (CMSSM)/Minimal Supergravity (mSUGRA)~\cite{msugra},
one can also define the Constrained NMSSM (CNMSSM)~\cite{Fowlie:2014faa,Kim:2013uxa,Kowalska:2012gs,Kobayashi:2012xv,Gunion:2012zd,Ellwanger:2010es,Ellwanger:2009dp,LopezFogliani:2009np,Belanger:2008nt,Ellwanger:2008ya,Djouadi:2008yj,Hugonie:2007vd,King:1995vk}, where the SUSY breaking (SSB) soft terms are: universal scalar mass $m_{0}$,
universal gaugino mass $M_{1/2}$, and universal trilinear coupling term $A_0$ 
 at the GUT scale $M_{GUT}$.  In the CNMSSM, in contrast
 to the unconstrained NMSSM, one needs small value of $\lambda$ but large value of
 $\tan\beta\equiv \frac{\langle {\hat H}_u \rangle}{\langle {\hat H}_d \rangle}$
 to get the SM-like Higgs mass around 125 GeV (For example, see \cite{Fowlie:2014faa}).

 In this paper, we point out that the NMSSM provides an excellent framework for super-natural SUSY
 since its superpotential can be scale invariant~\cite{Djouadi:2008uj}. In particular,
 we do not have the $\mu$ term issue any more. To satisfy
three conditions of super-natural SUSY (see Section~\ref{sec:3}) and give a soft mass to
the singlet, we shall consider the moduli dominant SUSY breaking (MDSB) 
and dilaton dominant SUSY breaking (DDSB) scenarios in M-theory
on $S^1/Z_2$~\cite{Horava:1996ma, Li:1997sk, Lukas:1997fg, Nilles:1998sx, Li:1998rn},
and propose the M-theory inspired CNMSSM (MCNMSSM).
In the MCNMSSM, SUSY is broken by one and only one $F$-term of moduli or dilaton.
The SUSY breaking  soft terms, such as $m_{0}$, $M_{1/2}$ and $A_{0}$, can be calculated explicitly
via the K\"ahler potential and superpotential from Calabi-Yau compactification of M-theory on $S^1/Z_2$,
and they are functions of the gravitino mass ($M_{3/2}$) and hidden/observable sector gauge couplings
at the GUT or string scale~\cite{Li:1998rn}
which should be determined after moduli stabilization.
And superpotential is scale invariant. Therefore,
according to the super-natural SUSY, the fine-tuning measure is
order of unity. In other words, there will be no EW fine-tuning problem at all in the MCNMSSM.
In the MDSB scenario,
we find that the minimal values for $m_{0}$ and $M_{1/2}$ consistent with sparticle mass bounds, B-physics bounds,
and the light CP-even Higgs mass bound of $125\pm2$ GeV are about 0.6 TeV and 1.4 TeV, respectively,
and the corresponding $A_{0}$ range is $[-4,~-2]$ TeV. We also find that the range of parameter $\lambda$
is $[0,0.1]$, and $\tan\beta$ is from 5 to 28. Moreover, we notice 
$m_{H_{2}} \approx m_{H^{\pm}} \approx m_{A_{1}}$ in most part of the parameter space, while 
we have $m_{H_{3}}\approx m_{H^{\pm}}$ in the mass range $[1.8,~2.7]$ TeV.  
The gluino mass $m_{\tilde g}$ is found to be relatively heavy $\gtrsim$ 3 TeV, and
the light stop is the lightest colored sparticle ($\gtrsim$ 2 TeV). The first two-generation squarks
are about 3 TeV but they are lighter than the gluinos. In the slepton sector, the first two-generation
sleptons have masses around 1 TeV or larger, while the light stau, which is mainly the right-handed stau,
can be as light as 560 GeV. The LSP neutralino are in the mass
range $[0.55,1.1]$ TeV while charginos are heavier than 1 TeV. We notice that despite the fact
that the LSP neutralino and light stau are almost degenerate, the
minimal values of DM relic density we get is about 0.2. In the DDSB scenario, the minimal values
for $m_{0}$ and $M_{1/2}$ consistent with various constraints are about 0.8 TeV and 1.6 TeV, respectively.
The ranges for $A_{0}$, $\lambda$, and $\tan\beta$ are respectively $[-8.8,~-2]$ TeV, $[0,~0.15]$ and $[2,~41]$.
Due to this slightly larger 
range of $\lambda$, the low mass values of the CP-even Higgs $m_{H_{2,3}}$ and CP-odd Higgs $m_{A_{1,2}}$
are somewhat smaller than the MDSB. So these Higgs particles can come closer in mass with the LSP
neutralino which can have mass in the range $[0.6,~4]$ TeV.
It is also observed that $m_{A_1}=m_{H^{\pm}}$, while $m_{A_{1}}\approx m_{A_{A_2}}\approx m_{H_{2}}\approx m_{H_{3}}$
in some portions of parameter space.
The light stop is still the lightest colored sparticle with mass $\gtrsim$ 2 TeV, the first two-generation
squark masses are $\gtrsim$ 3.4 TeV, while
the gluino mass is $\gtrsim$ 3.5 TeV. The first two-generation sleptons are heavier than 1 TeV
while the light stau can be as light as 600 GeV. 
The chargino masses are $\gtrsim$ 1.2 TeV. Even though we have the resonance conditions such as 
$2m_{\tilde \chi_{1}^{0}}\approx m_{H_{2},A_{1}}$ as well as the neutralino-stau coannihilation scenario,
the minimal relic density we get is still around 0.2.
We also present a couple of tables for benchmark points as examples of our findings. Furthermore,
 the minimal DM relic density is about 0.2 in both scenarios. To obtain the correct
 DM relic density, we can consider the dilution effect from supercritical string cosmology~\cite{Lahanas:2006xv}
 or introduce a LSP axino as the DM candidate. Especially, in the 
 supercritical string cosmology, the DM relic density can be diluted by a factor ten (${\cal O}(10)$)~\cite{Lahanas:2006xv}.

This paper is organized as follows. In Section~\ref{model}, we introduce the CNMSSM
as well as its SUSY breaking soft terms. In Section~\ref{sec:3}, we briefly review the super-natural SUSY
and address its subtle issues. We  give the SUSY breaking soft terms
from M-theory on $S^1/Z_2$ as well. We outline the detailed scanning procedure, and the relevant
experimental constraints in Section~\ref{sec:scan}.
We present in detail results of our scans in Section~\ref{results}.
A summary and conclusion are given in Section~\ref{summary}.


\section{The Constrained Next to Minimal Supersymmetric Standard Model}
\label{model}

The NMSSM is the simplest extension of the MSSM. In the NMSSM, we introduce an
SM singlet superfield $\hat S$, as well as a $Z_3$ symmetry which forbids
the $\mu$ term in the MSSM.  The scale-invariant
superpotential is 
\begin{eqnarray}
W_{NMSSSM}=(\textrm{MSSM Yukawa terms}) + \lambda \hat S \hat H_{u} \hat H_{d}+\frac{\kappa}{3}\hat S^3\,,\label{NMSSM}
\end{eqnarray}
where $\lambda$ and $\kappa$ are Yukawa couplings. The above two terms substitute
the $\mu \hat H_{u}\hat H_{d}$ term in the MSSM superpotential. After spontaneous EW gauge 
symmetry breaking, a non-vanishing VEV $v_{S}$ of ${\hat S}$ at the minimum of Higgs potential
generates an effective $\mu_{eff}$ term 
in the MSSM, {\it i.e.}, $\mu_{eff} \equiv \lambda v_{S}$. 
The SUSY breaking soft terms in the Higgs sector are then given by 
\begin{eqnarray}
{V}_{\rm{soft}}=m_{H_u}^2|H_u|^2+m_{H_d}^2|H_d|^2+m_{S}^2|S|^2+\left(\lambda A_{\lambda}
  S H_{u} H_{d}+\frac{1}{3}\kappa A_{\kappa}
  S^3+\textrm{h.c.}\right)\,,\label{soft} 
\end{eqnarray}
where $A_{\lambda}$ and $A_{\kappa}$ are soft trilinear terms associated with the
$\lambda$ and $\kappa$ terms in the superpotential.
The VEV $v_{S}$ of ${\hat S}$, determined by the minimization conditions of Higgs potential,
is effectively induced by the
SUSY breaking soft terms in Eq.~(\ref{soft}),  and is naturally set by $M_{SUSY}$. Thus,
the $\mu$ problem in the MSSM is solved.

In the CNMSSM, the SUSY breaking soft terms are universal gaugino mass $M_{1/2}$,
scalar mass $m_0$, and trilinear term $A_0$ at the GUT scale $M_{GUT}$.
Through the minimization of Higgs potential, $m_{S}^2$ can be traded for $\tan\beta$,
and $\kappa$ can be determined in terms of the other parameters for
a correct value of $M_Z$~\cite{Kowalska:2012gs,Djouadi:2008yj}. Moreover one can also chose either $\kappa$ or sgn$(\mu_{eff})$. For conventional reasons we chose sgn$(\mu_{eff})$.
In short, the CNMSSM can be defined in terms of 
five continuous input parameters and one sign as follows
\begin{eqnarray}
m_{0}, M_{1/2}, A_{0}, \tan\beta, \lambda, \textrm{sgn}(\mu_{eff})\, .\label{params}
\end{eqnarray}

\section{The Super-Natural SUSY and the M-Theory Inspired SUSY Breaking Soft Terms}
\label{sec:3}

To study the fine-tuning issue in the supersymmetric SMs, we need to define the
fine-tuning measures first. There are two kinds of definitions:
the low energy definition~\cite{Kitano:2005wc, Papucci:2011wy, Baer:2012mv},
and the high energy definition~\cite{Ellis:1986yg, Barbieri:1987fn}.
The low energy definition of fine-tuning measure does not give strong constraints
on the SSMs. In particular,  if we allow a few percent fine-tuning,
we can still have the viable parameter spaces in the MSSM and NMSSM,
which satisfy all the current experimental constraints including
the low bounds on the masses of the gluino, first/second generation squarks,
and sleptons,  etc,  from the LHC SUSY
searches~\cite{Aad:2014wea,Chatrchyan:2013wxa,Aad:2014nua,Aad:2014vma,CMS-PAS-SUS-13-006}.
However, the high energy definition of fine-tuning measure
is still a big challenge. To be concrete,
we can have the benchmark points which have
the low energy fine-tuning measure $\Delta_{\rm EW}$ around 20 while the high energy fine-tuning
measure $\Delta_{\rm EENZ}$ around 1,500. For instance, see the benchmark points 1 and 2
in Table 1 of Ref.~\cite{Li:2014dna}. Because
the fine-tuning measures for high energy definition in the viable SSMs are 
very large at the order of $10^3$ (${\cal O} (10^3)$),
we shall concentrate on it
in the following discussions. The typical quantitative measure $\Delta_{\rm EENZ}$ of 
SUSY EW fine-tuning is defined by the maximum of the logarithmic derivative of $M_Z$ with respect to
all fundamental parameters $a_i$  at the GUT scale~\cite{Ellis:1986yg,Barbieri:1987fn}
\begin{eqnarray}
\Delta_{\rm EENZ} ~=~ {\rm Max}\{\Delta_i^{\rm GUT}\}~,~~~
\Delta_i^{\rm GUT}~=~\left|
\frac{\partial{\rm ln}(M_Z)}{\partial {\rm ln}(a_i^{\rm GUT})}
\right|~.~\,
\label{eq:BG-EENZ}
\end{eqnarray}
So we would like to explore the supersymmetry breaking
scenario whose fine-tuning measure for high energy definition is automatically at
the order one (${\cal O} (1)$), {\it i.e.}, the fine-tuning measure in Eq.~(\ref{eq:BG-EENZ})
is exactly one in the dream case.  Interestingly, there exists a simple solution 
with $\Delta_{\rm EENZ}=1$. Assuming that there is one and only one mass parameter
$M_*$ in the SSMs, to be concrete,  we shall take  $M_*$ as the universal gaugino mass
$M_{1/2}$ for no-scale supergravity and gravitino mass $M_{3/2}$ for all the other supergravity
including the M-theory supergravity. Thus,  $M_{Z}$ will be  a trivial function of $M_*$, and
we have the following approximate scale relation
\begin{eqnarray}
M_Z^n ~=~ f_n \left( c_i \right) ~M_*^n~,~\,
\label{eq:fn}
\end{eqnarray}
where $f_n$ is a dimensionless parameter, and $c_i$ denote the dimensionless coupling parameters,
such as gauge and Yukawa couplings, as well as the ratio between $\mu$ and $M_{1/2}$ for the MSSM
with the GM mechanism, etc.

For the nearly constant $f_n$ of Eq. (\ref{eq:fn}), we have
\begin{eqnarray}
\frac{\partial M_Z^n}{\partial M_*^n} ~\simeq~ f_n~,~\, 
\label{eq:fnd}
\end{eqnarray}
 and therefore we obtain
\begin{eqnarray}
\frac{\partial{\rm ln}(M_Z^n)}{\partial {\rm ln}(M_{*}^n)}
~\simeq~  \frac{M_{*}^n}{M_Z^n} \frac{\partial M_Z^n}{\partial M_{*}^n}
~\simeq~  \frac{M_{*}^n}{M_Z^n} \frac{\delta M_Z^n}{\delta M_{*}^n} 
~\simeq~ \frac{1}{f_n} f_n~.~\
\label{fnpar}
\end{eqnarray}
Consequently, the fine-tuning measure is an order one constant
\begin{eqnarray}
\left|
\frac{\partial{\rm ln}(M_Z^n)}{\partial {\rm ln}(M_{*}^n)}
\right| \simeq {\cal O}(1)~.~\,
\label{eq:orderone}
\end{eqnarray}
Therefore, there is no electroweak fine-tuning problem in such kind of SSMs.
This conclusion has been confirmed numerically in the ${\cal F}$-$SU(5)$ model
and MSSM with no-scale supergravity and GM mechanism~\cite{Leggett:2014mza, Leggett:2014hha, Du:2015una}. 


Based on the above discussions, we proposed the super-natural SUSY with
$\Delta_{\rm EENZ} \simeq 1$~\cite{Du:2015una}.
The three necessary conditions for super-natural SUSY are~\cite{Du:2015una}

\begin{itemize}

\item {The K\"ahler potential and superpotential can be calculated in principle or at least inspired 
from a fundamental theory such as string theory with suitable compactification. In other words, one cannot
add arbitrary high-dimensional terms in the K\"ahler potential and superpotential.  }

\item {There is one and only one chiral superfield or modulus whose F-term breaks
  supersymmetry. And all the supersymmetry breaking soft terms are obtained from
  the above K\"ahler potential and superpotential.}

\item {All the other mass parameters, if there exist such as the $\mu$ term in the MSSM,
 must arise from supersymmetry breaking.}

\end{itemize}
Therefore, all the SUSY breaking soft terms and mass parameters in the SSMs 
are linearly proportional to the gravitino mass, and then the fine-tuning measure
$\Delta_{\rm EENZ} $ from high energy definition is order of unity.  

For the first time, we would like to address a few subtle issues publicly in the
super-natural SUSY as follows

\begin{itemize}

\item {The EW Symmetry Breaking and Determination of $M_*$ from $Z$ Boson Mass}

  Assuming that the SSMs arise from string theories with suitable compactification
  and moduli stabilization, and there is one and only one F-term of moduli or dilaton
  whose F-term breaks SUSY, we can calculate the corresponding K\"ahler potential
  and superpotential, and then all the SUSY breaking soft terms can be determined
  in terms of $M_*$.
  Also, we can calculate the corresponding gauge couplings and Yukawa couplings at the
  GUT or string scale in principle, which should be required to be consistent with the low energy
  experimental values via renormalization group equation (RGE)
  running. For any set of the gauge couplings,
  Yukawa couplings, and SUSY breaking soft terms at the GUT or string scale,
  because the only free parameter is $M_*$, we might have three cases:
  (1) No RGE solution. (2) No EW gauge symmetry breaking, for example,
  stau is tachyonic. (3) The EW gauge symmetry breaking.
  In particular, for case (3), the observed $Z$ boson mass $M_Z$ as a low energy input
  will determine the corresponding $M_*$ since it is the only dimensionful free
  parameter. Of course, if the RGEs have several solutions,
  we may have a few corresponding  $M_*$ values.

\item {New $\mu$ Problem in the  MSSM and ${\cal F}$-$SU(5)$ Model}

  In the MSSM and ${\cal F}$-$SU(5)$ model with no-scale supergravity,
  to solve the $\mu$ problem, we employ the
  GM mechanism~\cite{Giudice:1988yz}. Thus, we have $\mu \propto M_{1/2} \propto M_{3/2}$,
  and $M_*$ is assumed to be $M_{1/2}$.
  The ratio $c\equiv \mu/M_{1/2}$ is an order one constant but we cannot determine the
  exact value of $c$ from the GM mechanism since we cannot determine the coefficient
   of the high-dimensional operator up to order one which generates the  $\mu$ term.
  This new $\mu$ problem was pointed out to us not only by referees but also by audiences.

  We have considered it in details, and confirmed that there is no gap
  in our previous studies~\cite{Leggett:2014mza, Leggett:2014hha, Du:2015una}. From top-down
  approach, $c$ is a fixed real number at the order one, and it can be determined
  from our above string model assumptions in principle. So the low energy 
  $Z$ boson mass $M_Z$ is predicted from the high energy fundamental theory. From the
  phenomenological point of view,  the observed value of
  $Z$ boson mass $M_Z$ determines the gaugino mass $M_{1/2}$ at the GUT scale for some
  narrow range of $c$. By the way, for the other numerical values of $c$, we
  do not have the correct $M_Z$ value, or the EW gauge symmetry breaking, or
  the RGE solution for  no-scale boundary conditions.
  To be concrete, from Fig. 2 of Ref.~\cite{Leggett:2014hha}, we found that for a fixed $c$,
  there is one to one correspondence between $M_Z$ and $M_{1/2}$ clearly.

\item{Symmetry for Super-Natural Supersymmetry}

  In the super-natural supersymmetry, the fine-tuning measure is exact one for the perfect scenario.
  So it is naive to think that there may exist a symmetry behind it. This symmetry is the
  scale invariance: for the fixed dimensionless coefficients at the unification scale from
  the top-down approach, we define the mass ratio $r_{\phi} \equiv M_{\phi}/M_*$, where $\phi$ is
  a supersymmetric particle (sparticle) and $M_{\phi}$ is its mass at low energy. We found $r_{\phi}$ is scale
  invariant, i.e., $r_{\phi}$ does not depend on $M_*$. This has been confirmed numerically by the previous
  studies in the  MSSM and ${\cal F}$-$SU(5)$ model with no-scale supergravity
  and GM mechanism~\cite{Leggett:2014mza, Leggett:2014hha, Du:2015una}.
  In other words, the sparticle mass spectra for different $M_*$ are correlated by a overall rescale.

  Similar to the low energy definition of the fine-tuning, we may require extra naturalness conditions at the
  GUT or string scale.
  In the MSSM, with the one-loop effective potential contributions to the tree-level 
Higgs potential, we get the $Z$-bosom mass $M_Z$~\footnote{The following comment is based
on the private discussions with Daniel Chung.}
\begin{equation}
\frac{M_Z^2}{2} =
\frac{m_{H_d}^2-m_{H_u}^2\tan^2\beta}{\tan^2\beta-1} -\mu^2 \; .
\label{eq:mssmmu}
\end{equation}
For moderate large $\tan\beta$, we have 
\begin{equation}
\frac{M_Z^2}{2} \simeq -m_{H_u}^2 -\mu^2 \; .
\end{equation}
At the GUT or string scale, although we do not have the EW gauge symmetry breaking,
{\it i.e.}, $M_Z=0$, to be natural, one might still require
\begin{equation}
\frac{m_{H_u}^2}{\mu^2} \sim \frac{|m_{H_u}^2 -\mu^2|}{m_{H_u}^2 + \mu^2} \sim {\cal O} (1)~.~\,
\end{equation}
In the no-scale supergravity, the above requirement cannot be satisfied since
the universal scalar mass $m_0$ vanish, {\it i.e.}, $m_0=0$.
However, our models, such as the MSSM and ${\cal F}$-$SU(5)$ model with
no-scale supergravity and GM mechanism,
are indeed technically natural since $m_0=0$ arises from the 
$SU(N,1)/SU(N)\times U(1)$ symmetry or a Heisenberg symmetry
in the K\"ahler potential~\cite{Gaillard:1995az}.

\item {Multi F-Term SUSY Breakings}  

  If SUSY is broken by two or more F-terms of moduli and/or dilaton, we should define the
  corresponding fundamental mass parameters as $M_*^i \equiv F_i/{\sqrt 3} M_{\rm Pl}$, and
  calculate the corresponding fine-tuning measures $\Delta_{M_*^i}^{\rm GUT}$. In other words,
  $M_*$ cannot be the gravitino mass.  Such kind of
  scenarios should be studied in details as well since there might
  exist the corresponding super-natural SUSY, which is different from our current study.
  For example, if the K\"ahler potential and superpotential are determined from string constructions
  and the moduli and dilaton are stabilized properly,
  the super-natural SUSY can still be valid.
  
\item { Effective Super-Natural SUSY}

  The above definition for super-natural SUSY is very strong, so we can relax the conditions,
  in addition to the above multi F-term SUSY breakings.
  In fact, to solve the SUSY EW fine-tuning problem, we only require that the dimensionful
  parameters at the GUT scale, which have large fine-tuning measures $\Delta_{\rm EENZ}$,
  are related to the fundamental mass parameter  $M_*$~\cite{Ding:2015epa}. Similar to
  the natural SUSY or more effective SUSY  where only the third generation
  sfermions like stops  need  to be light while the first two-generation sfermions
  can be very heavy, we shall call it the effective super-natural SUSY~\cite{Ding:2015epa}.
  Furthermore, for the super-natural SUSY, we can make small
  perturbations to the leading order SUSY breaking soft terms. Obviously, the solution to the
  SUSY EW fine-tuning problem is still valid.
  Interestingly, although it might only change the particle spectra a little bit, it will have
  big effects on DM candidate and DM relic density, which will be studied elsewhere.

\end{itemize}

In this paper, we shall study the scale invariant NMSSM. Because  ${\hat S}$ is an SM singlet,
its scalar mass can only be generated via two-loop effects via RGE running and then is
too small for no-scale supergravity. To solve this problem, we consider the SUSY breaking soft terms
from M-theory on $S^1/Z_2$~\cite{Horava:1996ma, Li:1997sk, Lukas:1997fg, Nilles:1998sx, Li:1998rn}.
As we know, in the weakly coupled heterotic
string theory, there exist two simplified scenarios: (1) The moduli dominant SUSY breaking scenario or
say no-scale scenario~\cite{Cremmer:1983bf, JLDVN} with  $m_0=A=0$; (2) The dilaton  dominant SUSY breaking 
scenario~\cite{ABIM, VSKJL} with $M_{1/2}=-A={\sqrt 3}\, m_0$, which can also escape the above problem.
Generically speaking,  the M-theory on $S^1/Z_2$ seems to be
a better candidate
than the weakly coupled heterotic string theory to explain the low energy phenomenology and high energy
unification of all the fundamental interactions. In particular, we can have
the next-to-leading order corrections to the K\"ahler potential and gauge kinetic functions
and then to the SUSY breaking soft terms as well~\cite{Horava:1996ma, Li:1997sk, Lukas:1997fg, Nilles:1998sx, Li:1998rn}. To parametrize the  next-to-leading order corrections to
the SUSY breaking soft terms, we define~\cite{Li:1998rn}
\begin{eqnarray}
x\equiv {{\alpha ( T + \bar T )} \over\displaystyle { S + \bar S }} 
={{\alpha_{GUT}^{-1} \alpha_H - 1}\over\displaystyle 
{\alpha_{GUT}^{-1} \alpha_H + 1}} ~,~\,
\end{eqnarray}
where $\alpha$ is related to the extra space dimensions and defined
in Refs.~\cite{Lukas:1997fg, Nilles:1998sx, Li:1998rn},
$S$ and $T$ are dilaton and moduli fields, and
$\alpha_{GUT}$ and $\alpha_{H}$ are the gauge couplings at the GUT scale
in the observable and  hidden sectors, respectively.
With the assumption $\alpha_{H} \ge \alpha_{GUT}$ and
 to avoid  $\alpha_{H}$ to be infinity, we obtain
\begin{eqnarray}
0 \leq x \leq 1 ~.~ \,
\end{eqnarray}
In the super-natural SUSY, there exists one and only one moduli or dilaton field whose F-term breaks
the SUSY. Thus, we will consider 
the moduli dominant SUSY breaking (MDSB) and
dilaton dominant SUSY breaking (DDSB) scenarios as follows~\cite{Li:1998rn} \\

(I) Moduli dominant SUSY breaking scenario. The SUSY breaking soft terms are:
\begin{eqnarray}
m_0&=& {x\over\displaystyle {3+x}} M_{3/2} ~,~\,
\label{mod_m0}
\end{eqnarray}
\begin{eqnarray}
M_{1/2}&=&{x\over\displaystyle {1+x}} M_{3/2} ~,~ \,
\label{mod_mhf}
\end{eqnarray}
\begin{eqnarray}
A&=&- {{ 3x } \over\displaystyle {3+x}} M_{3/2} ~.~\,
\label{mod_A0}
\end{eqnarray}
\\

(II) Dilaton dominant SUSY breaking scenario. The SUSY breaking soft terms are
\begin{eqnarray}
m_0^2&=& M_{3/2}^2 - {{ 3  M_{3/2}^2} \over\displaystyle 
{(3+x)^2}} ~x ~(6+x) ~,~  \,
\label{dia_m0}
\end{eqnarray}
\begin{eqnarray}
M_{1/2}&=&{{\sqrt 3  M_{3/2}} \over\displaystyle {1+x}} ~,~\,
\label{dia_mhf}
\end{eqnarray}
\begin{eqnarray}
A&=&- {{\sqrt 3  M_{3/2}} \over\displaystyle 
{3+x }} (3-2x) ~.~\,
\label{dia_A0}
\end{eqnarray}
From the requirement $m_0^2 > 0$, we obtain that
$x$ is smaller than about 0.67423. Choosing $x=0$, we obtain the relation
$M_{1/2}=-A={\sqrt 3}\,m_0$ in the weakly coupled heterotic string theory.

In short, in the  M-theory motivated CNMSSM with MDSB and DDSB scenarios,
all the SUSY breaking soft
mass parameters have fixed relations with gravitino mass $M_{3/2}$ after
the moduli stabilization which determine $\alpha_{GUT}$ and  $\alpha_{H}$ as well.
According to the super-natural SUSY, the fine-tuning measure
is automatically of order one. In other words, such kind of models are super-natural,
even though their particle spectra are heavy.

\section{Phenomenological Constraints}
\label{sec:scan}

We use the publicly available code MicrOmegas3.5.5~\cite{Belanger:2013oya} for random scans over the parameters space 
given in Eq.~(\ref{input_param_range}). 
We consider $\mu > 0$, $m_t = 173.3\, {\rm GeV}$~\cite{ATLAS:2014wva} and $m_b^{\overline{DR}}(M_{\rm Z})=2.83$ GeV. 
We do random scans on the following parameter space:
\begin{align}
0\leq  x  \leq 1~,~ \nonumber \\
0\leq    M_{3/2}   \leq 5 \, \rm{TeV} ~,~\nonumber \\ 
2\leq  \tan\beta  \leq 60~,~ \nonumber \\
0\leq  \lambda  \leq 0.7~.~
\label{input_param_range}
\end{align}



After collecting the data, we require the following bounds 
on sparticle masses from the LEP2 experiment
\begin{eqnarray} 
m_{\tilde t_1},m_{\tilde b_1} \gtrsim 100 \; {\rm GeV} ~,~\\
m_{\tilde \tau_1} \gtrsim 105 \; {\rm GeV}  ~,~\\
m_{\tilde \chi_{1}^{\pm}} \gtrsim 103 \; {\rm GeV}~.~
\end{eqnarray}
We implement the following B-physics constraints
\begin{eqnarray}
1.6\times 10^{-9} \leq{\rm BR}(B_s \rightarrow \mu^+ \mu^-) 
  \leq 4.2 \times10^{-9} \;(2\sigma)~~&\cite{CMS:2014xfa} ~,~& 
\\ 
2.99 \times 10^{-4} \leq 
  {\rm BR}(b \rightarrow s \gamma) 
  \leq 3.87 \times 10^{-4} \; (2\sigma)~~&\cite{Amhis:2014hma}~,~ &  
\\
0.70\times 10^{-4} \leq {\rm BR}(B_u\rightarrow\tau \nu_{\tau})
        \leq 1.5 \times 10^{-4} \; (2\sigma)~~&\cite{Amhis:2014hma}~.~&  
\end {eqnarray}
In addition, we impose the following bounds from the LHC SUSY searches as well
\begin{eqnarray}
m_{H_{1}}  = 123-127~{\rm GeV}~~&\cite{ATLAS, CMS}& ~,~ \\ 
m_{\tilde{g}}\gtrsim 1.7 \, {\rm TeV}\ ({\rm for}\ m_{\tilde{g}}\sim m_{\tilde{q}}) &
\cite{Aad:2014wea,Chatrchyan:2013wxa}~,~\\
m_{\tilde{g}}\gtrsim 1.3 \, {\rm TeV}\ ({\rm for}\ m_{\tilde{g}}\ll m_{\tilde{q}}) &
\cite{Aad:2014wea,Chatrchyan:2013wxa}~.~\\
\end{eqnarray}
For the muon anomalous magnetic moment $a_{\mu}$, we require that the benchmark
points be at least as consistent with the data as the SM.
\begin{figure}[htp!]
\centering

\subfigure{
\includegraphics[totalheight=10.5cm,width=12.cm]{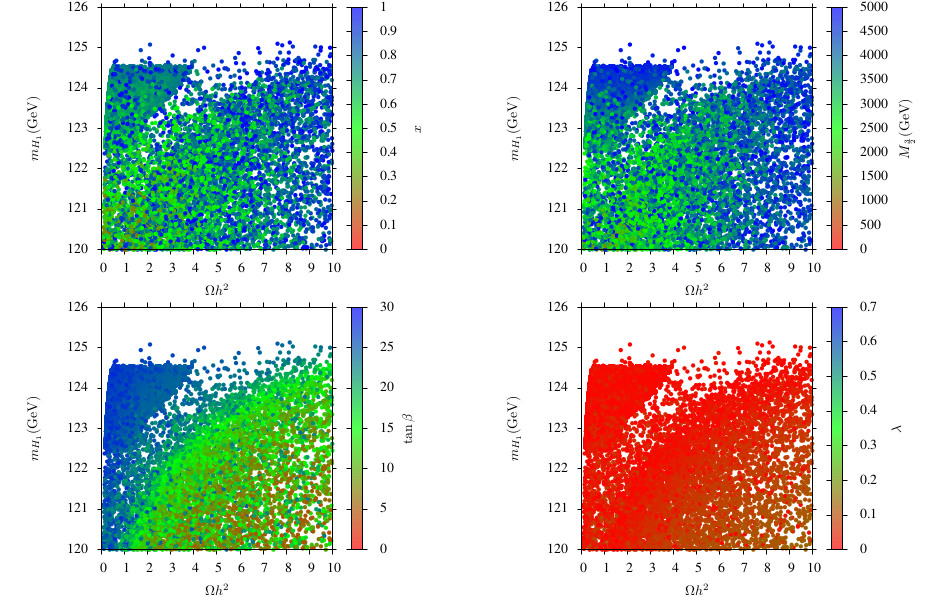}
}
\caption{Plots in $\Omega h^{2}-m_{H_{1}}$ plane for moduli dominant SUSY breaking scenario. The ranges of input parameters given in
Eq.~(\ref{input_param_range}) are shown in vertical bars.
}
\label{mhoh2mod}
\end{figure}
\begin{figure}[htp!]
\centering
\subfigure{
\includegraphics[totalheight=5.5cm,width=7.cm]{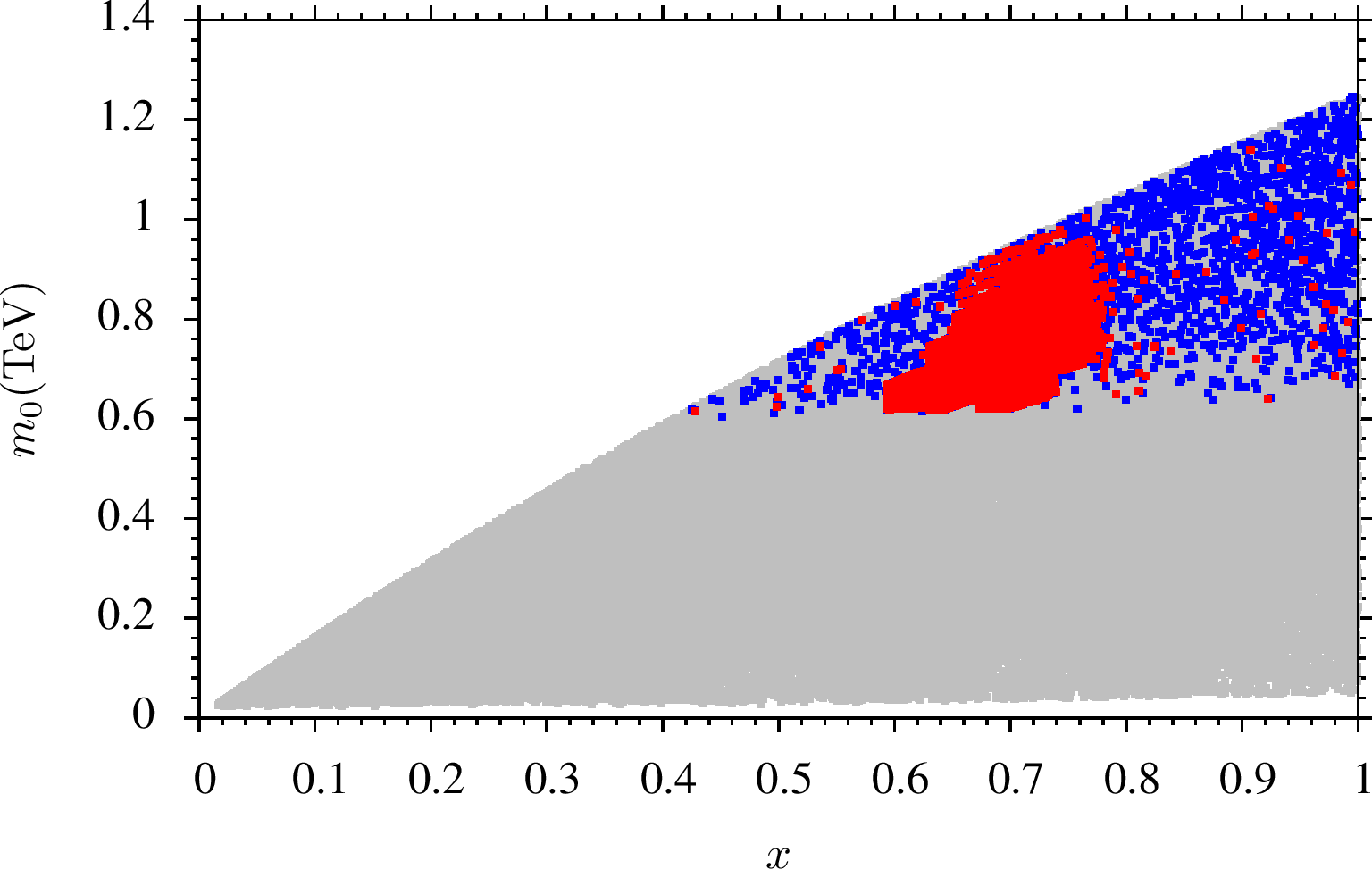}
}
\subfigure{
\includegraphics[totalheight=5.5cm,width=7.cm]{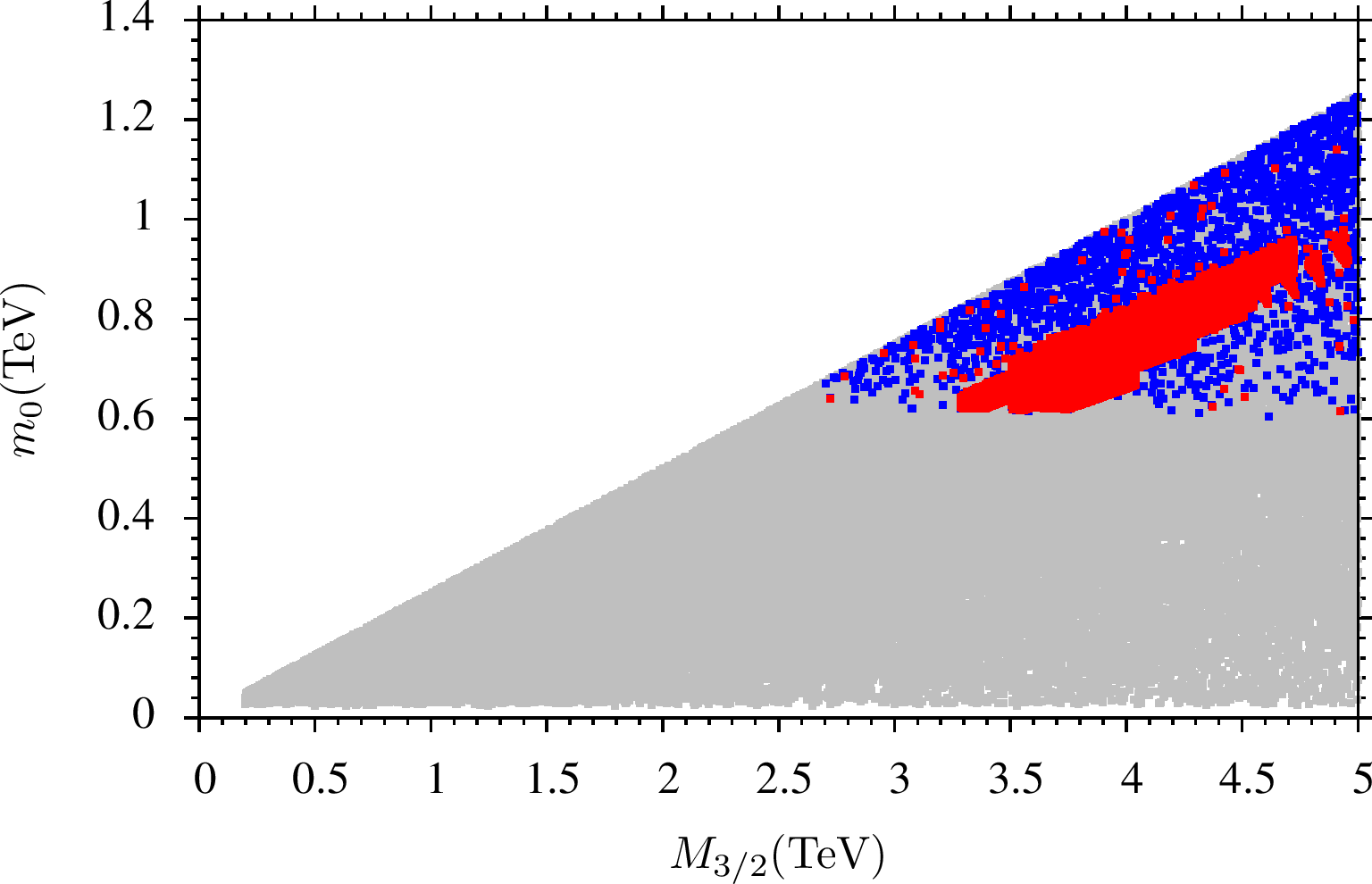}
}
\subfigure{
\includegraphics[totalheight=5.5cm,width=7.cm]{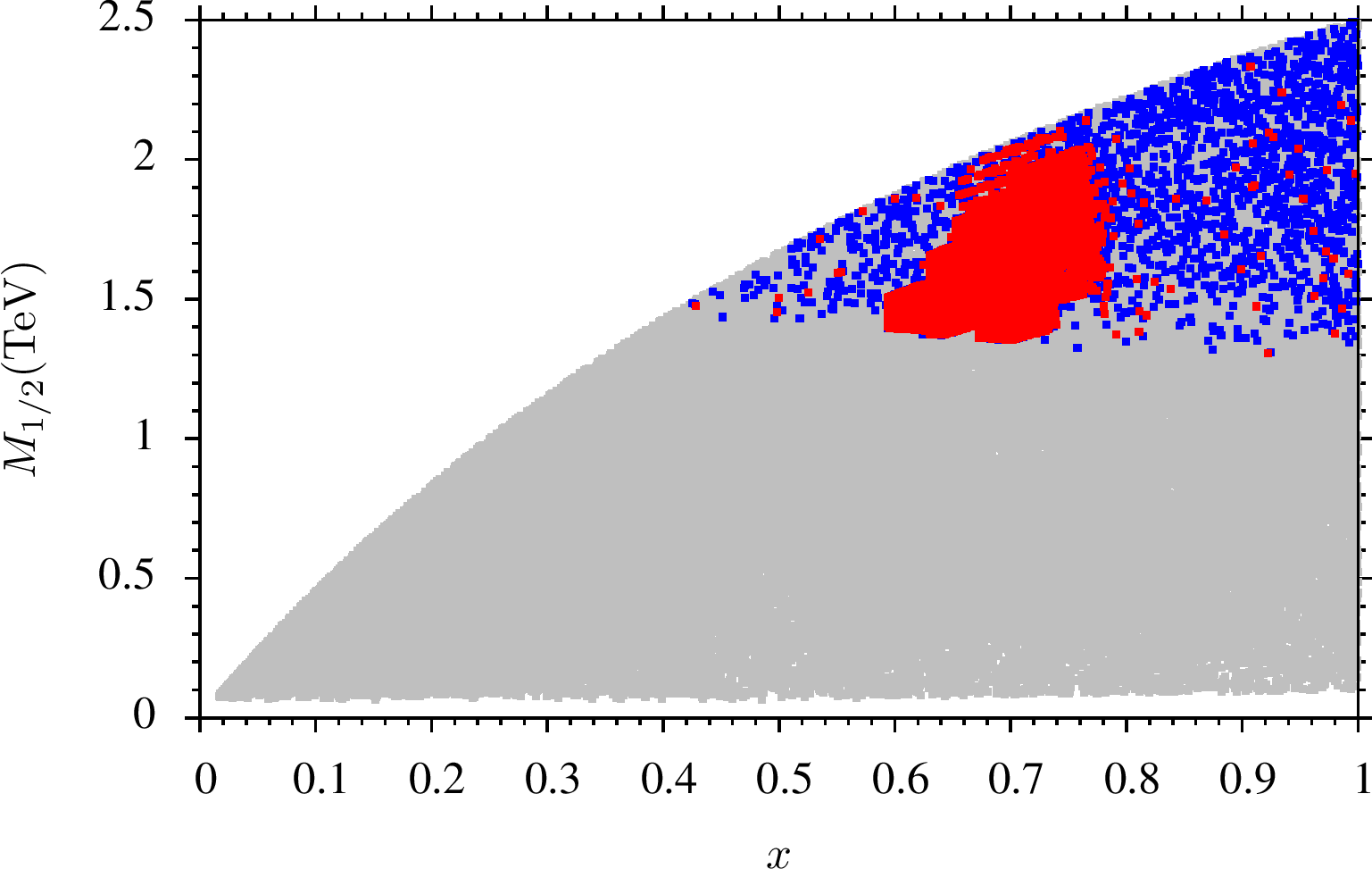}
}
\subfigure{
\includegraphics[totalheight=5.5cm,width=7.cm]{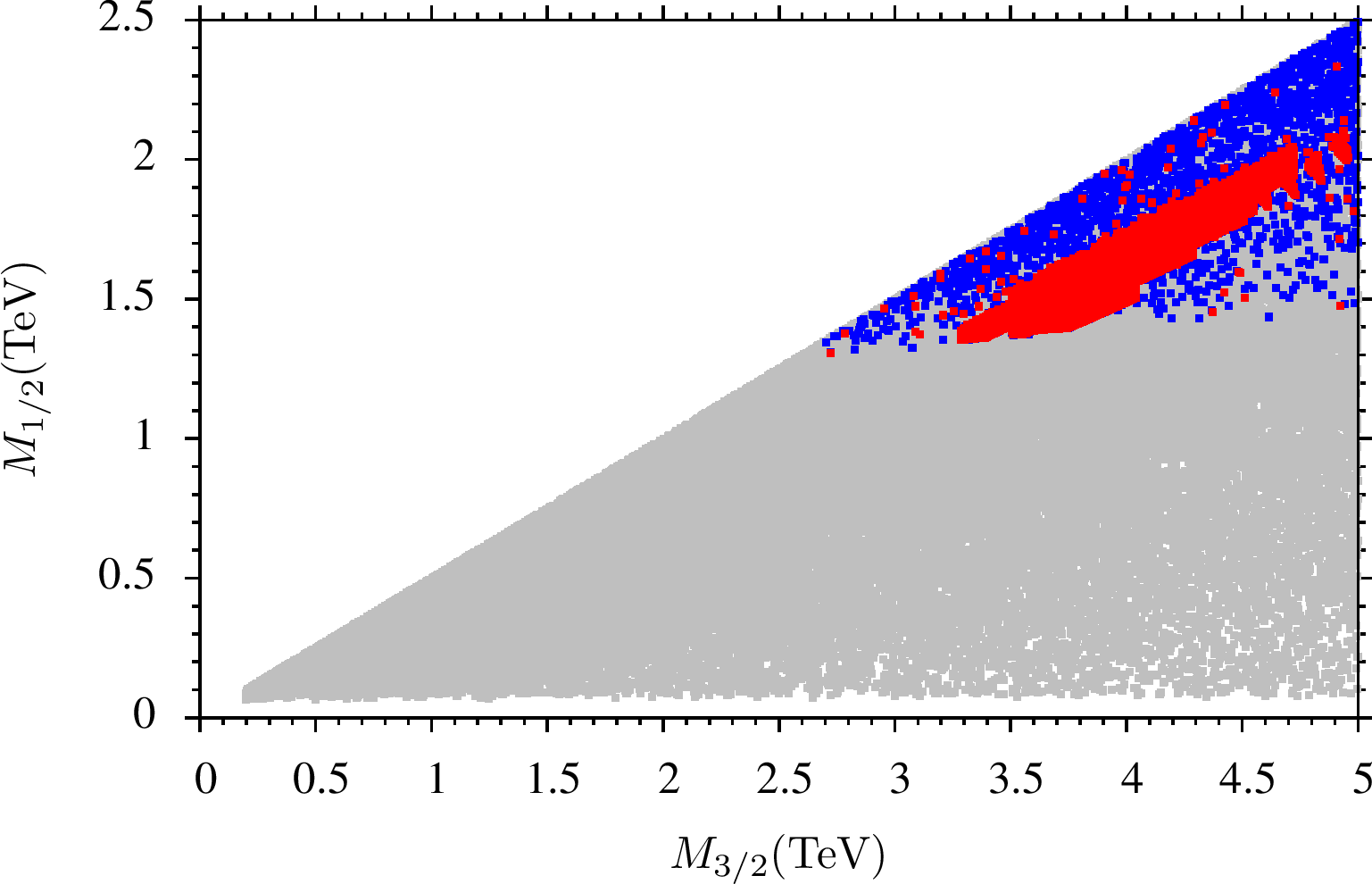}
}
\subfigure{
\includegraphics[totalheight=5.5cm,width=7.cm]{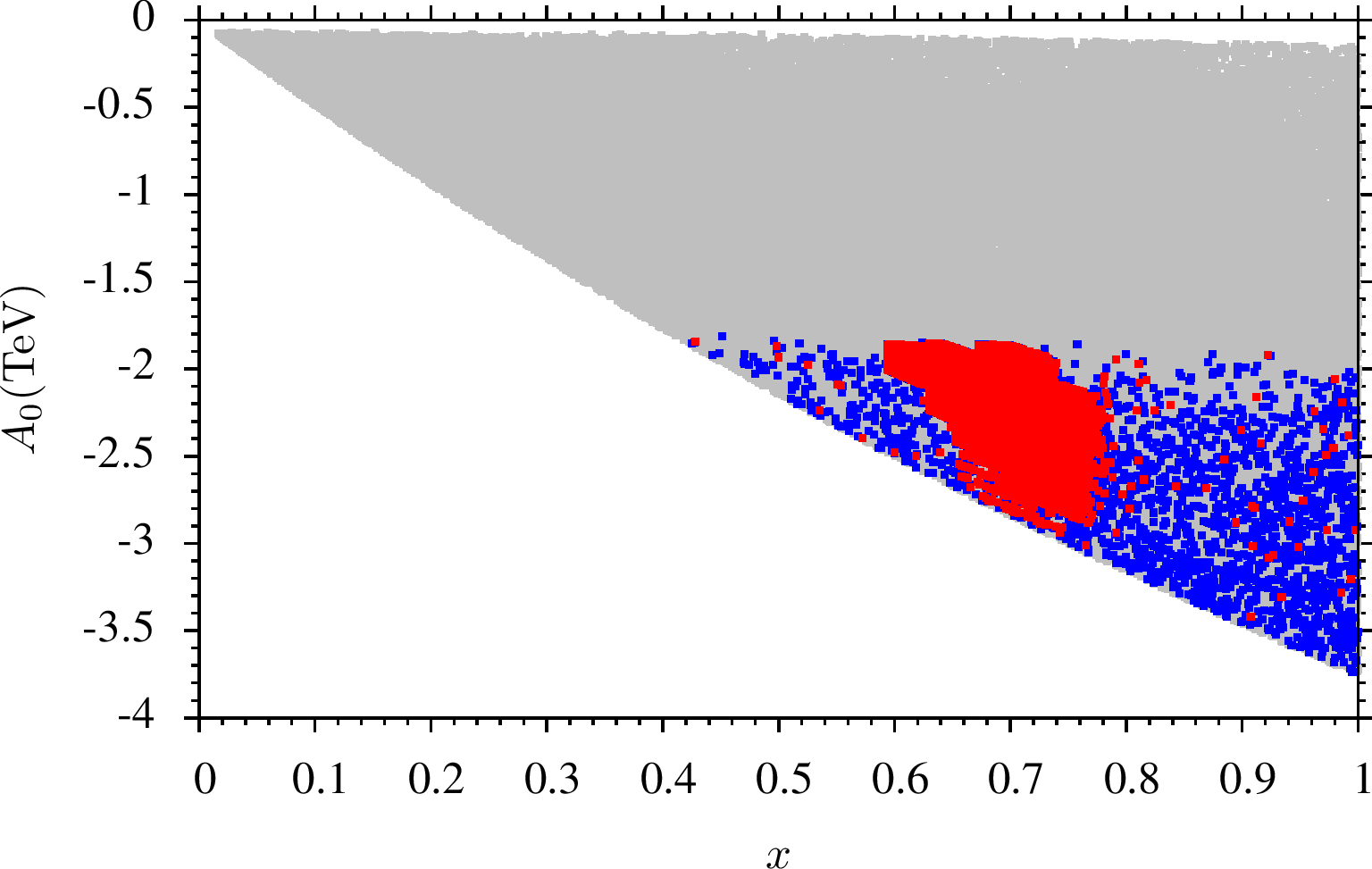}
}
\subfigure{
\includegraphics[totalheight=5.5cm,width=7.cm]{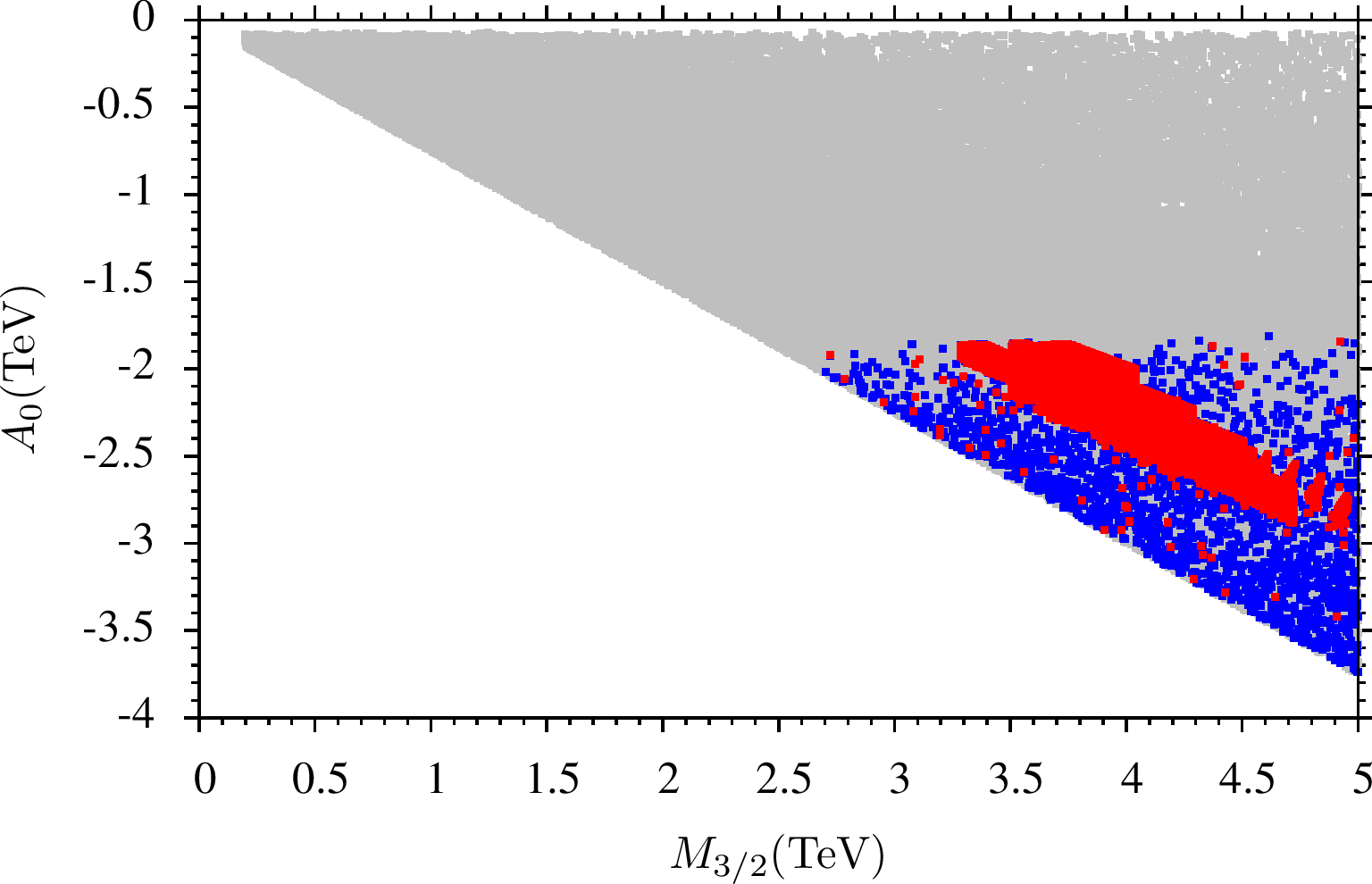}
}

\caption{Plots of $m_{0}$, $M_{1/2}$ and $A_{0}$ as functions of $x$ and $M_{3/2}$ for moduli dominant SUSY breaking scenario.
Gray points satisfy successful radiative electroweak symmetry breaking. Blue points form a subset of gray points and satisfy particle mass bounds, B-physics bounds and Higgs mass bounds. Red points further form a subset of blue points and satisfy $\Omega h^2 \lesssim$ 1.
}
\label{fundamod}
\end{figure}
\section{Numerical Results}
\label{results}
In the following, we will present our results for moduli dominant SUSY breaking (MDSB) and
dilaton dominant SUSY breaking (DDSB) scenarios.
\subsection{Moduli Dominant SUSY Breaking Scenario}
In Fig.~\ref{mhoh2mod}, we present our graphs in $\Omega h^{2}-m_{H_{1}}$ plane. The ranges of input parameters given in 
Eq.~(\ref{input_param_range}) are also displayed in vertical bars. In this figure, we show the CP-even Higgs mass $m_{H_{1}}$ larger than 120 GeV
and neutralino dark mater relic density $\Omega h^2$ between 0 and 10 to give a broader picture.
These plots show that in order to have $m_{H_{1}}$ in the range [123,~127] GeV,
the input parameters $x$ should be greater than 0.2 and gravitino mass $M_{3/2}$ should be greater than 2 TeV.
From the left-bottom panel,
we see that the Higgs mass constraint pushes $\tan\beta \gtrsim$ 5, and demanding $\Omega h^2 \lesssim$ 1
further pushes $\tan\beta$ values above 20. In the right bottom  panel, we see that all the points have $\lambda \lesssim$ 0.2.
Moreover, these plots show that the Higgs mass constraint alone
severely restricts the input parameter space, and we will study this scenario in more detail below. 
\begin{figure}[htp!]
\centering
\subfigure{
\includegraphics[totalheight=5.5cm,width=7.cm]{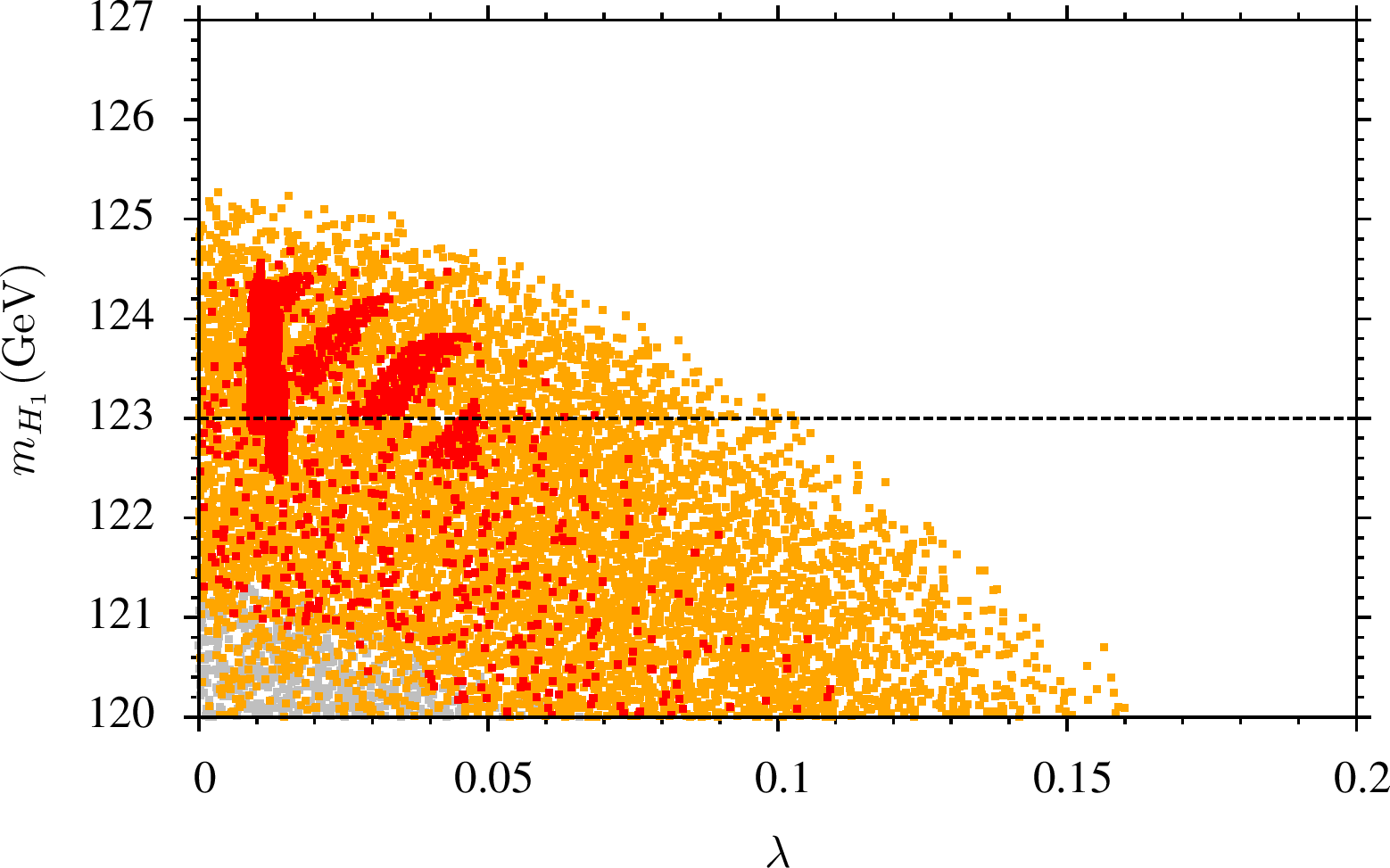}
}
\subfigure{
\includegraphics[totalheight=5.5cm,width=7.cm]{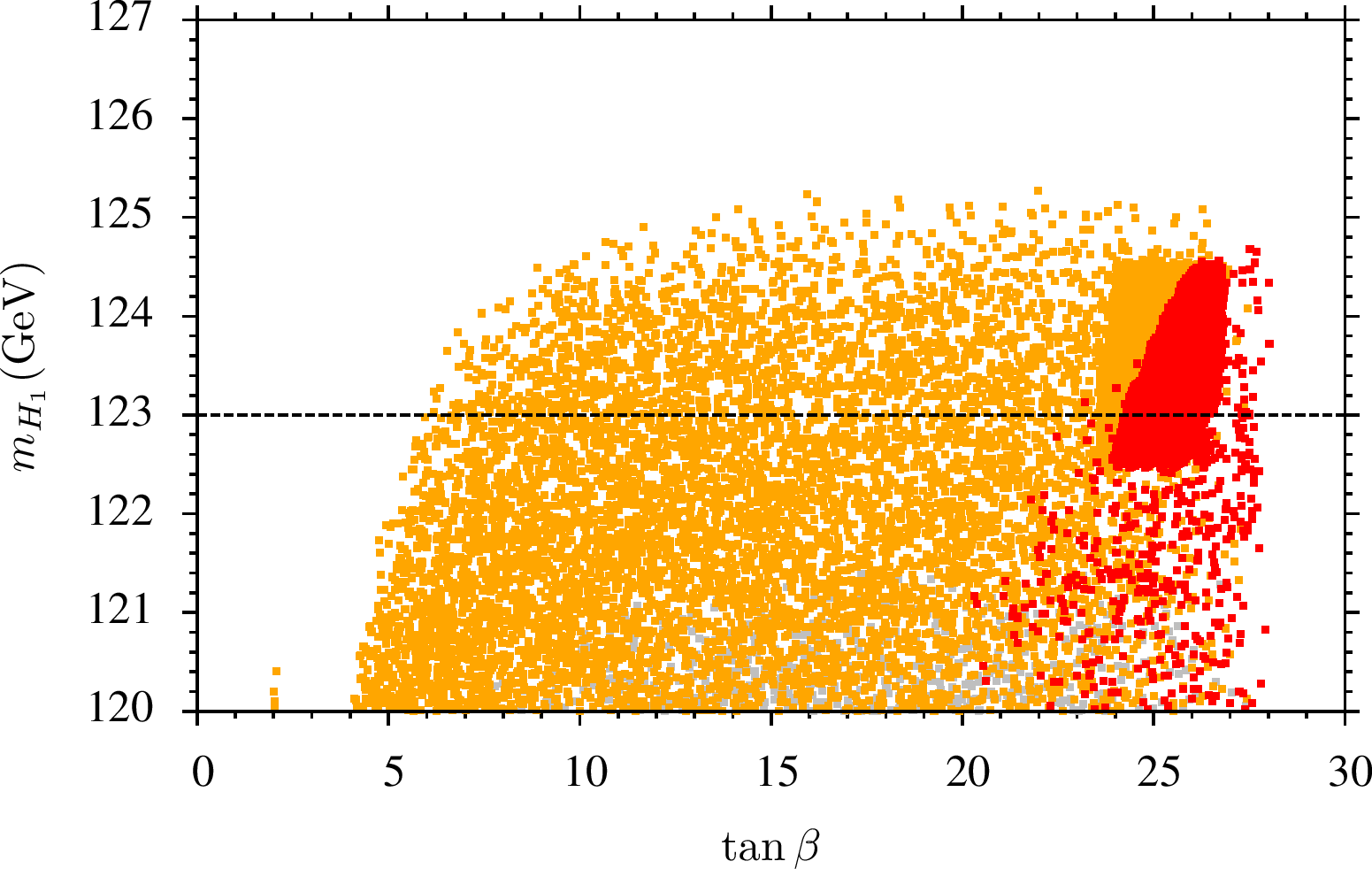}
}

\caption{
Plots in $\lambda-m_{H_{1}}$ and $\tan\beta$-$m_{H_{1}}$ planes for moduli dominant SUSY breaking scenario.
Gray points satisfy successful radiative electroweak symmetry breaking.
Orange points, which form a subset of gray points, satisfy particle mass bounds and B-physics bounds.
We do not apply Higgs mass bounds here. Red points further form a subset of orange points and satisfy $\Omega h^2 \lesssim$ 1.
 }
\label{lammh_mod}
\end{figure}
\begin{figure}[htp!]
\centering
\subfigure{
\includegraphics[totalheight=5.5cm,width=7.cm]{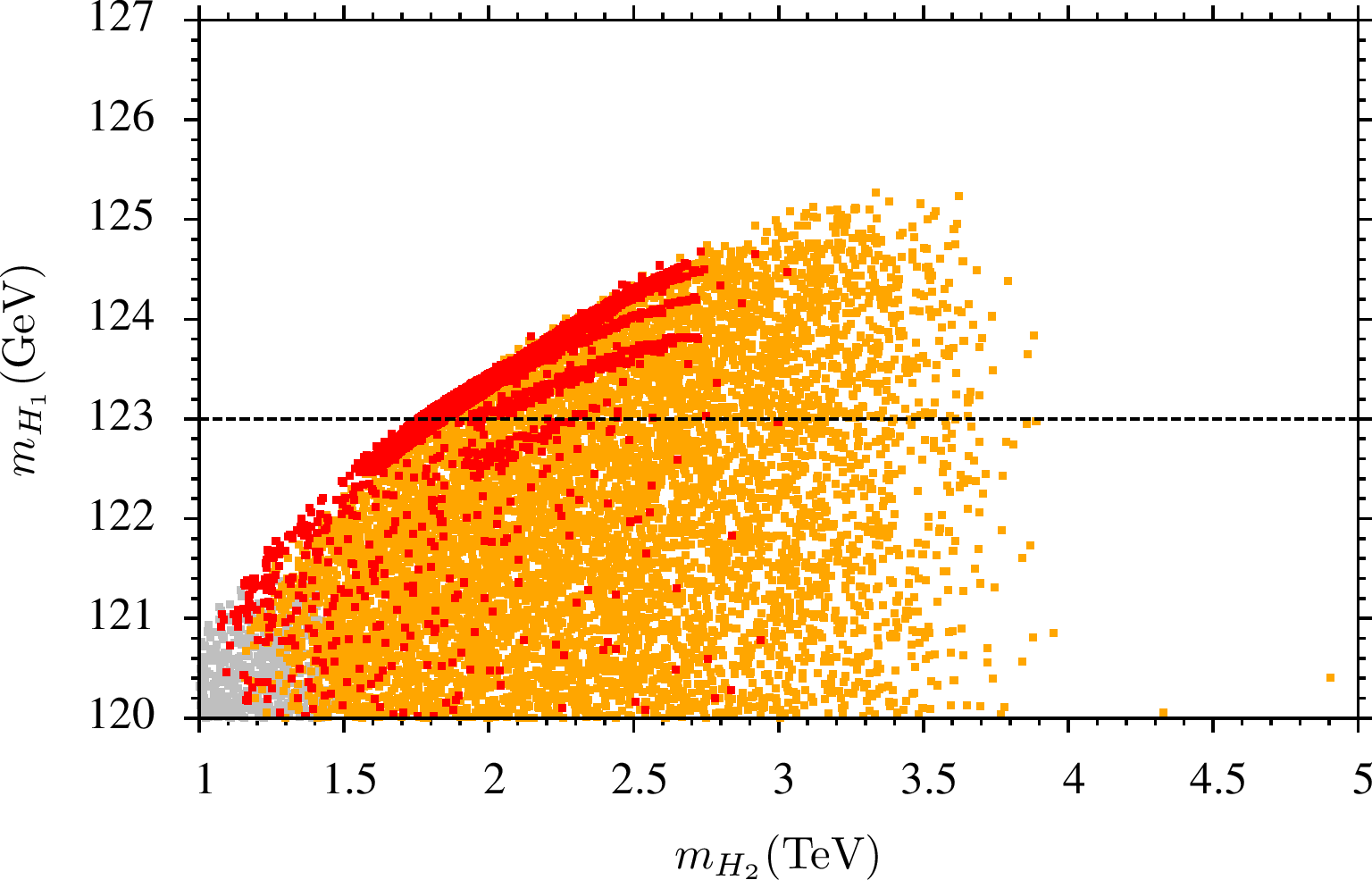}
}
\subfigure{
\includegraphics[totalheight=5.5cm,width=7.cm]{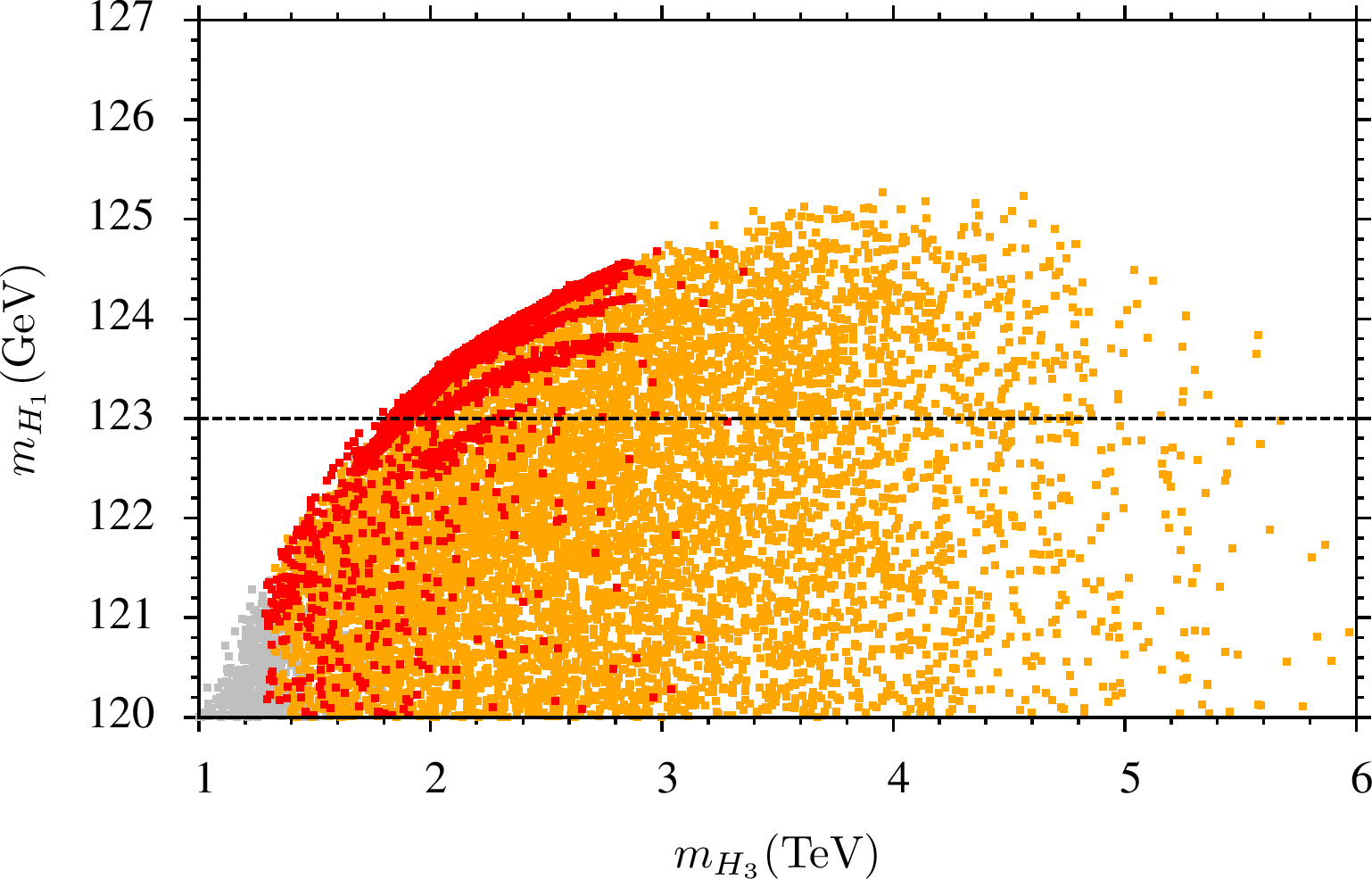}
}
\subfigure{
\includegraphics[totalheight=5.5cm,width=7.cm]{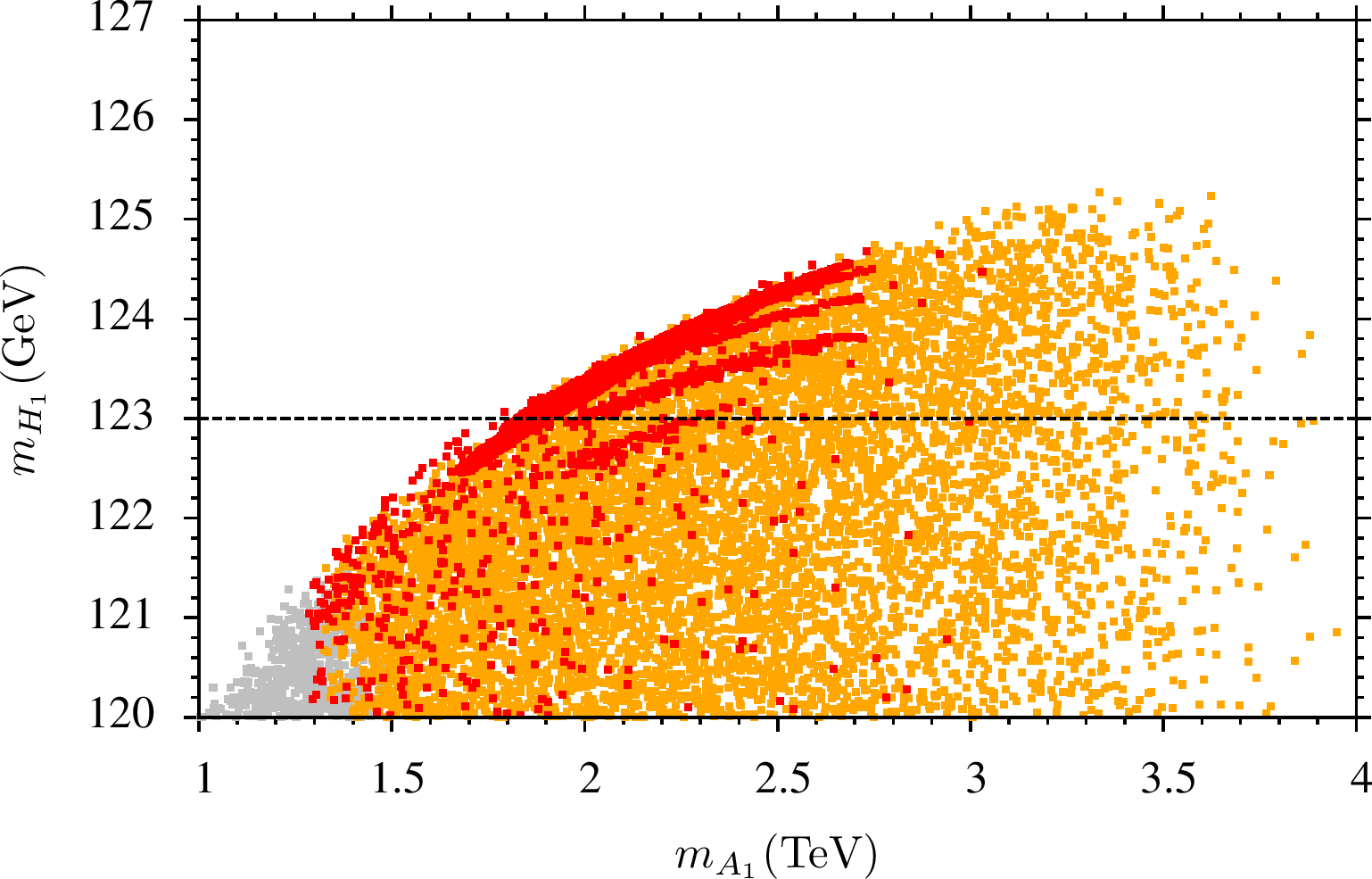}
}
\subfigure{
\includegraphics[totalheight=5.5cm,width=7.cm]{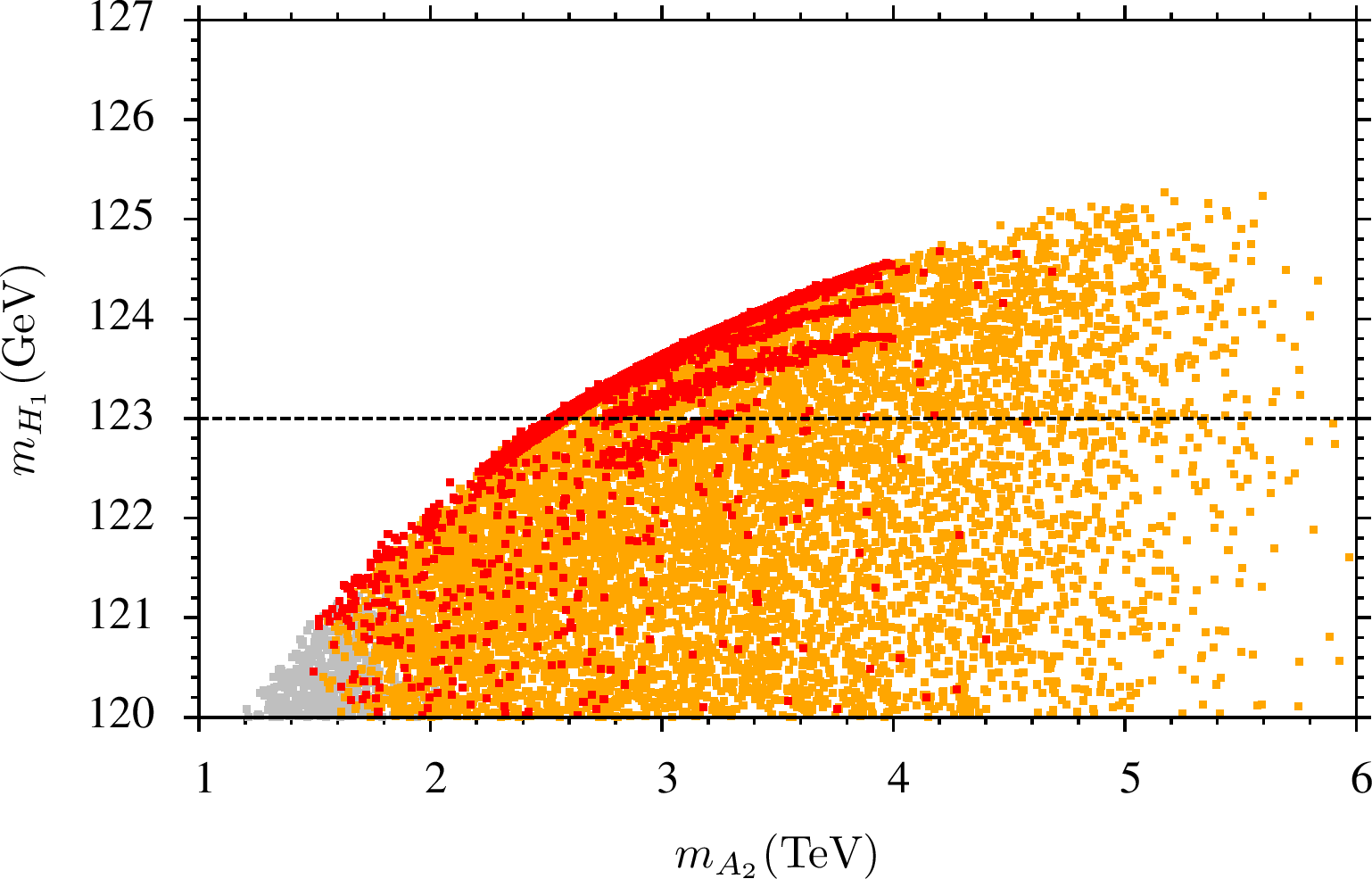}
}
\subfigure{
\includegraphics[totalheight=5.5cm,width=7.cm]{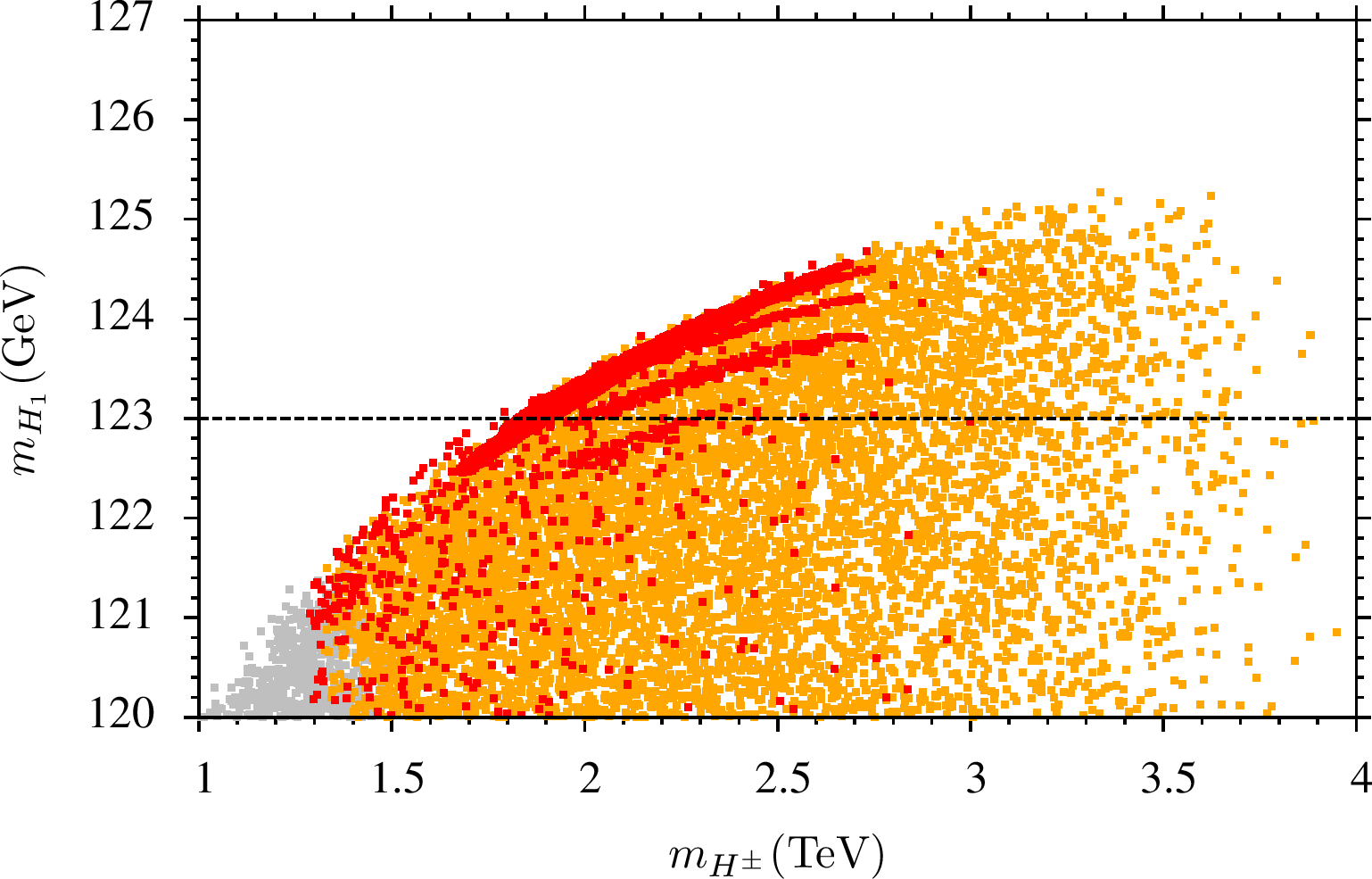}
}

\caption{
Plots in  $m_{H_{2}}-m_{H_{1}}$, $m_{H_{3}}-m_{H_{1}}$, $m_{A_{1}}-m_{H_{1}}$,  and
$m_{A_{2}}-m_{H_{1}}$ and $m_{H^{\pm}}-m_{H_{1}}$ planes for moduli dominant SUSY breaking scenario. The color coding is the same as in Fig.~\ref{lammh_mod}.
 }
\label{Higgs_mod}
\end{figure}
\begin{figure}[htp!]
\centering
\subfigure{
\includegraphics[totalheight=5.5cm,width=7.cm]{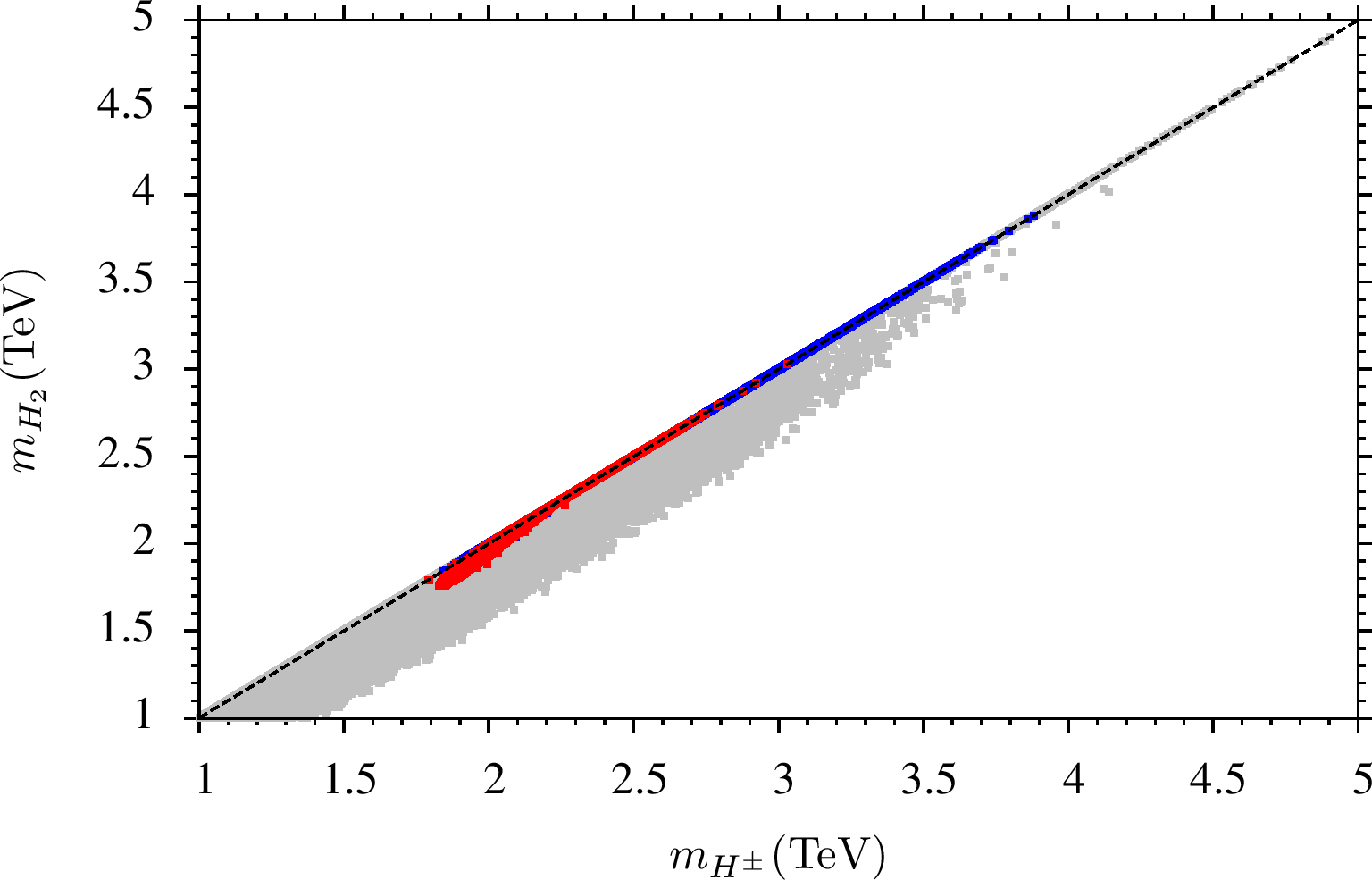}
}
\subfigure{
\includegraphics[totalheight=5.5cm,width=7.cm]{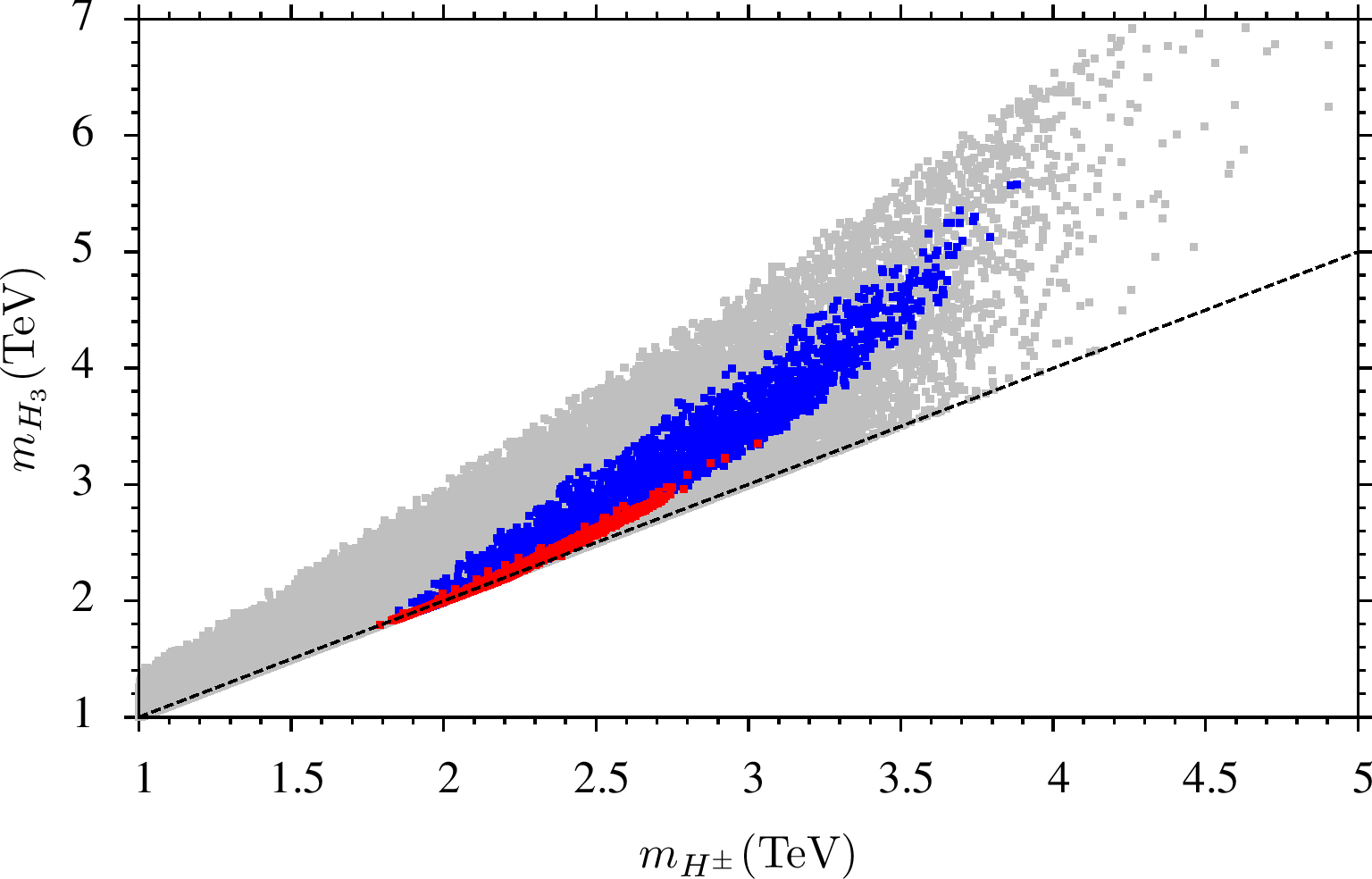}
}
\subfigure{
\includegraphics[totalheight=5.5cm,width=7.cm]{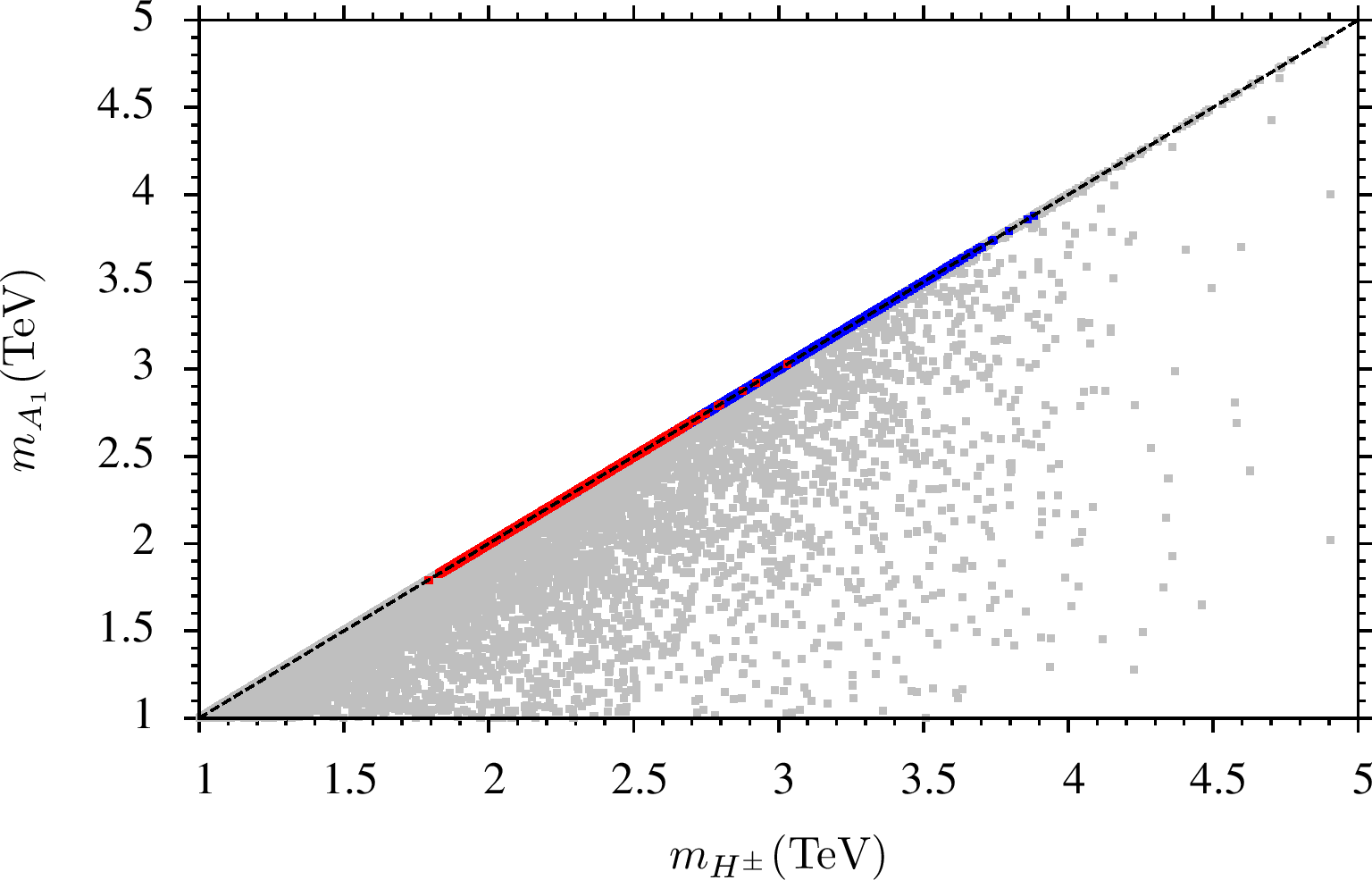}
}
\subfigure{
\includegraphics[totalheight=5.5cm,width=7.cm]{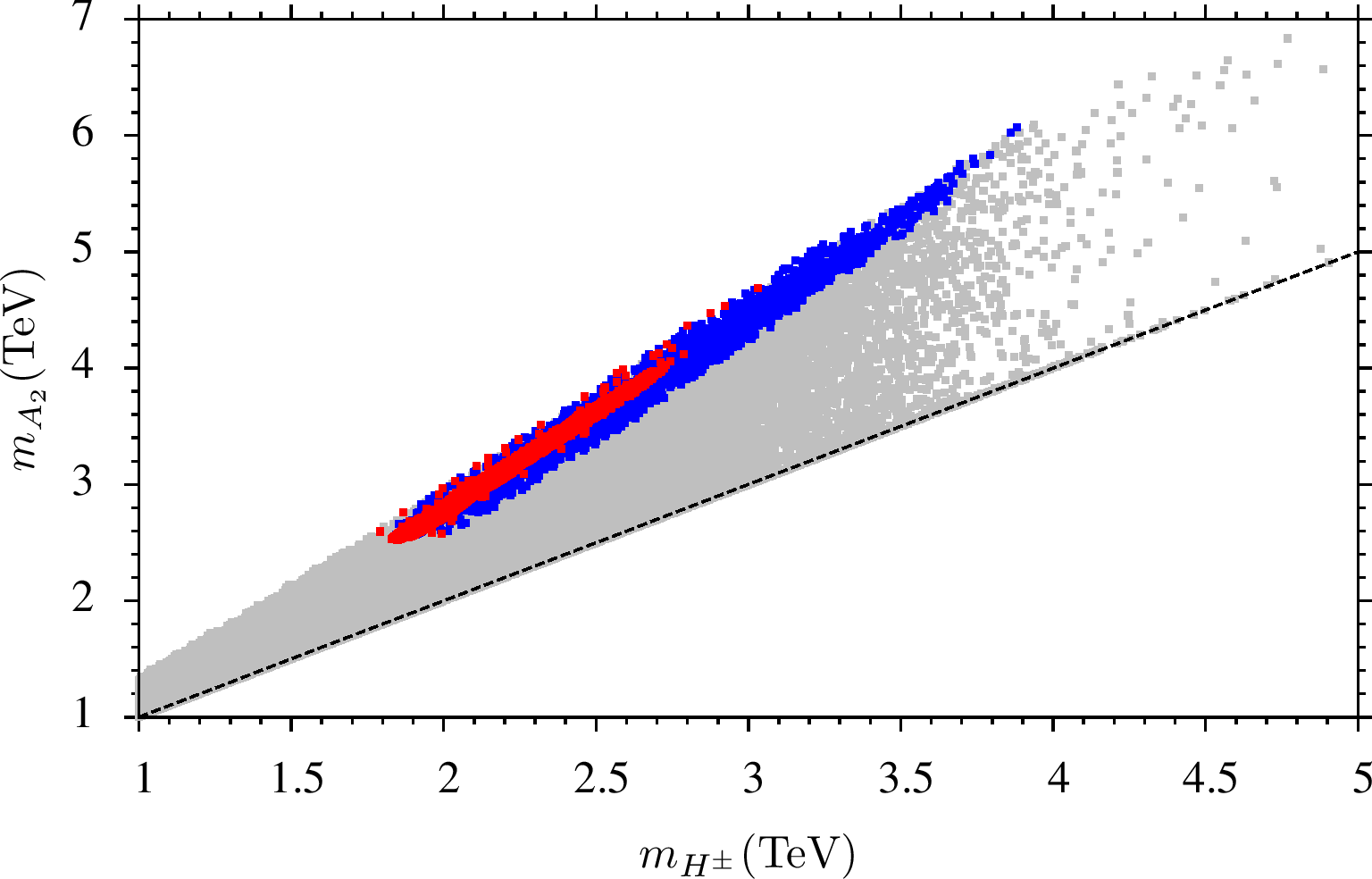}
}
\caption{
Plots in  $m_{H^{\pm}}-m_{H_{2}}$, $m_{H^{\pm}}-m_{H_{3}}$, $m_{H^{\pm}}-m_{A_{1}}$,  and
$m_{H^{\pm}}-m_{A_{2}}$ planes for moduli dominant SUSY breaking scenario. The color coding is the same as in Fig.~\ref{fundamod}.
 }
\label{4mh_mod}
\end{figure}
\begin{figure}[htp!]
\centering
\subfigure{
\includegraphics[totalheight=5.5cm,width=7.cm]{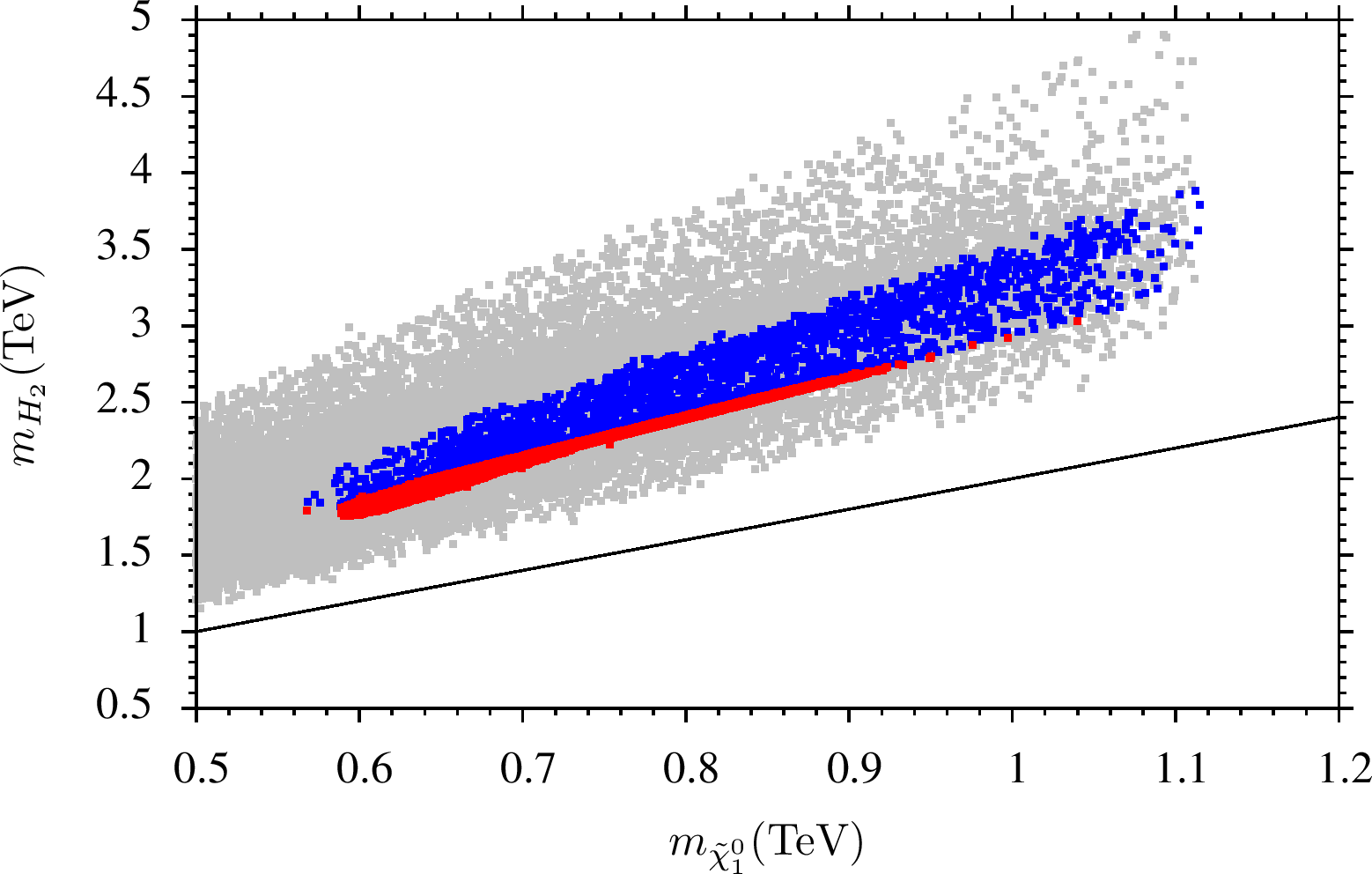}
}
\subfigure{
\includegraphics[totalheight=5.5cm,width=7.cm]{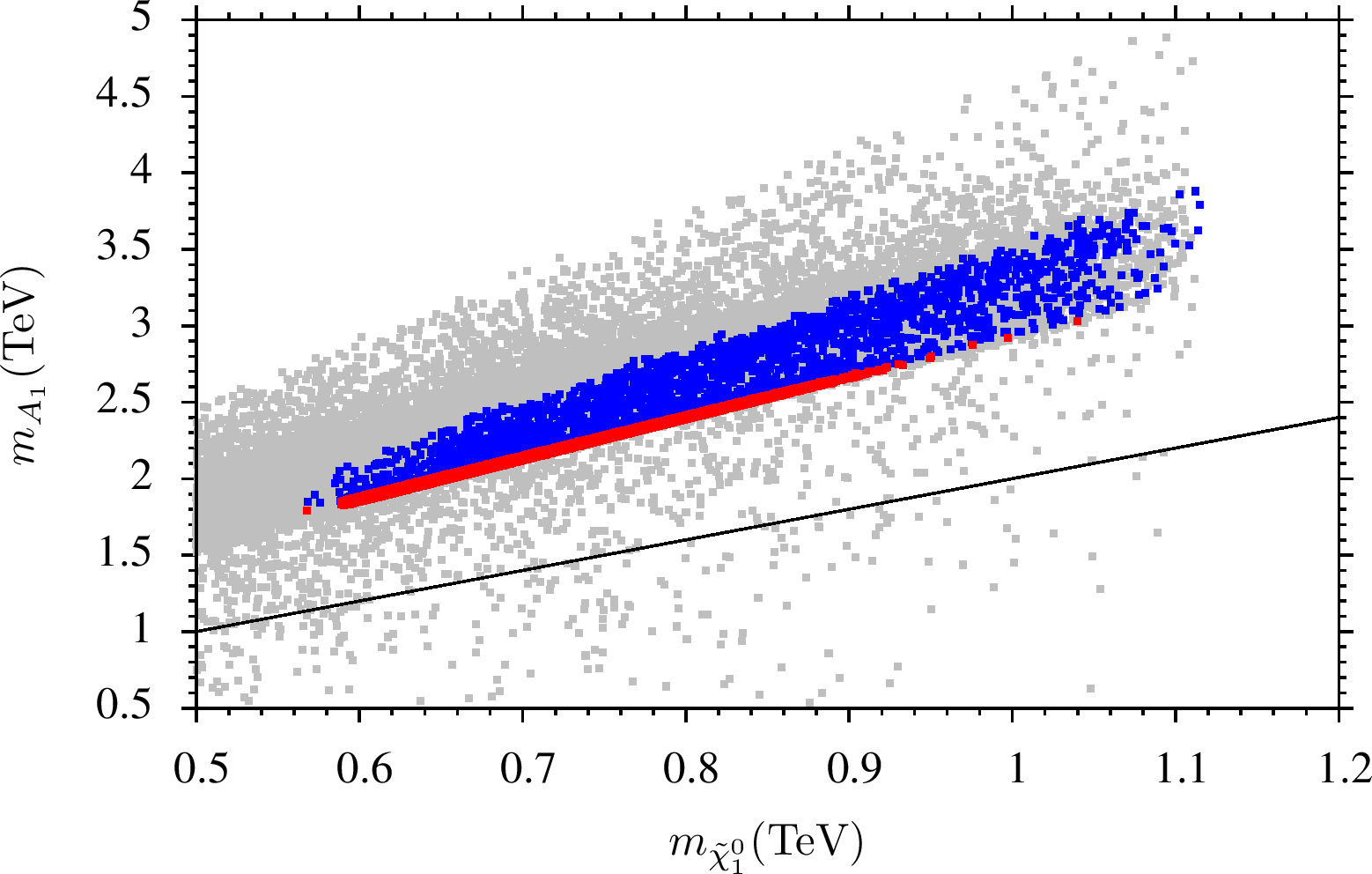}
}
\subfigure{
\includegraphics[totalheight=5.5cm,width=7.cm]{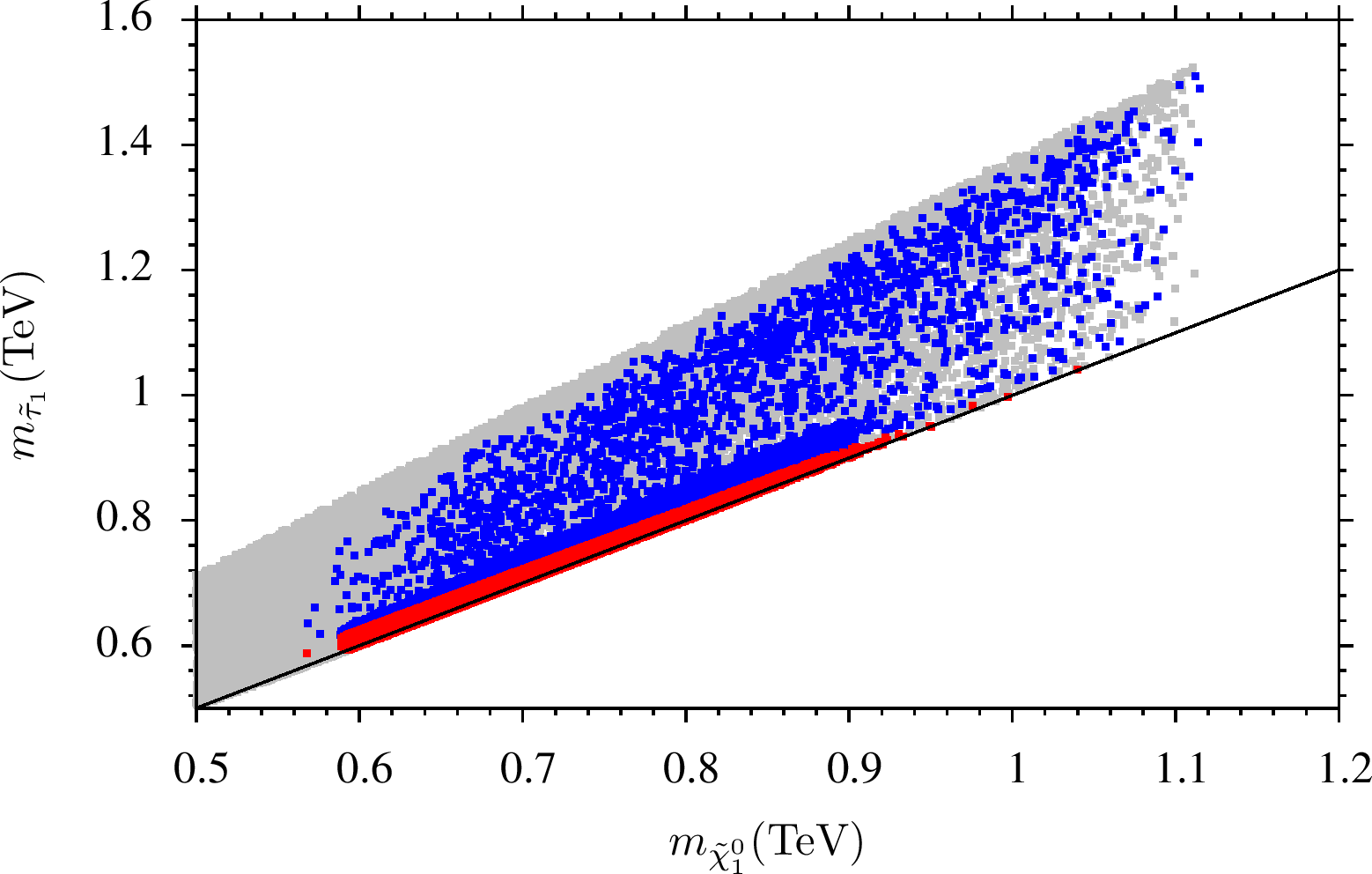}
}

\caption{
Plots in $m_{\tilde \chi_{1}^{0}}-m_{H_2}$, $m_{\tilde \chi_{1}^{0}}-m_{A_{1}}$ and $m_{\tilde \chi_{1}^{0}}-m_{\tilde \tau_{1}}$ planes for moduli dominant SUSY breaking scenario.
The color coding is the same as in Fig.~\ref{fundamod}. }
\label{spec_mod}
\end{figure}

Since the SUSY breaking soft terms $m_{0}$, $M_{1/2}$ and $A_{0}$ are
functions of input parameters $x$ and $M_{3/2}$, we calculate them using Eqs.~(\ref{mod_m0})-(\ref{mod_A0})
and show our results in Fig.~\ref{fundamod}. In these plots, gray points satisfy successful
radiative electroweak symmetry breaking. Blue points, which form a subset of gray points,
satisfy particle mass bounds, B-physics bounds and
Higgs mass bounds. We further constrain the parameter space by demanding $\Omega h^2 \lesssim$ 1, which
is shown by red points. As we have
already observed that the light CP-even Higgs mass ranges $[123,~127]$ GeV constrain
the input parameter space a lot. This constraint already
makes the spectra too heavy so that the viable points satisfy various above mentioned bounds.
From the first row of Fig.~\ref{fundamod}, we see that
the minimum value of $m_{0}$ consistent with all the constraints is about 0.6 TeV, corresponding to
$x\approx$ 0.4 and $M_{3/2} \approx$ 2.5 TeV while the maximum allowed values of $m_{0}$ is about 1.3 TeV.
The plots in the second row display dependence of $M_{1/2}$ on $x$ and $M_{3/2}$.
Here, we see that the minimum and maximum values of $M_{1/2}$ consistent with all the above mentioned constraints
are about 1.4 TeV and 2.5 TeV, respectively.
Finally, the plots in the third row depict that the allowed ranges of universal trilinear scalar coupling $A_{0}$ are 
$[-4,~-2]$ GeV. This indicates that the scalar top quarks are not highly mixed. So in order to have
the CP-even SM-like Higgs boson mass around 125 GeV, we need
heavy squarks/stops. The large values of $m_{0}$ and $M_{1/2}$ indicate heavy spectra at low energy.
One can write the EW-scale masses of squarks and 
sleptons in terms of $m_0$ and $M_{1/2}$ as follows~\cite{Baer:2015fsa}
\begin{eqnarray}
m_{\tilde q}^2 &\simeq& m_{0}^2 + (5 - 6) M_{1/2}^2, \label{squark} \\  
m_{\tilde e_{L}}^2 &\simeq& m_{0}^2 + 0.5 M_{1/2}^2, \\ \label{lslepton}
m_{\tilde e_{R}}^2 &\simeq& m_{0}^2 +  0.15 M_{1/2}^2. \label{stau}
\end{eqnarray}
By plugging in the minimum values of $m_{0}$ and $M_{1/2}$ in the above semi-analytical expressions
we see that the squarks may be around 3 TeV,
the left-handed sleptons can be around 1 TeV while the right-handed slepton can be relatively light around 650 GeV.
We will see that this indeed is the case and the right-handed staus are light.
One can also observe this trend in Table~\ref{table1}. 

In order to have the SM-like Higgs mass around 125 GeV in the NMSSN,
the Yukawa coupling $\lambda$ also plays a very crucial role.
In the unconstrained NMSSM, one needs
large $\lambda$ values (but less than 0.7 to avoid the  Landau pole problem
in GUT models) and small $\tan\beta \lesssim$ 10. However, in the
CNMSSM, the requirement is almost reversed. One usually needs small values of
$\lambda$ and large values of $\tan\beta$~\cite{Fowlie:2014faa}. This can be seen
in the Fig.~\ref{lammh_mod} where we display plots in $\lambda-m_{H_1}$ plane
(left panel ) and $\tan\beta-m_{H_1}$ plane (right panel). In these plots,
gray points satisfy the successful radiative electroweak symmetry breaking,
orange points satisfy all the above mentioned constraints except the Higgs mass constraints,
and red points further form a subset of orange points and satisfy the bound $\Omega h^{2}\lesssim$ 1 from
supercritical string cosmology~\cite{Lahanas:2006xv}.
The horizontal black line indicates the lower bounds on Higgs mass of 123 GeV. Here,
we see that orange points with $m_{H_{1}}\approx$ 123 GeV have maximum value of $\lambda$ is about 0.1. The maximum
value of $\lambda$  further
shrinks to  about 0.08 when we demand $\Omega h^{2}\lesssim$ 1. Also,  the plot in $\tan\beta-m_{H_1}$ plane shows that
the allowed range of $\tan\beta$ is $[5,28]$.

In the NMSSM, due to the presence of an additional gauge singlet $\hat{S}$, we have an extra CP-even Higgs $H_3$
and a pseudo-scalar $A_2$ as compared to MSSM. The approximate tree-level Higgs boson masses in the NMSSM are
given in Ref.~\cite{Miller:2003ay}. From there we see that these masses are 
proportional to $v_{S}$. From Fig.~\ref{lammh_mod}, we see that the minimum value
of $\lambda$ consistent with all constraints is about 0.1. Note that
$\mu_{eff}\equiv \lambda v_{S}$ and $\mu_{eff}\approx \chi_{1}^{\pm} \gtrsim$ 100 GeV from the LEP bound on chargino mass,
we obtain $v_{S}\gtrsim$ 1 TeV.
Such a large value of $v_{S}$ in turn implies heavy masses of Higgs bosons. In Fig.~\ref{Higgs_mod},
we display relations among the CP-even Higgs 
$H_{2,3}$, CP-odd Higgs $A_{1,2}$, and charged Higgs $H^{\pm}$. The color coding is
the same as in Fig.~\ref{fundamod}. It is very clear  that 
$m_{H_{2}} \approx m_{H^{\pm}} \approx m_{A_{1}}$ in the MDSB scenario.
On the other hand, we see $m_{H_{3}}\approx m_{H^{\pm}}$ in the mass range $[1.8,~2.7]$ TeV. 

The addition of a gauge singlet also affects the neutralino sector of the NMSSM. Now
 one can have the singlino-type neutralino in addition to 
the bino-type, wino-type and higgsino-type neutralinos. For dark matter relic density,
one can try to have $m_{{H_{1,2,3}}/A_{1,2}}\simeq 2m_{\tilde \chi_{1}^{0}}$ resonance solutions.
Moreover, in the CMSSM, small values of $m_{0}$ give rise to a stable charged slepton LSP.
While in the CNMSSM, this problem can be
evaded due to the presence of the extra singlino-like neutralino~\cite{Hugonie:2007vd}.
We would like to remind readers  
that in a good approximation one can show that
$m_{\tilde \chi_{1,2}^{0}}$ are proportional to gaugino masses
$M_{1,2}$, $m_{\tilde \chi_{3,4}^{0}}$ are proportional to $\mu_{eff}$, and singlino mass $m_{\tilde \chi_{5}^{0}}$ is
directly proportional to $\kappa$ and $\mu_{eff}$ but inversely proportional to $\lambda$~\cite{Accomando:2006ga}.
From Fig.~\ref{fundamod}, we see that the
minimum allowed value of $M_{1/2}$ is about 1.4 TeV. Since $m_{\tilde \chi_{1}^{0}}\approx 0.44 M_{1/2}$, the lightest neutralino
should be much heavier than the SM-like Higgs boson. Moreover, from Fig.~\ref{4mh_mod} we find that
$m_{H_{2,3}}$ and $m_{A_{1,2}}$ are heavier
than 1.5 TeV. So no resonance solutions can be realized here. As we already discussed in this case,
$\lambda$ is small and $\mu_{eff}$ is about 1 TeV. Thus, the
singlino is also heavy (as $m_{\tilde \chi_{5}^{0}}\propto \kappa,\mu_{eff}/\lambda$), and the LSP
neutralino in the MDSB scenario is bino-like. On the other hand, because $|A_{0}|$ is not large enough,
top squark masses must be heavy to achieve $m_{H_{1}}\sim$ 125 GeV. And then we do not have the
LSP neutralino-stop coannihilation channel. 
The focus point SUSY or Hyperbolic SUSY cannot be realized as well due to
$m_{0}< M_{1/2}$ from Fig.~\ref{fundamod}. We have mentioned earlier in Eq.~\ref{stau} 
that the right-handed slepton can be relatively light for relatively small values of $m_{0}$,
thus we can expect the LSP neutralino-stau coannihilation. From Fig.~\ref{spec_mod} it is evident that we do have
neutralino-stau coannihilation region. The color coding for this figure is the same as in Fig.~\ref{fundamod}.
Here, for the red points, the minimum masses for the light stau and LSP neutralino are respectively
580 GeV and 570 GeV  while the light stau and LSP neutralino can be as heavy as $\approx$ 1400 GeV.
We notice here that the best point we have here in this plot have $\Omega{h^2} \approx$ 0.2. However, we
are not able to get the points with relic density within 5$\sigma$ of WMAP9 bounds~\cite{Hinshaw:2012aka}.
Moreover, we present the plots in 
$m_{\tilde \chi_{1}^{0}}-m_{H_2}$ and $m_{\tilde \chi_{1}^{0}}-m_{A_{1}}$ planes to show that there is no resonance
solutions in our present scans for the MDSB scenario.

We would like to comment here that since relic density calculations are highly sensitive to
sparticle spectra and slight change in sparticle masses may change relic density
a lot. It is, therefore, $\Omega h^2 \approx$ 0.2 is not that bad value. 
To obtain the correct
 DM relic density, we can consider the dilution effect from supercritical string cosmology or introduce a LSP axino
 as the DM candidate.


\begin{figure}[htp!]
\centering

\subfigure{
\includegraphics[totalheight=10.5cm,width=15.cm]{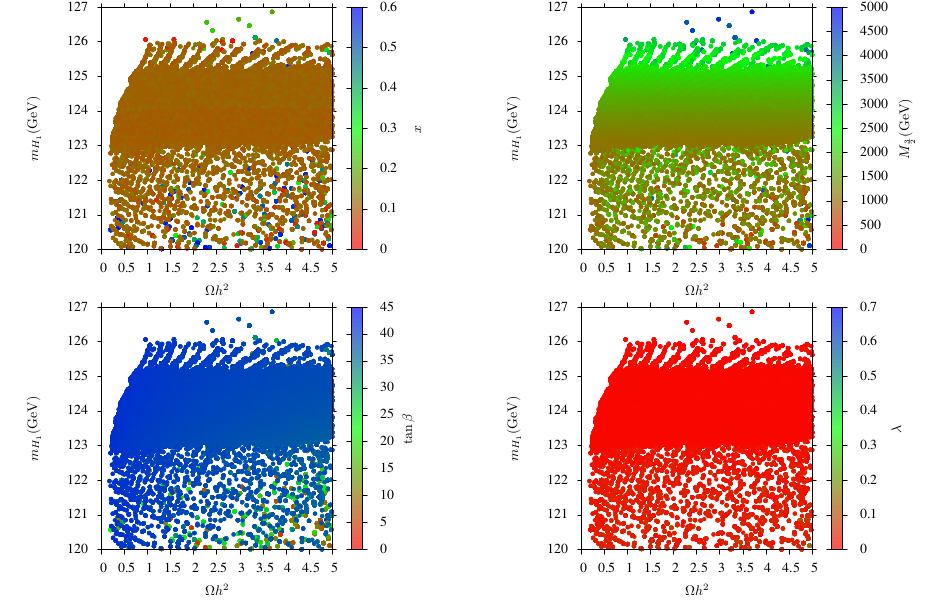}
}
\caption{
Plots in $\Omega h^{2}-m_{H_{1}}$ plane for dilaton dominant SUSY breaking scenario. The ranges of input parameters given in
Eq.~(\ref{input_param_range}) are shown in vertical bars.
}
\label{mhoh2dia}
\end{figure}
\begin{figure}[htp!]
\centering
\subfigure{
\includegraphics[totalheight=5.5cm,width=7.cm]{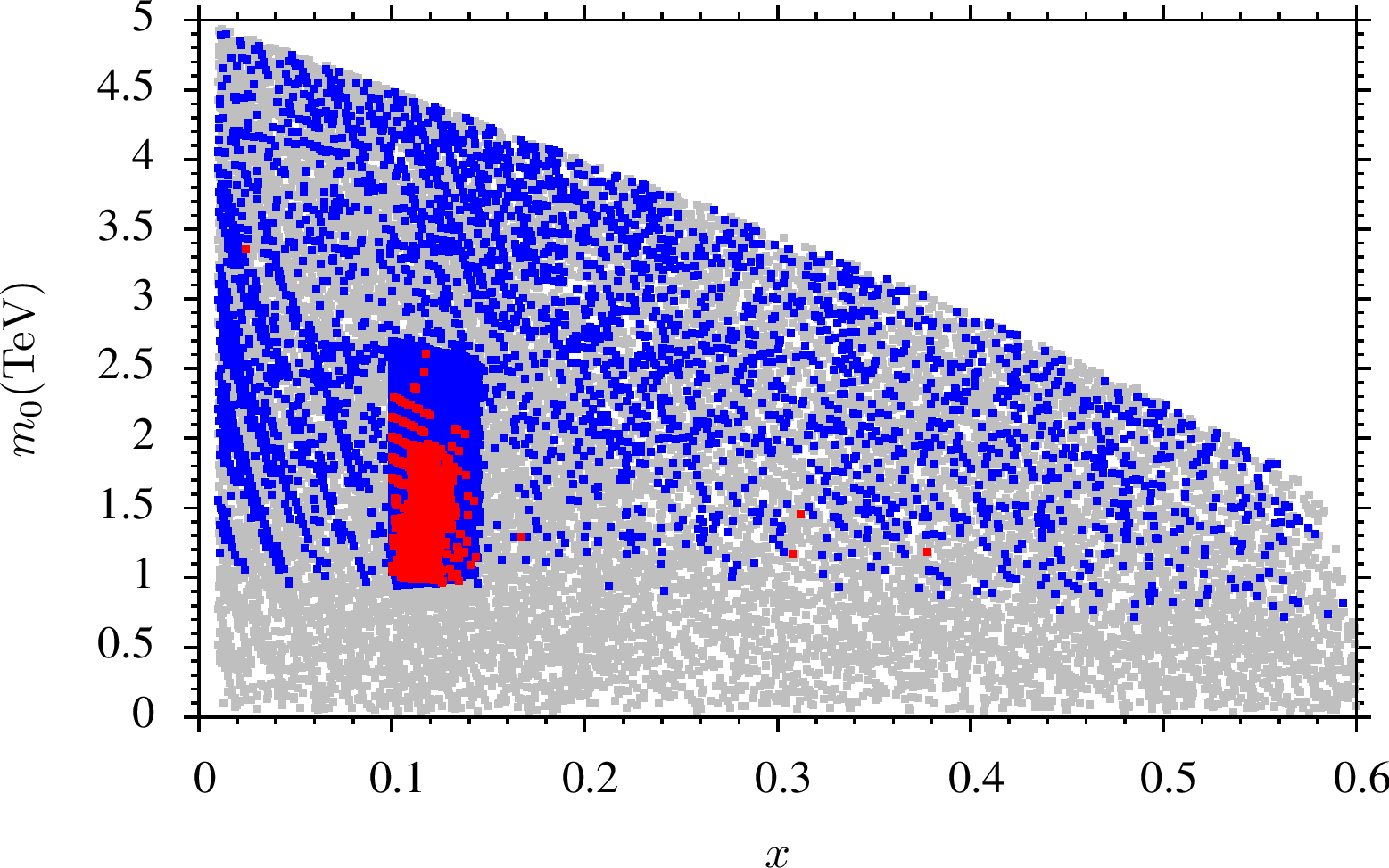}
}
\subfigure{
\includegraphics[totalheight=5.5cm,width=7.cm]{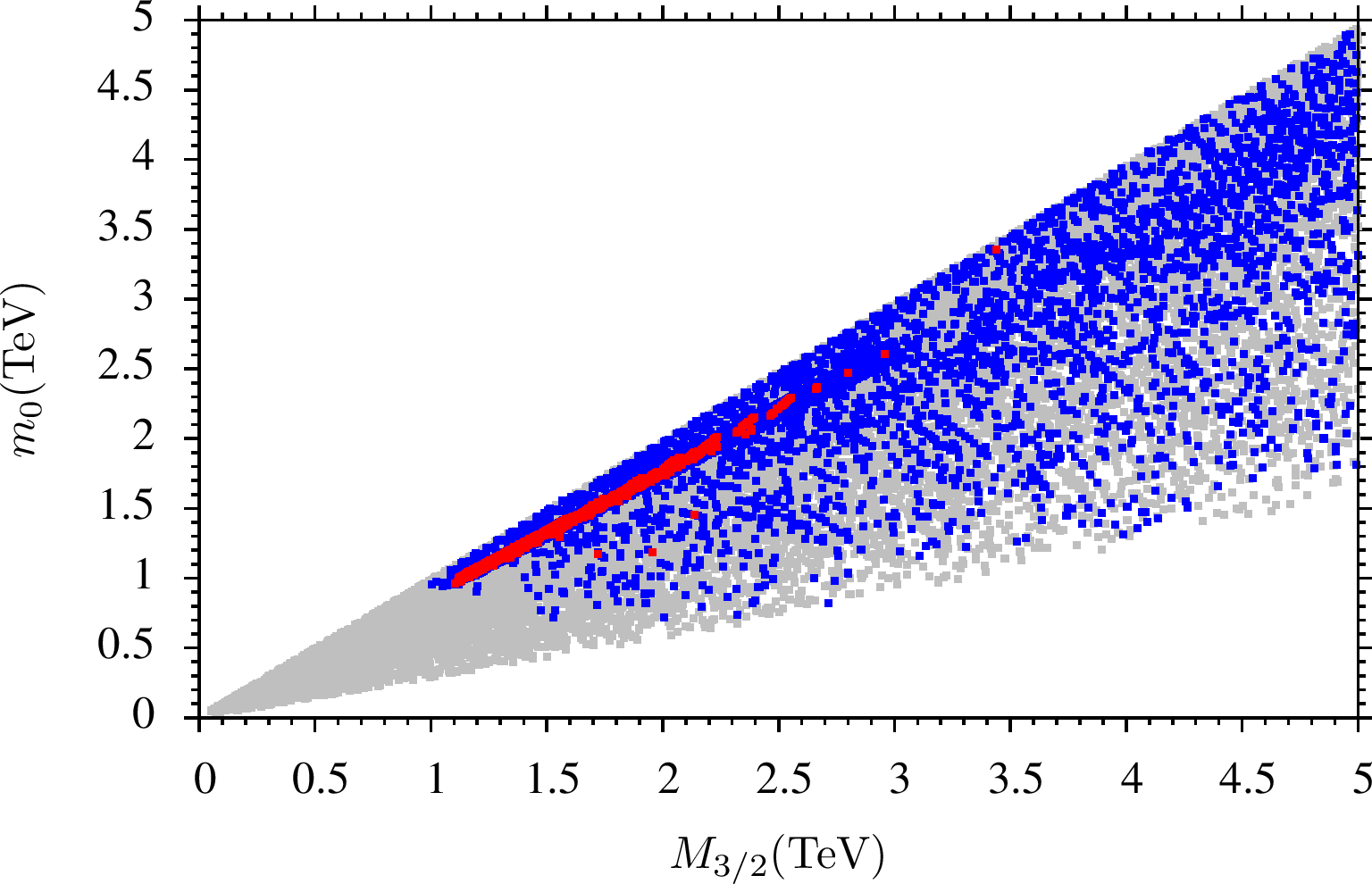}
}
\subfigure{
\includegraphics[totalheight=5.5cm,width=7.cm]{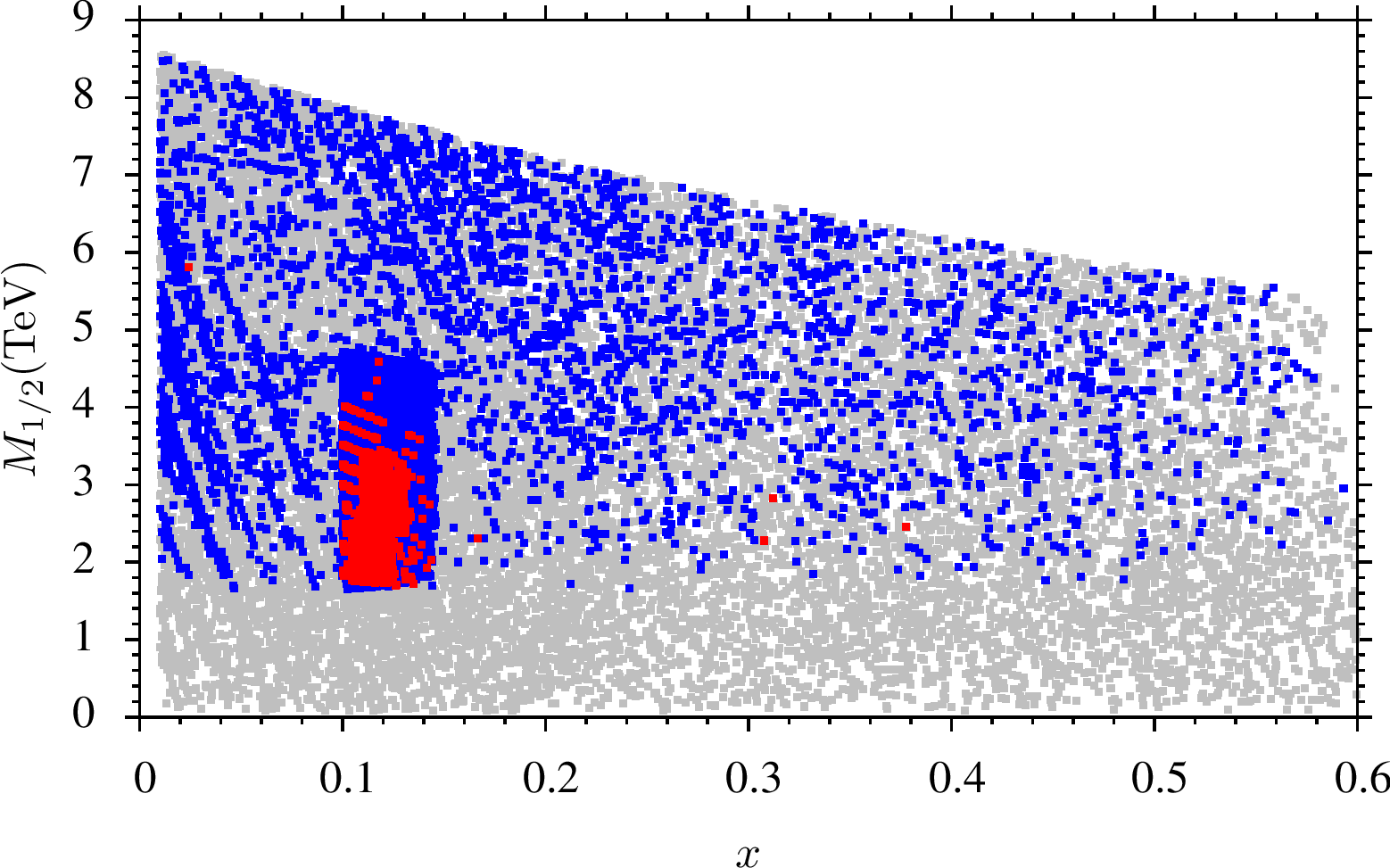}
}
\subfigure{
\includegraphics[totalheight=5.5cm,width=7.cm]{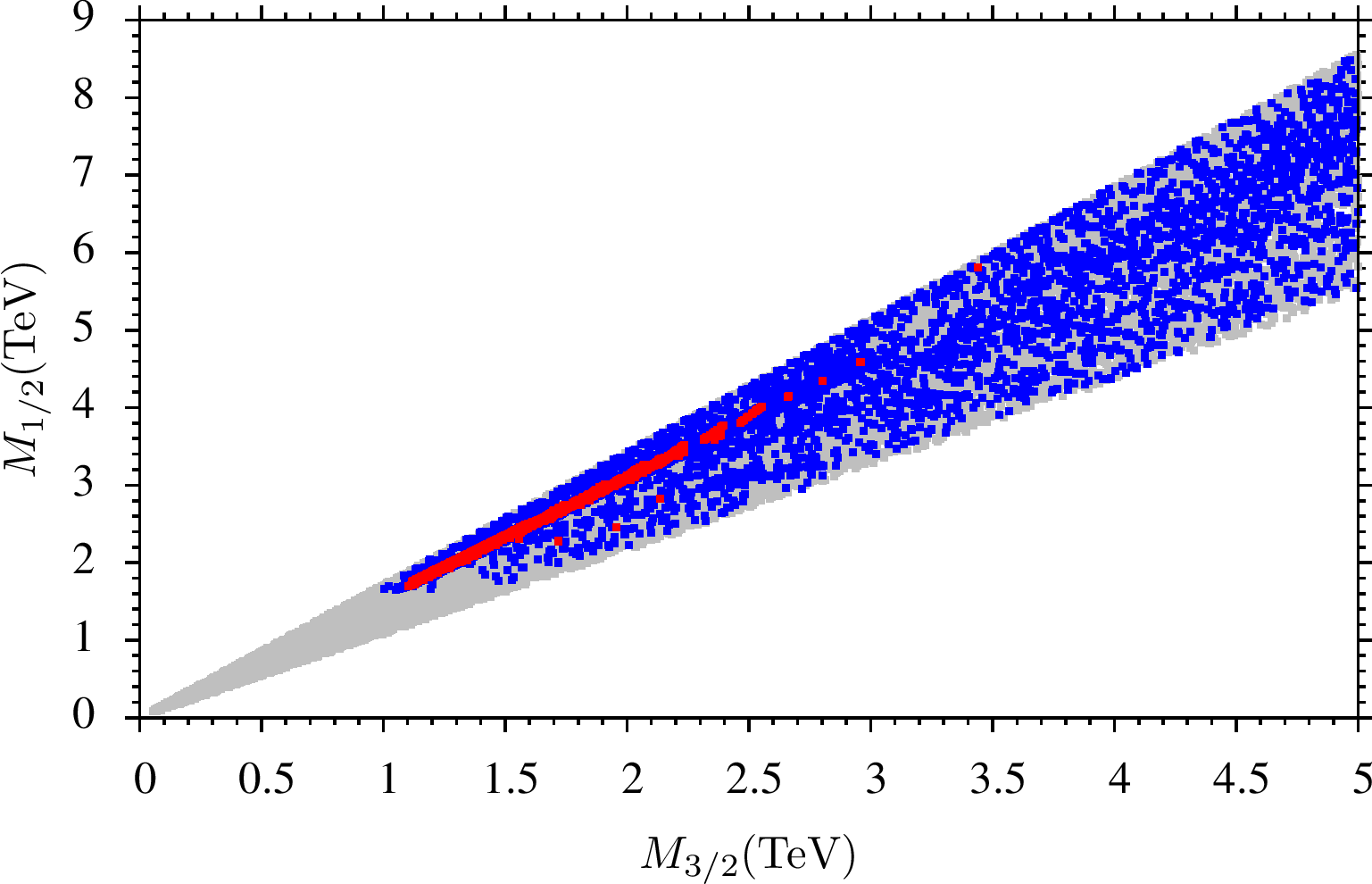}
}
\subfigure{
\includegraphics[totalheight=5.5cm,width=7.cm]{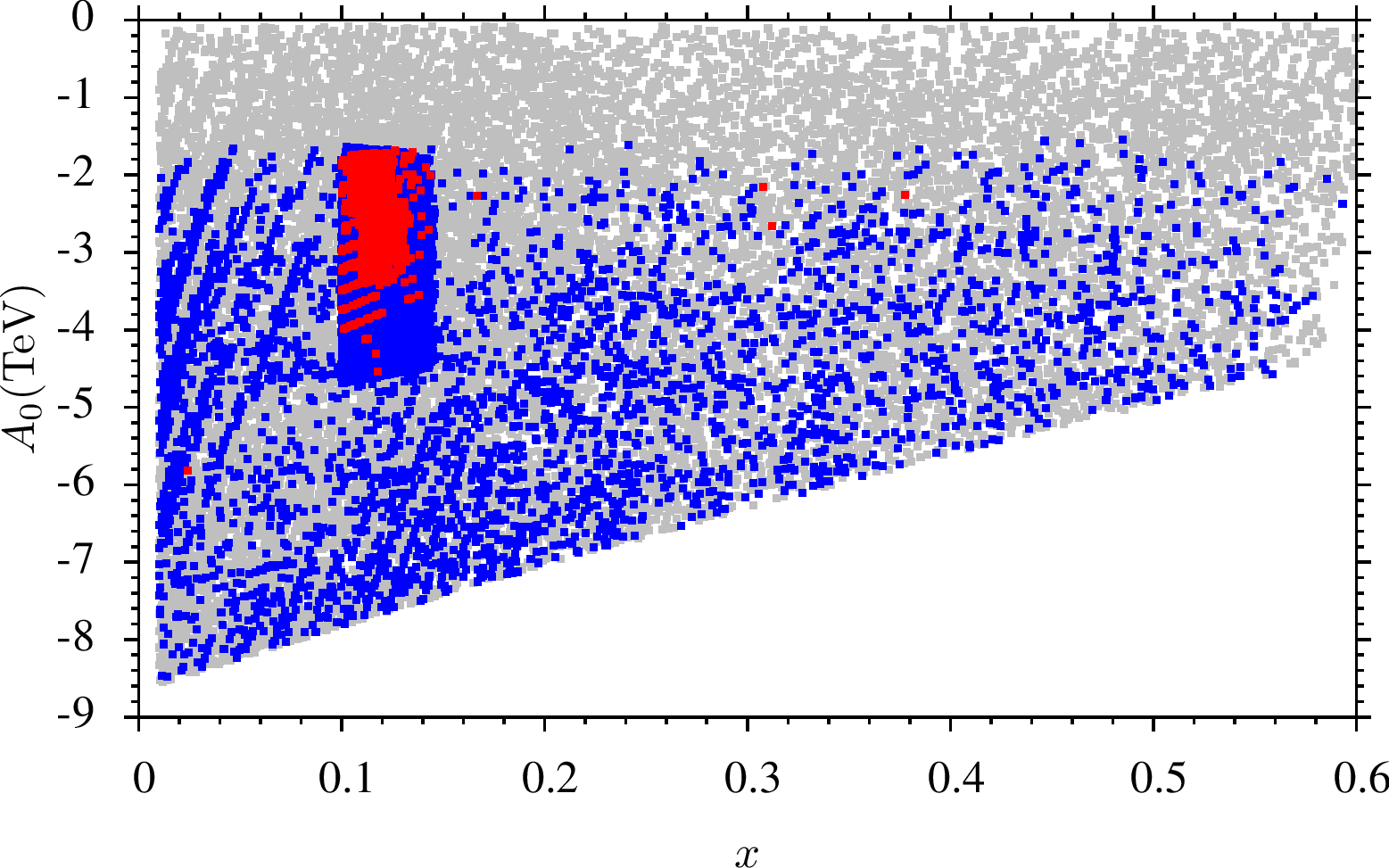}
}
\subfigure{
\includegraphics[totalheight=5.5cm,width=7.cm]{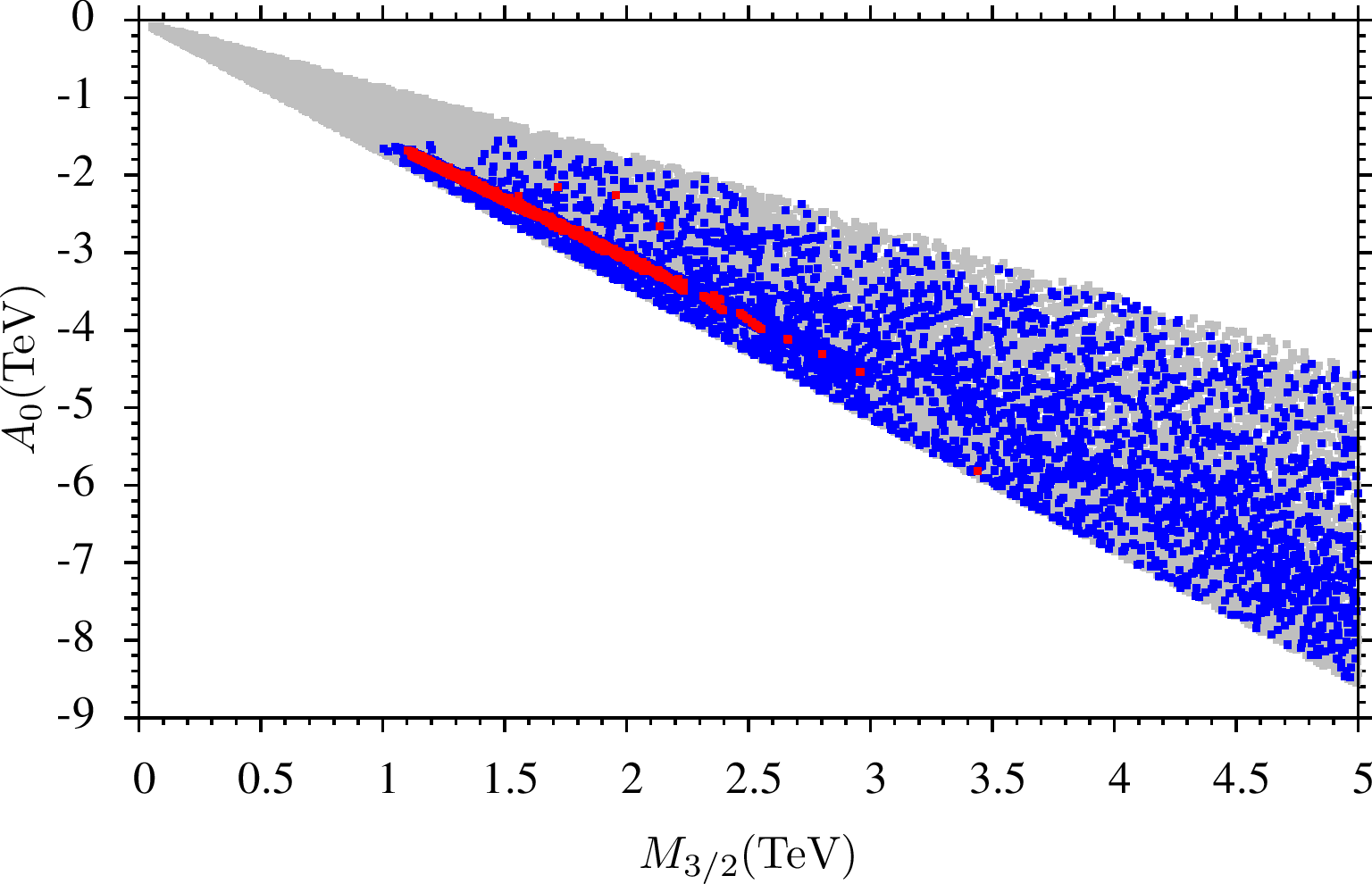}
}

\caption{
  Plots of $m_{0}$, $M_{1/2}$ and $A_{0}$ as functions of $x$ and $M_{3/2}$ for dilaton dominant SUSY breaking scenario.
  The color coding is the same as in Fig.~\ref{fundamod} }
\label{fundadia}
\end{figure}
\begin{figure}[htp!]
\centering
\subfigure{
\includegraphics[totalheight=5.5cm,width=7.cm]{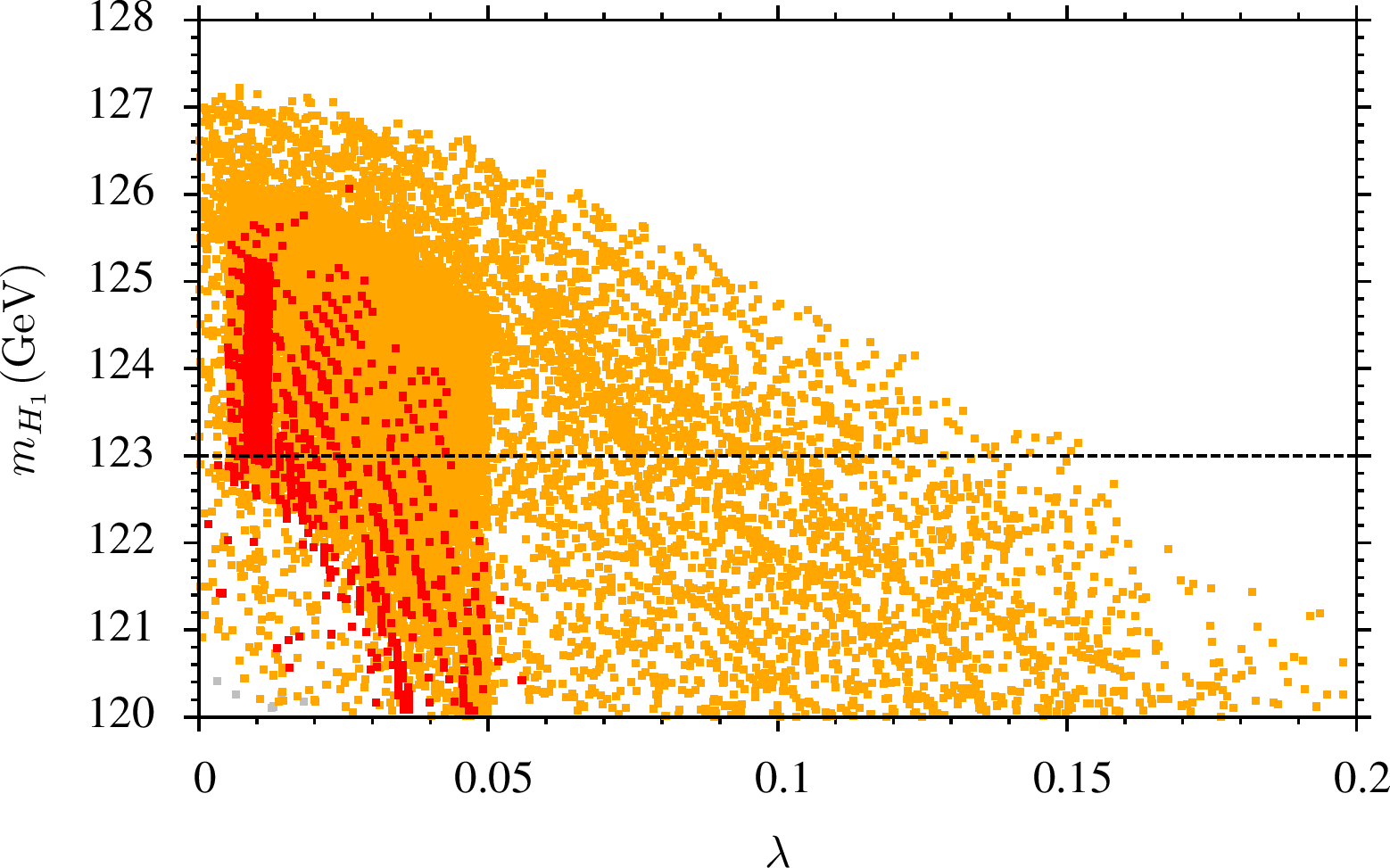}
}
\subfigure{
\includegraphics[totalheight=5.5cm,width=7.cm]{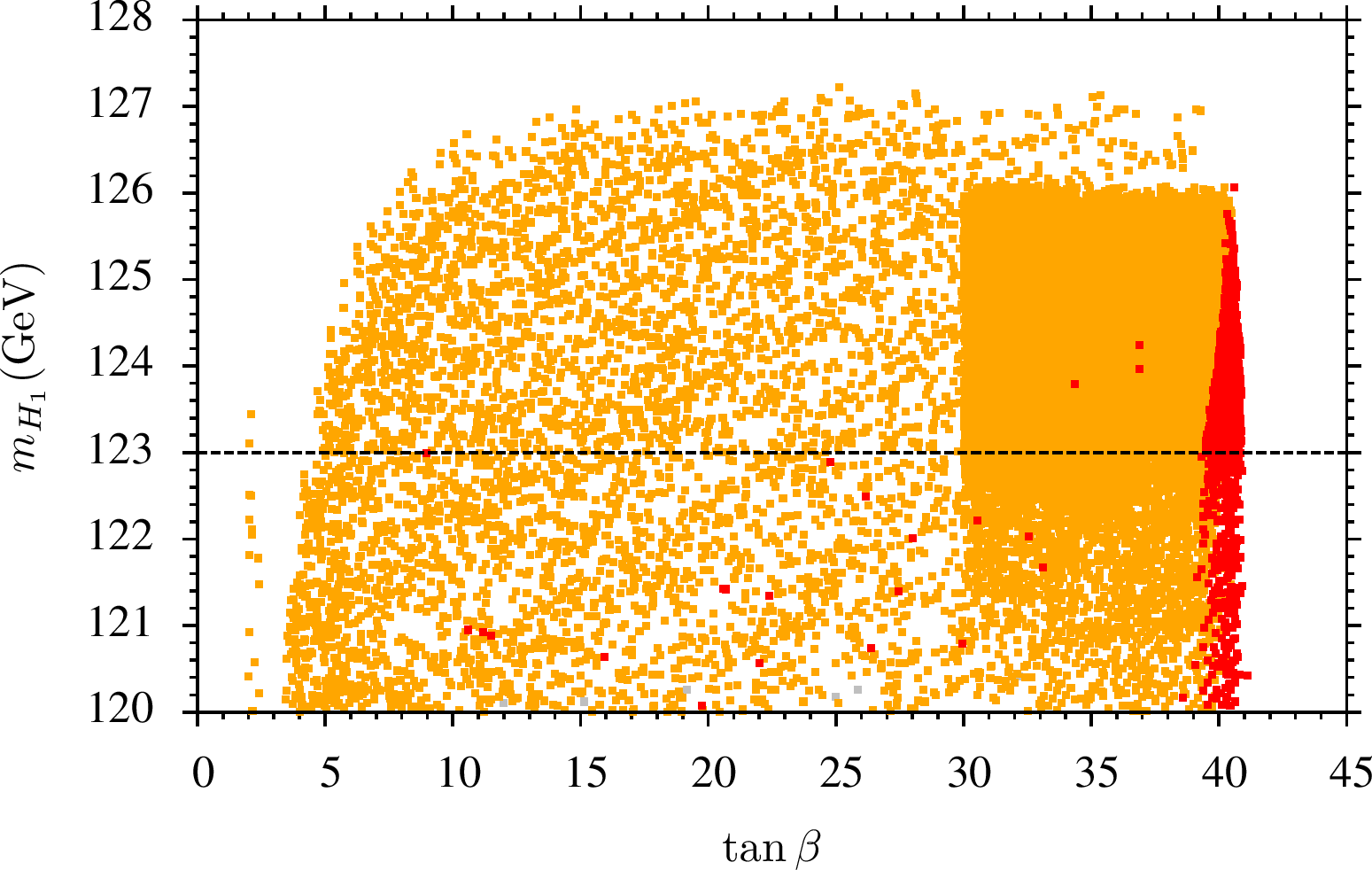}
}

\caption{
  Plots in $\lambda-m_{H_{1}}$ and $\tan\beta$-$m_{H_{1}}$ planes for dilaton dominant SUSY breaking scenario.
  The color coding is the same as in Fig.~\ref{lammh_mod} }
\label{mhmdia}
\end{figure}

\subsection{Dilaton Dominant SUSY Breaking Scenario}

In Fig.~\ref{mhoh2dia}, we present plots for Dilaton Dominant SUSY Breaking (DDSB) scenario
in $\Omega h^{2}-m_{H_1}$ plane. We also display the ranges of input parameters given in Eq.~(\ref{input_param_range})
in vertical bars. In the top left panel we see that $x$ should be in the range around $[0.1,~0.2]$
to have $m_{H_{1}} \gtrsim$ 123 GeV, while the DM relic density can be anywhere between 0 to 10.
In the plots of this figure, the viable points with Higgs mass above 123 GeV tend to have more
 or less one particular color and hence show the narrow ranges of input parameters. This is because  
 of our dedicated searches: if we generated more data around some good points, the corresponding ranges
 of those input parameter's color dominate (this is very much true for $x$ and $\tan\beta$).
 These dedicated search effects will also appear in Fig.~\ref{fundadia}. In the top right panel of
 Fig.~\ref{mhoh2dia}, we see that the Higgs mass larger than 123~GeV requires $M_{3/2}\gtrsim$ 1 TeV.
 In the bottom left panel, for  $m_{H_1}\gtrsim$ 123 GeV, we need $\tan\beta \gtrsim$ 35,
 but we can see some green points at the top of the figure which shows that
 the low bound on $\tan\beta$ can be relaxed. The appearance of only blue points is just
 an artifact of dedicated searches. In Fig.~~\ref{mhmdia} one will
see that the actual $\tan\beta$ lower limit consistent with 123 GeV Higgs mass is about 5.

We use Eq.~(\ref{dia_m0})-(\ref{dia_A0}) to calculate $m_{0}$, $M_{1/2}$ and $A_{0}$ as functions of input parameters $x$ and $M_{3/2}$. We show our results in 
Fig.~\ref{fundadia}. The color coding is the same as in Fig.~\ref{fundamod}. As compared to the MDSB scenario, in the
DDSB scenario the input parameters $x$ and $M_{3/2}$ are less constrained under various bounds we mentioned earlier.
The patches of points
correspond to our dedicated searches. From the plots in the first row of Fig.~\ref{fundadia}, we observe that 
the minimum value of $m_{0}$ consistent with all the constraints is about 0.7 TeV, corresponding
to $x\approx$ 0.6 but the maximum value of
$m_{0}\sim$ 5 TeV occurs at very small values of $x$. On the other hand, the minimum value of $m_{0}$ is correlated 
to $M_{3/2}$ and increases linearly with $M_{3/2}$ up to 5 TeV. We also notice here that the red points
have $m_{0}\lesssim$ 3.5 TeV.
The plots in  the second row display dependence of $M_{1/2}$ on $x$ and $M_{3/2}$. The minimum and maximum values
of $M_{1/2}$ consistent with all the above mentioned constraints is about 2 TeV and 8.5 TeV, respectively.
Finally, the plots in the third row display that the allowed range of universal trilinear soft term
$A_{0}$ is $[-2,-8.5]$ GeV. Such relative large values of $|A_{0}|$ show the top squarks have larger mixing in the
DDSB scenario than the MDSB scenario.
This implies that now $A_{0}$ will share the burden of achieving the SM-like Higgs mass around 125 GeV with top squarks.
Using Eqs.~(\ref{squark})-(\ref{stau}), we see that in the DDSB scenario, we have heavier spectra as compared to
the MDSB scenario.
\begin{figure}[htp!]
\centering
\subfigure{
\includegraphics[totalheight=5.5cm,width=7.cm]{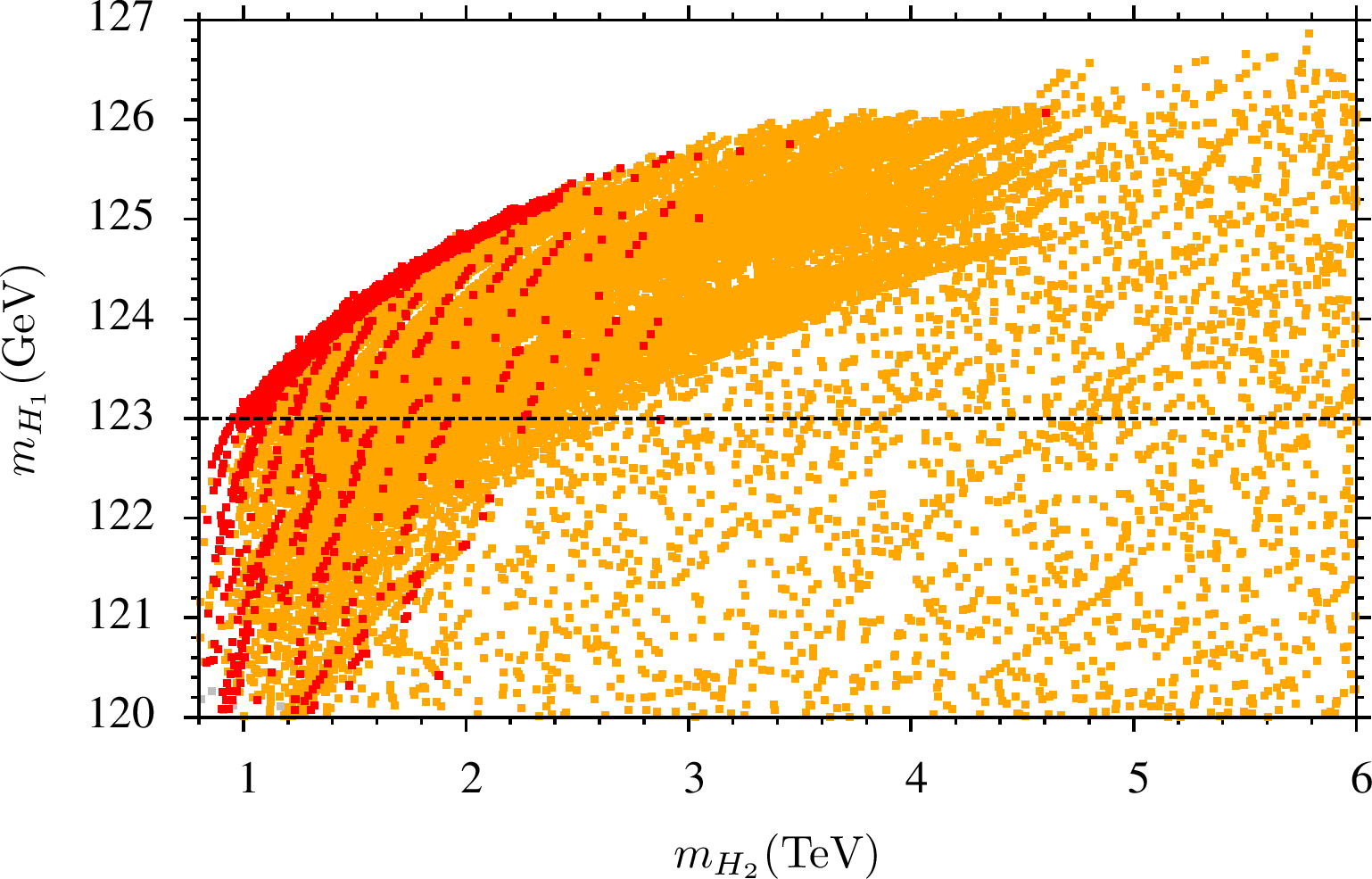}
}
\subfigure{
\includegraphics[totalheight=5.5cm,width=7.cm]{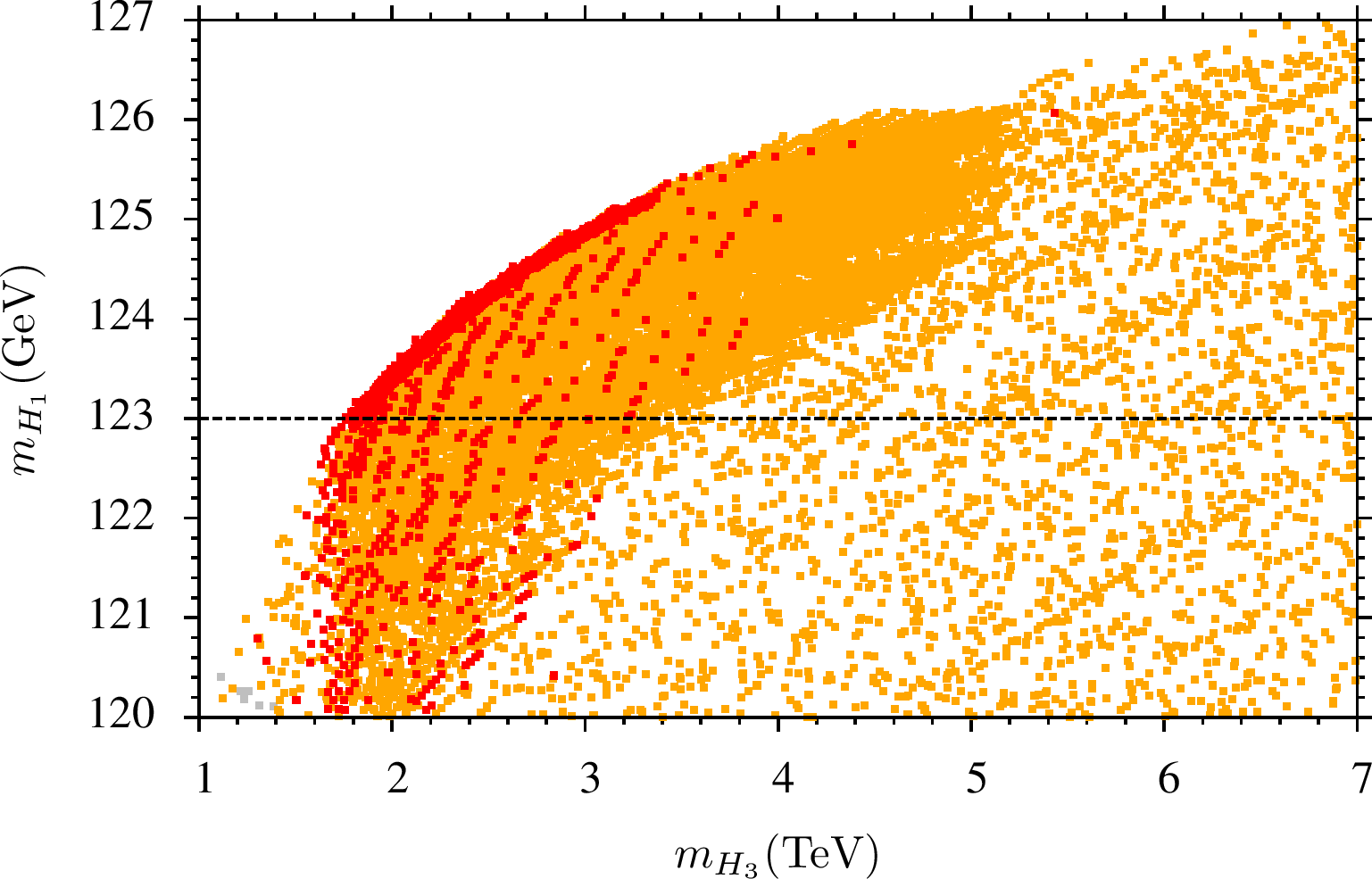}
}
\subfigure{
\includegraphics[totalheight=5.5cm,width=7.cm]{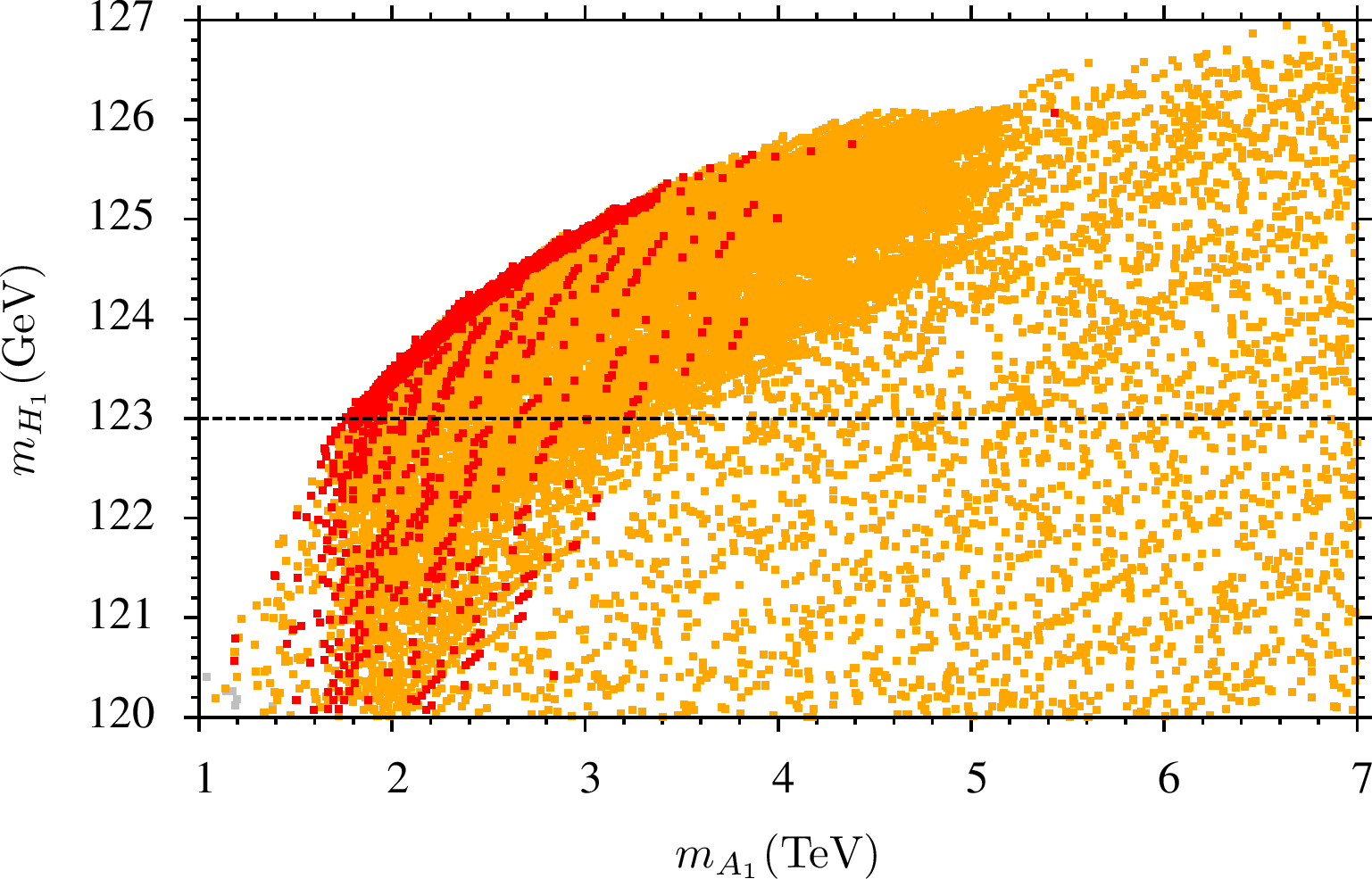}
}
\subfigure{
\includegraphics[totalheight=5.5cm,width=7.cm]{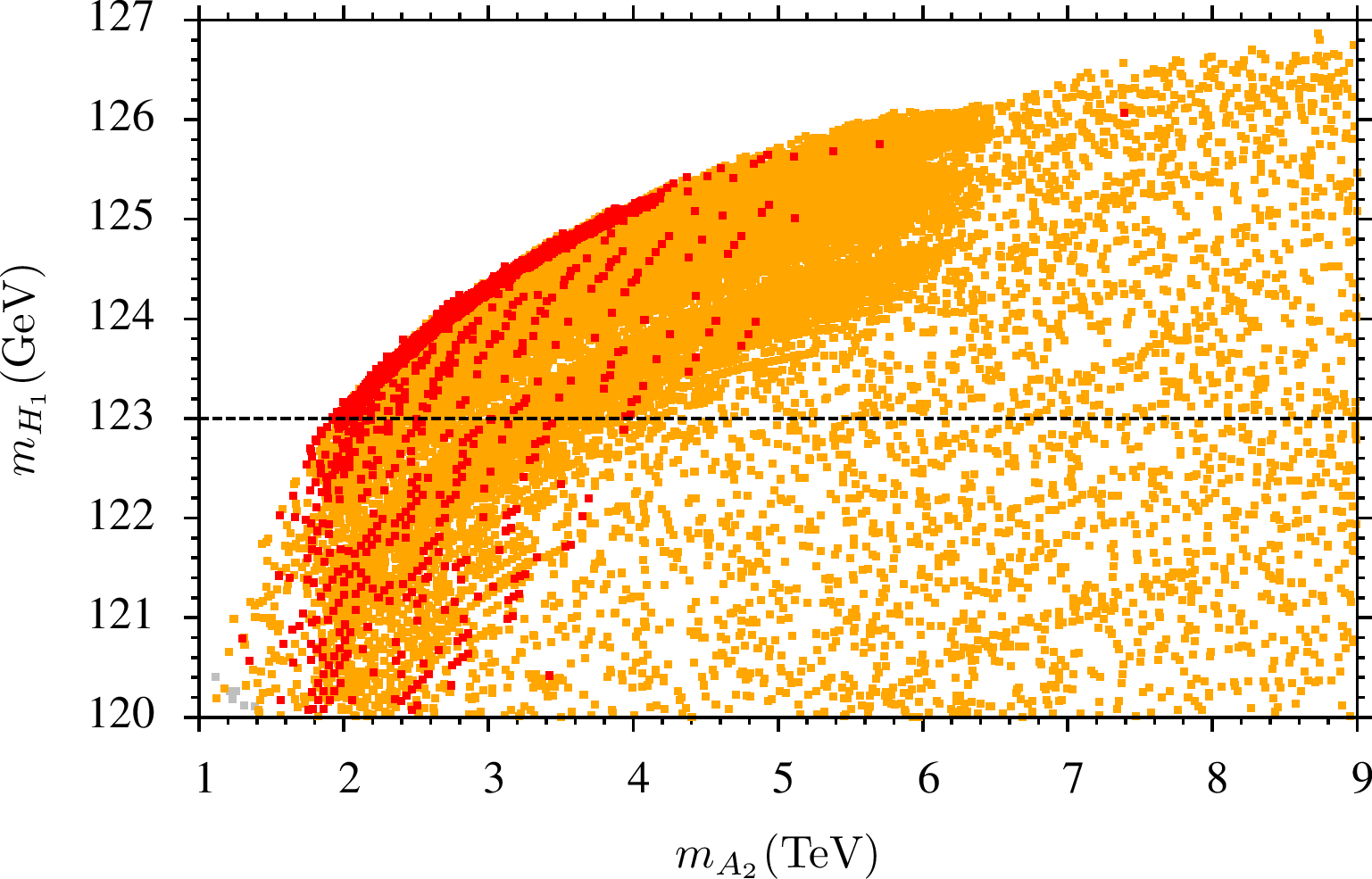}
}
\subfigure{
\includegraphics[totalheight=5.5cm,width=7.cm]{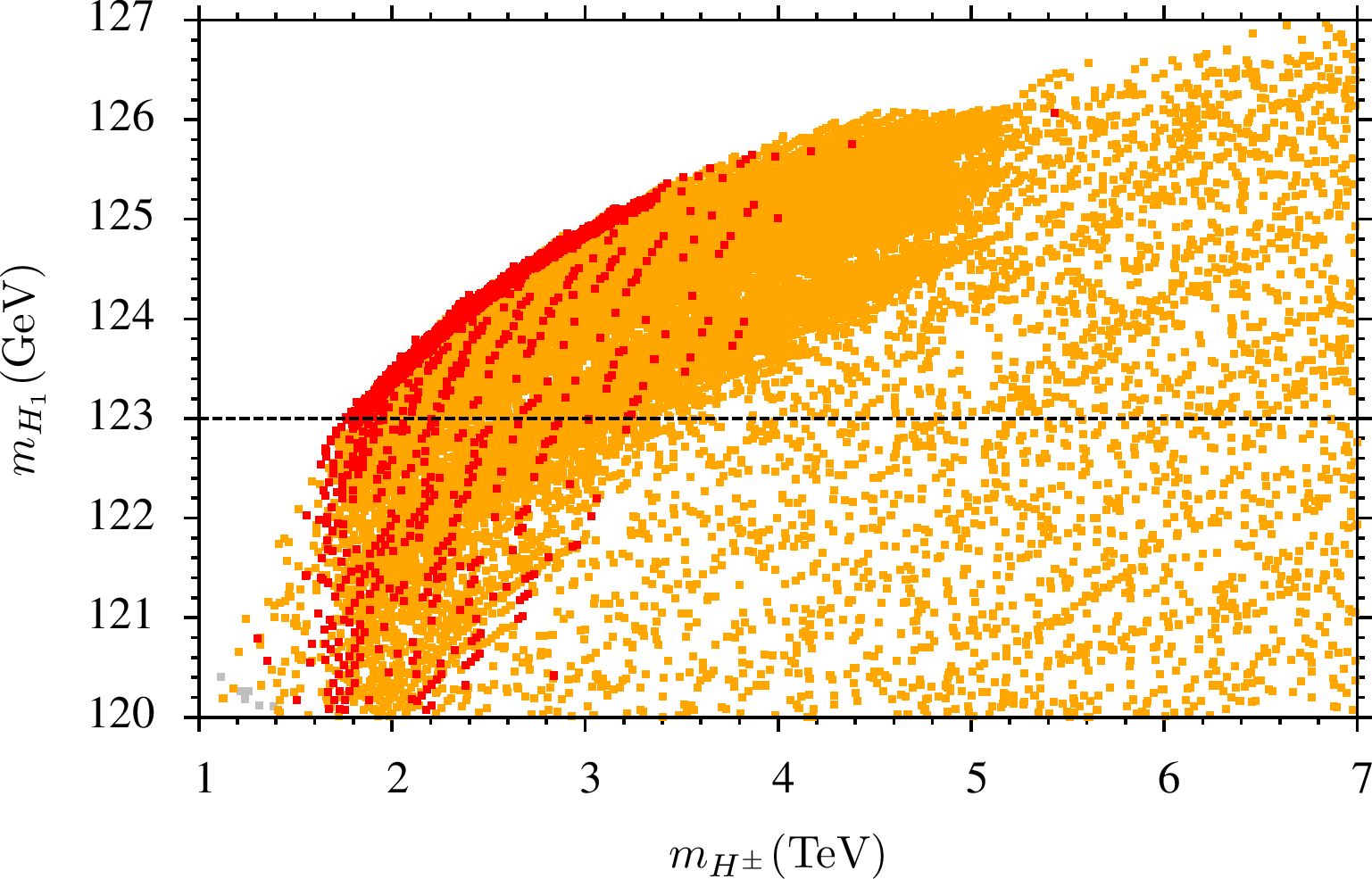}
}
\caption{
Plots in  $m_{H_{2}}-m_{H_{1}}$, $m_{H_{3}}-m_{H_{1}}$, $m_{A_{1}}-m_{H_{1}}$  and
$m_{A_{2}}-m_{H_{1}}$ and $m_{H^{\pm}}-m_{H_{1}}$ planes for dilaton dominant SUSY breaking scenario. The color coding is
the same as in Fig.~\ref{lammh_mod}.}
\label{Higgs_dia}
\end{figure}
\begin{figure}[htp!]
\centering
\subfigure{
\includegraphics[totalheight=5.5cm,width=7.cm]{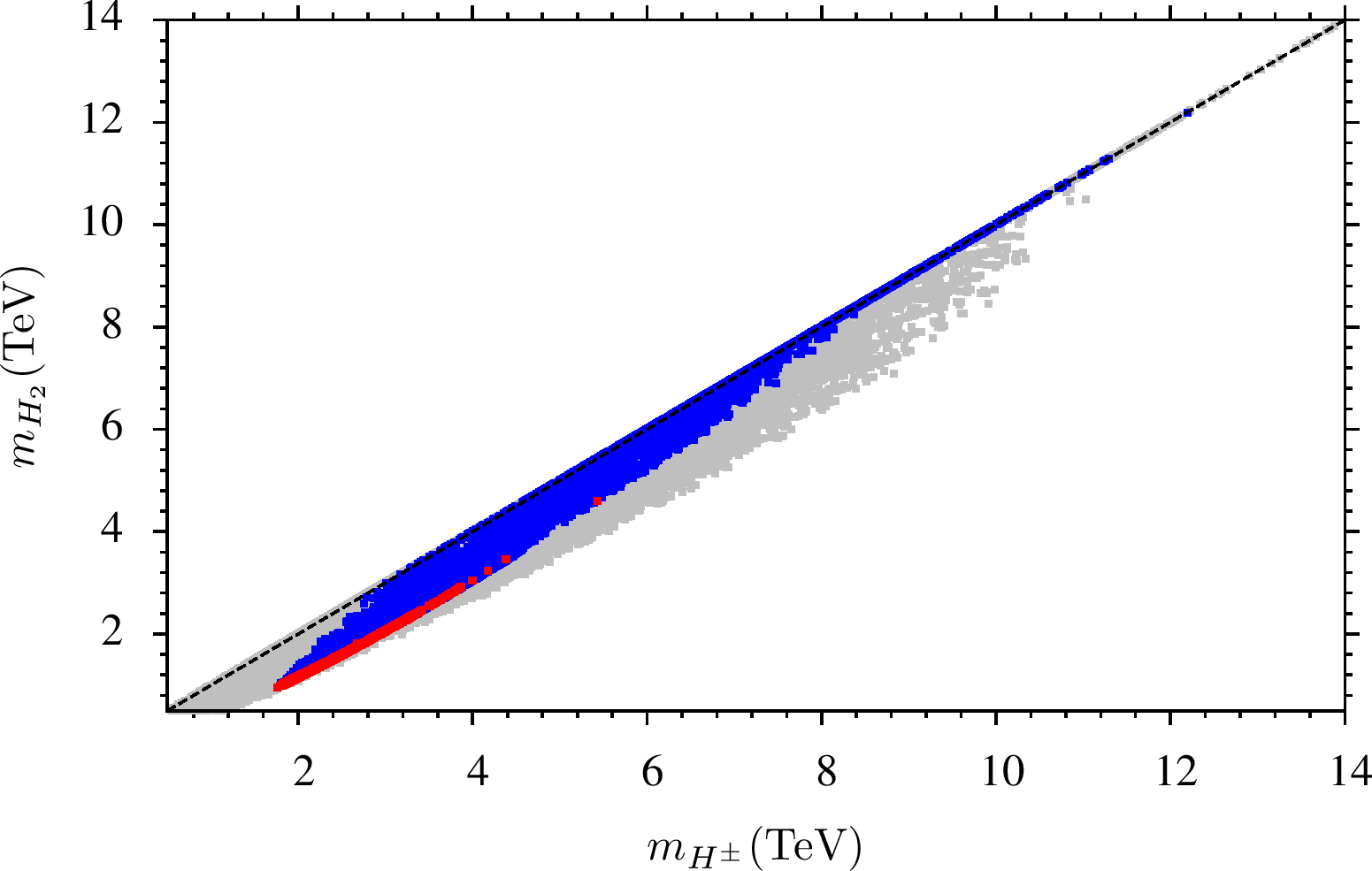}
}
\subfigure{
\includegraphics[totalheight=5.5cm,width=7.cm]{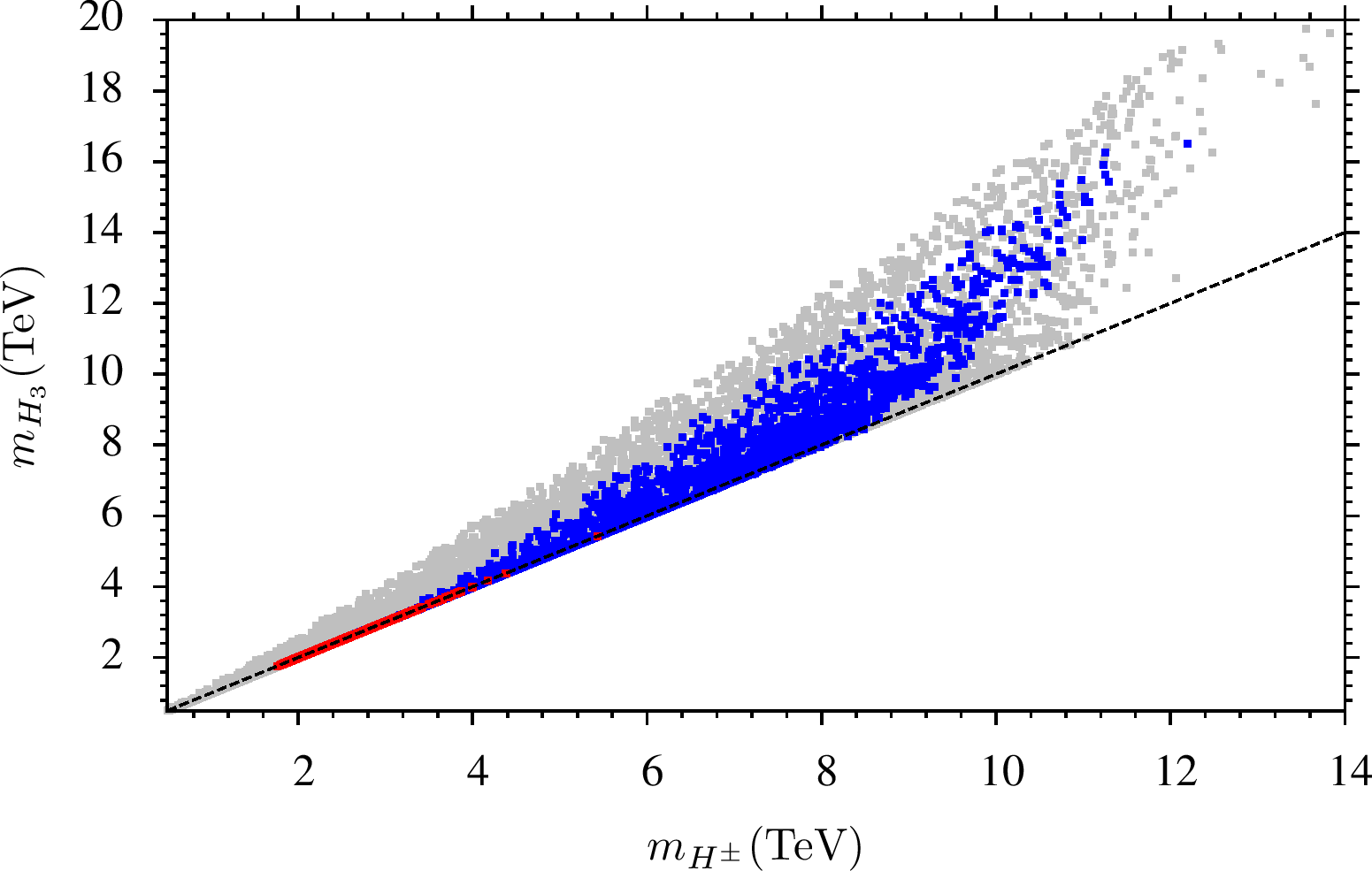}
}
\subfigure{
\includegraphics[totalheight=5.5cm,width=7.cm]{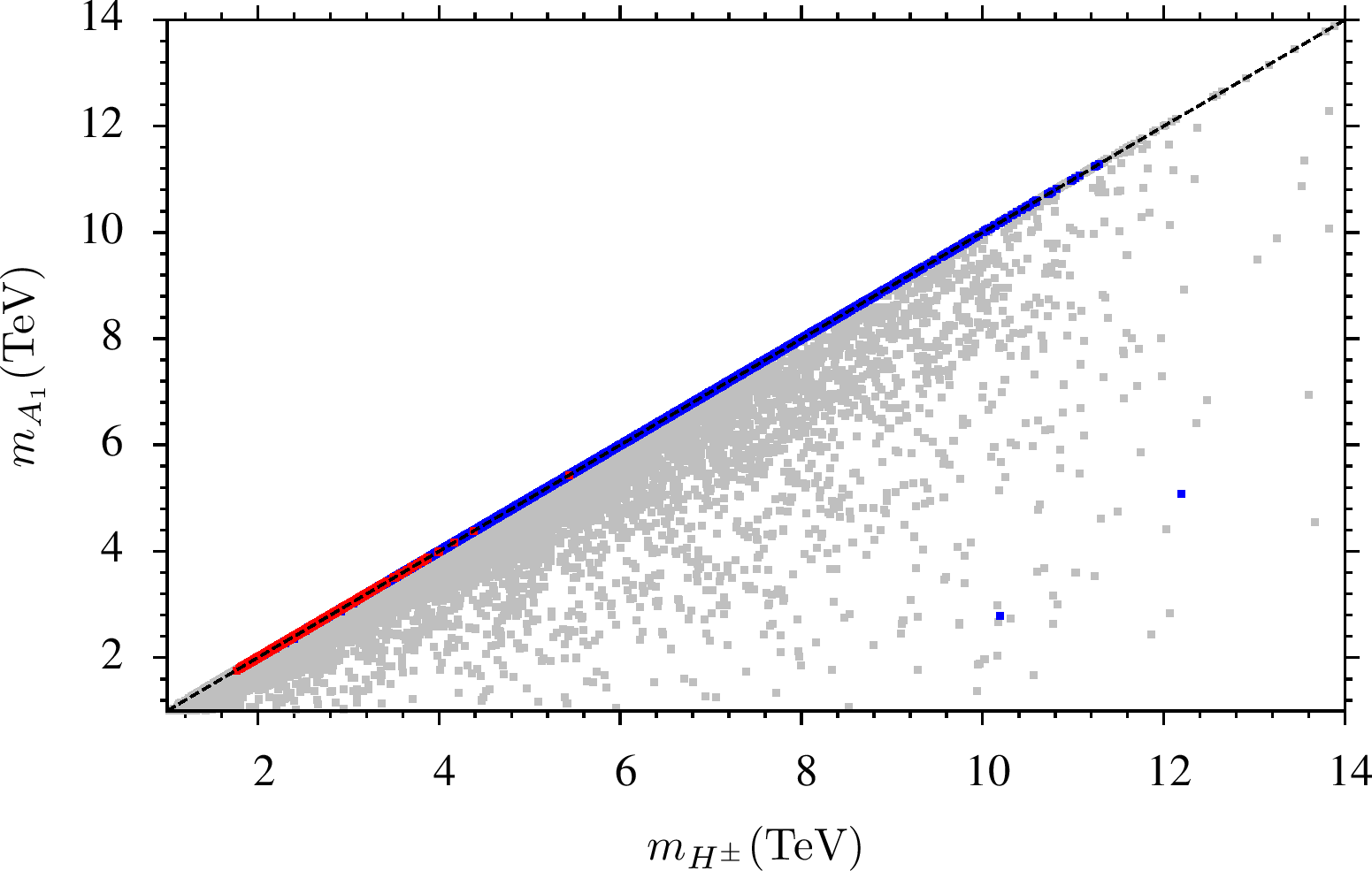}
}
\subfigure{
\includegraphics[totalheight=5.5cm,width=7.cm]{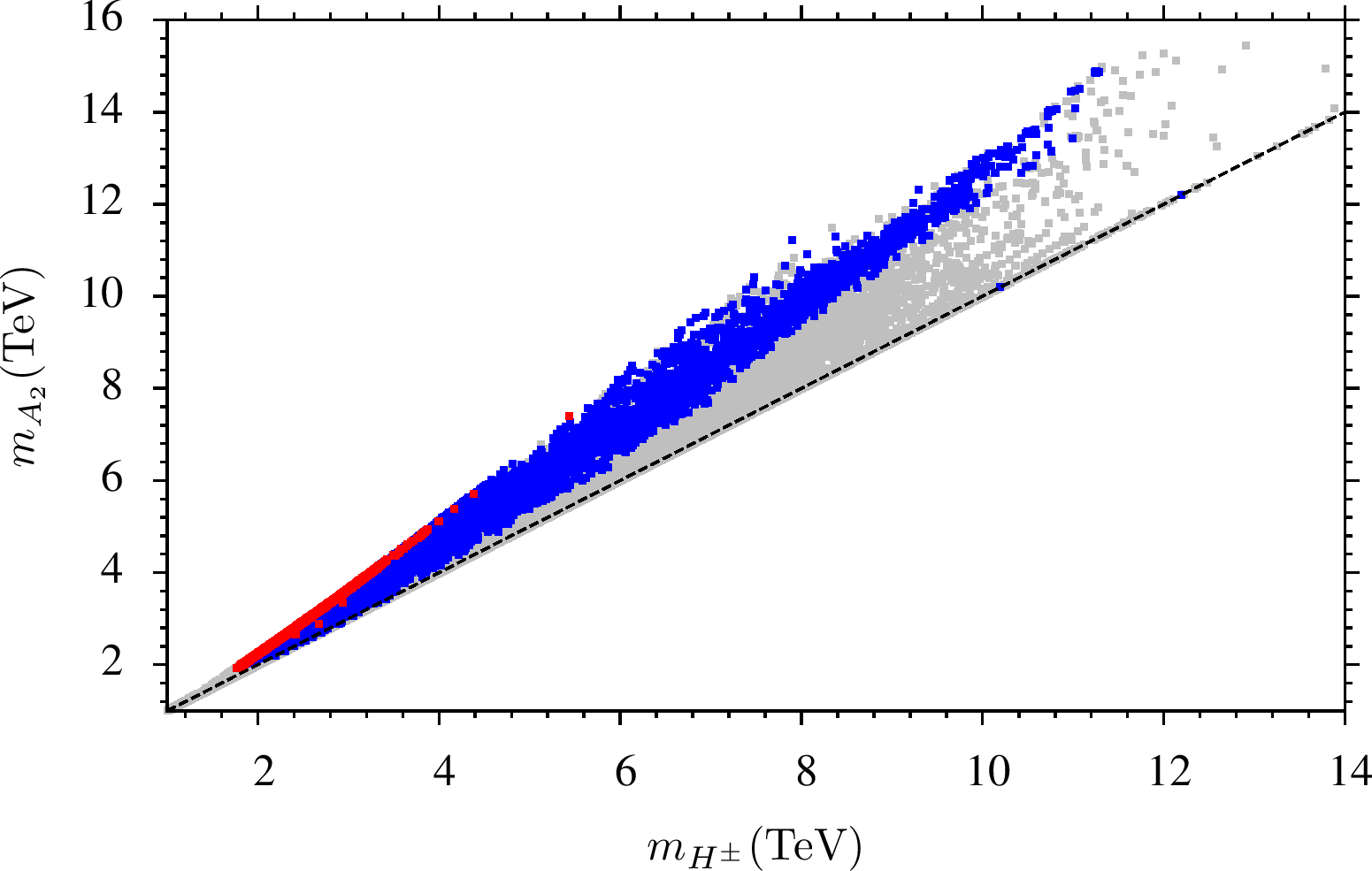}
}

\caption{
Plots in Plots in  $m_{H^{\pm}}-m_{H_{2}}$, $m_{H^{\pm}}-m_{H_{3}}$, $m_{H^{\pm}}-m_{A_{1}}$  and
$m_{H^{\pm}}-m_{A_{2}}$ planes for dilaton dominant SUSY breaking scenario. The color coding is the same as in Fig.~\ref{fundamod}. }
\label{4mh_dia}
\end{figure}
\begin{figure}[htp!]
\centering
\subfigure{
\includegraphics[totalheight=5.5cm,width=7.cm]{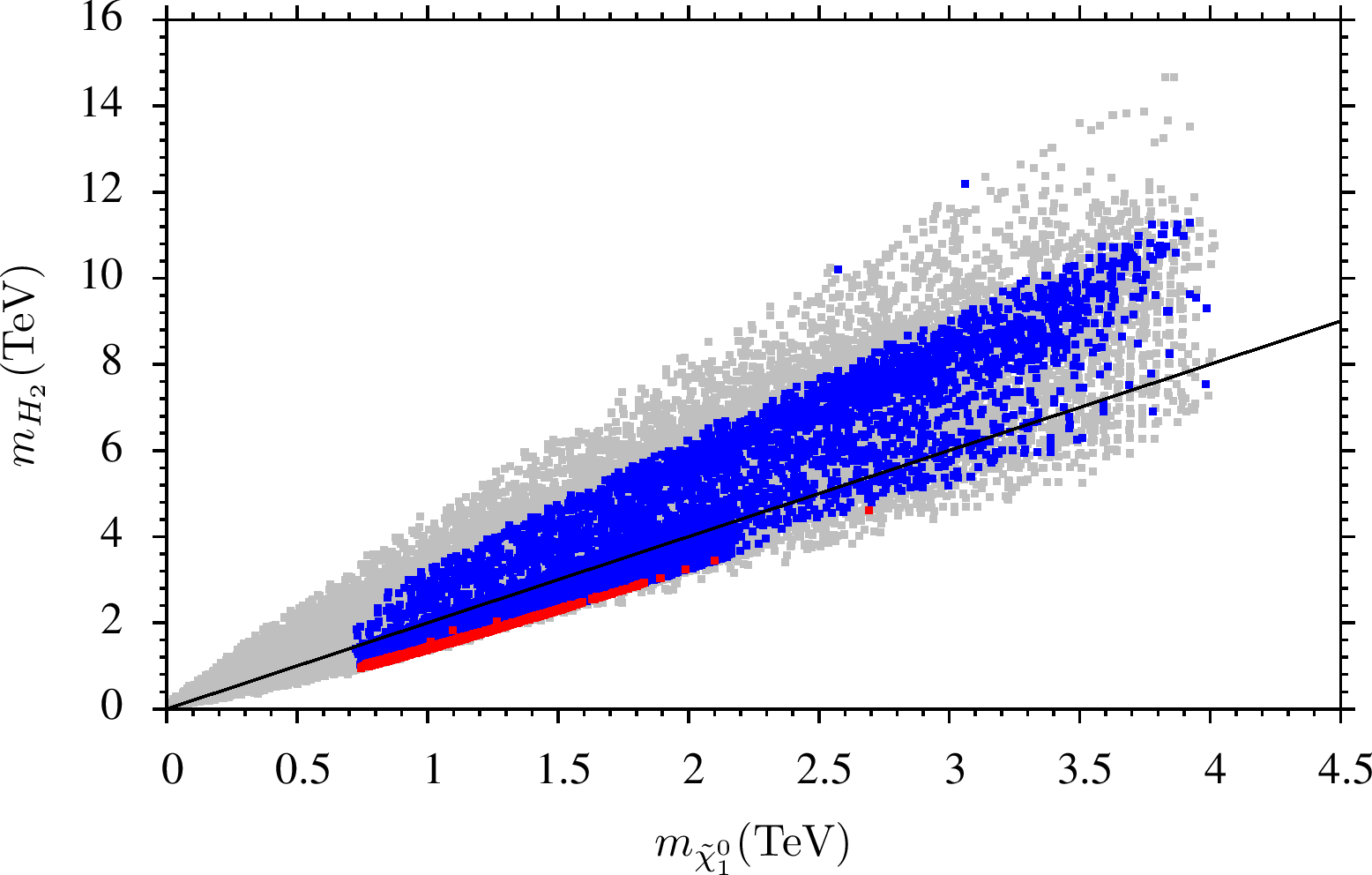}
}
\subfigure{
\includegraphics[totalheight=5.5cm,width=7.cm]{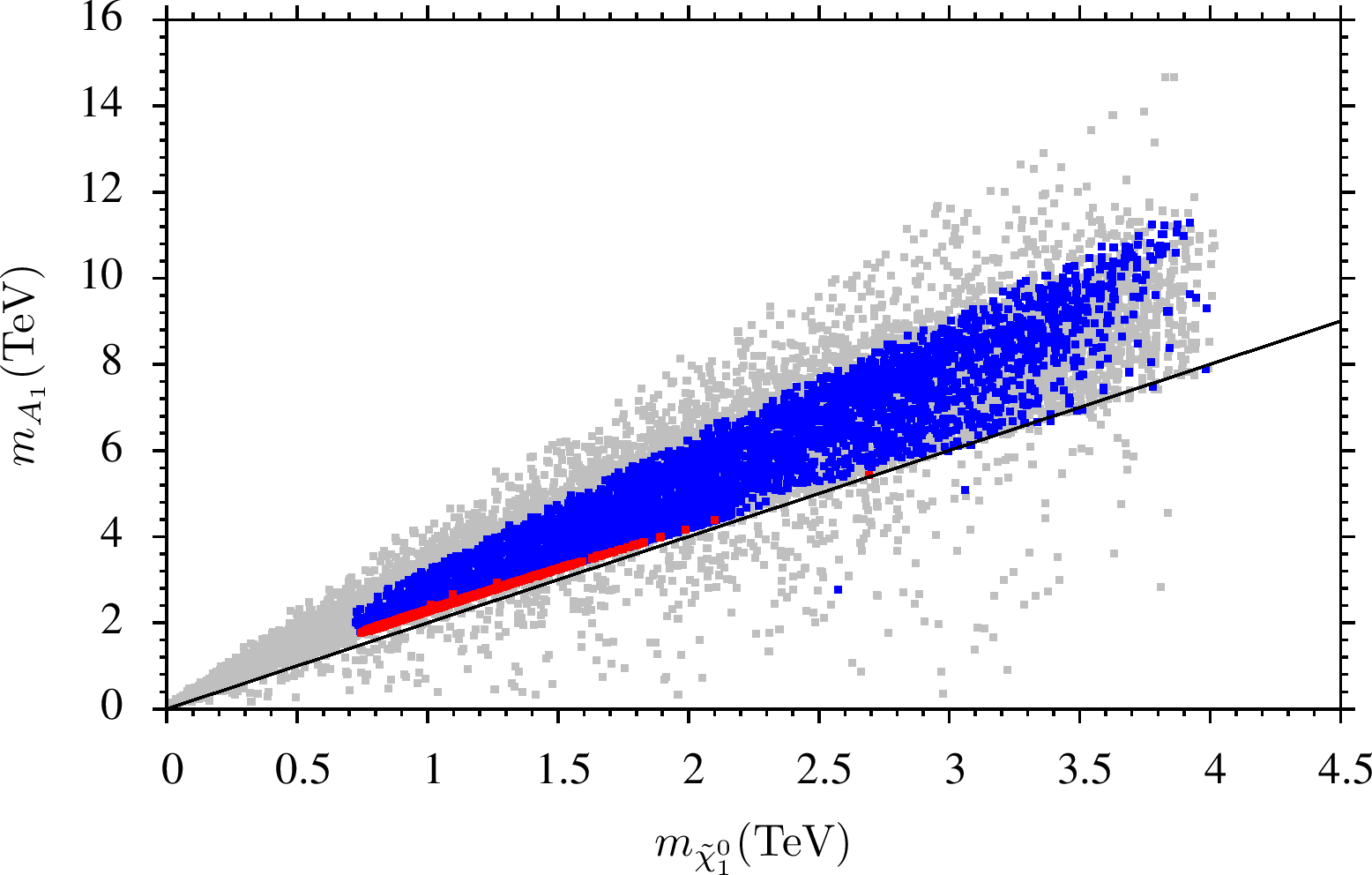}
}
\subfigure{
\includegraphics[totalheight=5.5cm,width=7.cm]{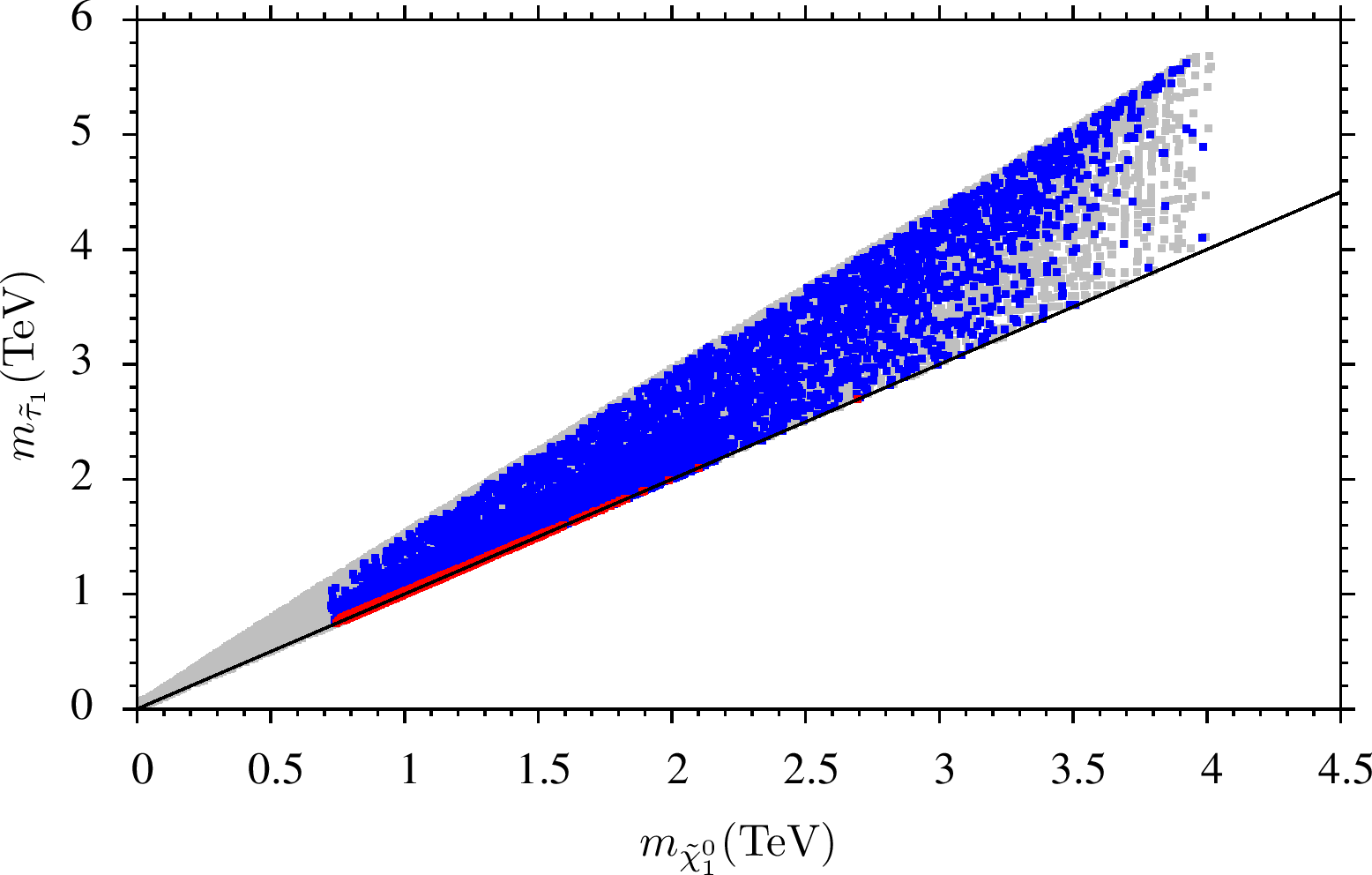}
}
\caption{
Plots in $m_{\tilde \chi_{1}^{0}}-m_{H_2}$, $m_{\tilde \chi_{1}^{0}}-m_{A_{1}}$ and $m_{\tilde \chi_{1}^{0}}-m_{\tilde \tau_{1}}$ planes for dilaton dominant SUSY breaking scenario.
The color coding is the same as in Fig.~\ref{fundamod}.}
\label{spec_dia}
\end{figure}


In Fig.~\ref{mhmdia}, we depict our results in $\lambda-m_{H_1}$ (left panel) plane and
$\tan\beta-m_{H_1}$ (right panel) plane. 
In the left panel we immediately see that the allowed values of $\lambda$ consistent with Higgs mass bounds as well
is about 0.15, which is 
slightly larger than what we got in the MDSB scenario ($\lambda \sim$ 0.1).
This slightly larger value of $\lambda$ has very important
consequences on Higgs sector. Similar to the above discussions,
with $\mu_{eff}=\lambda v_{S}$,  $v_{S}$ should be larger than 666 GeV (taking 
$\mu$ = 100 GeV). Thus, we have relatively small $v_{S}$, and then
the masses of the CP-even Higgs $H_2$, $H_3$ and CP-odd Higgs $A_1$ and $A_2$ can have 
relatively smaller values as compared to the MDSB scenario. We will see in Fig.~\ref{Higgs_dia} that
this is indeed the case. In the right panel
 of Fig.~\ref{mhmdia}, we see that  $\tan\beta$ can have any value between 2 to 41.

 In Fig.~\ref{Higgs_dia}, we display graphs for Higgs sector. The color coding is
 the same as in Fig.~\ref{lammh_mod}. Here, we clearly see that it is easier to achieve
 $m_{H_1}\sim$ 125 GeV. In fact, the SM-like Higgs mass can go up to 127 GeV. As we mentioned above, 
 the lower mass values of $H_2$ and $A_2$ are relatively smaller as compared to the MDSB scenario.
 This is useful as we will show latter. Like Fig.~\ref{4mh_mod}, we
 plot $m_{H_{2,3}}$, $m_{A_{1,2}}$ and $m_{H^{\pm}}$ in Fig.~\ref{4mh_dia}.
 The color coding is the same as Fig.~\ref{fundamod}. Here, we have
$m_{A_1}=m_{H^{\pm}}$, while $m_{A_{1}}\approx m_{{A_2}}\approx m_{H_{2}}\approx m_{H_{3}}$ in some portions of parameter space.

In Fig.~\ref{spec_dia}, we display plots in $m_{\tilde \chi_{1}^{0}}-m_{H_{2}}$, $m_{\tilde \chi_{1}^{0}}-m_{A_{1}}$ and 
$m_{\tilde \chi_{1}^{0}}-m_{\tilde \tau_{1}}$. The color coding of this figure is
the same as in Fig.~~\ref{fundamod}. The black solid lines indicate 
 $2 m_{\tilde \chi_{1}^{0}}=m_{H_{2},A_{1}}$ in the first row and $m_{\tilde \chi_{1}^{0}}=m_{\tilde \tau_{1}}$ in the second row. 
In the top two panels, we find $m_{A_1} \approx m_{H_{2}}$, which is consistent with
our above discussions.  Remember that in Fig.~\ref{4mh_dia} we showed that the red points have
$m_{A_1} \simeq m_{H^{\pm}}\simeq m_{H_{3}}$.
Here, we see that for red points, the neutralino mass range is about $[0.75,2.7]$ TeV while $\sim [1.5,5.5]$ TeV
is the corresponding mass range of $m_{A_{1}}$. In the bottom panel,
we present the LSP neutralino-stau coannihilation scenario. We see that for red points $m_{\tilde \tau_1}$ is in the
mass range $\sim [0.75,2.7]$ TeV while without the 
$\Omega h^2 \lesssim$ 1 bound from supercritical string cosmology~\cite{Lahanas:2006xv}
(blue points) $m_{\tilde \tau_{1}}$
can be as heavy as 5.8 TeV.
It is very clear that in such parameter space the gluino and the first two-generation squarks/sleptons can
not be probed at the 14 TeV LHC, which will provide a strong motivation for 33 TeV and 100 TeV proton-proton colliders.
It is shown in Ref.~\cite{future_colider} that the squarks/gluinos of 2.5 TeV, 3 TeV and 6 TeV may be probed
by the LHC14, High Luminosity (HL)LHC14 and High Energy (HE) LHC33, 
respectively.  Thus, our models have testable predictions.  If we have the collider
facility with even higher energy in the future, we will be able to probe over even larger values of sparticle masses.

\subsection{The Benchmark Points for the MDSB and DDSB Scenarios}

In Table~\ref{table1}, we display two benchmark points for the MDSB scenario. The first point is an example of
relatively light sparticle spectrum. Here, $\lambda \sim 9.9\times 10^{-3}$ and $\tan\beta \sim 26$
while the light CP-even Higgs $m_{H_{1}}\sim$ 123 GeV. This point is also 
an example of solutions where $m_{H_{3}}\approx m_{A_{1}}\approx m_{H^{\pm}}$. The first two generation squarks are
about 2.5 TeV or heavier. Since gluino is about 3 TeV, the light stop $\tilde t_{1} $ with mass around
 2 TeV is the lightest colored sparticle.
The $\tilde t_{2}$ and $\tilde b_{1,2}$ have comparable masses of about 2.3 TeV. The 
slepton masses are $\lesssim$ 1 TeV. We also notice that the LSP neutralino and NLSP stau are
almost degenerate $\approx$ 596 GeV. The neutralino-proton spin independent and 
spin dependent cross sections are very small for this point that is $\sim$ $10^{-12}$ and $10^{-9}$, respectively.
The dark matter relic density is about 0.2. For the second benchmark 
point, the input parameters have relatively large values and then implies heavier sparticle spectrum.
For example, $x\approx 0.93$, $M_{3/2}\approx$ 4329 GeV, $\tan\beta\approx$ 28,
and $\lambda \approx 1.6 \times 10^{-2}$. The light CP-even Higgs $m_{H_{1}}\approx$ 125 GeV. This point represents the part
of the parameter space with $m_{H_{2}}\approx m_{A_{1}}\approx m_{H^{\pm}}$. Here, the light stop is also
the lightest colored sparticle with mass around 2.8 TeV while gluino mass is around 4.3 TeV.
The first two generations of squarks have masses about 4 TeV, while $\tilde t_{2}$ and $\tilde b_{1,2}$ have comparable masses
about 3.4 TeV. The slepton masses are about 1 TeV or heavier. Although the LSP neutralino and the lighter stau are
almost degenerate and respectively have masses 923 GeV and 928 GeV but dark matter relic density is still about 0.65.    

In Table~\ref{table2}, we display three benchmark points for the DDSB scenario. Because we have already seen
in Fig.~\ref{fundadia} that the minimum required
values for $m_{0}$, and $M_{1/2}$ are large as compared to the MDSB scenario,
all these three points have heavier spectra. Point 1 is 
relatively light as compared to point 2 and point 3. For Point 1, since $x$ and $M_{3/2}$ have smaller values,
this translates into relatively
small values of $m_{0}$, $M_{1/2}$ and $A_{0}$ as 998.06 GeV, 1755.4 GeV and -1738.8 GeV, respectively.
The light CP-even Higgs boson is about
123 GeV. In these three points, we have $m_{H_{3}}\approx m_{A_{1}}\approx m_{H^{\pm}}$.
As in the MDSB scenario, the gluino is 
heavier than the light stop, and they are about 3.7 TeV and 2.6 TeV, respectively.
The first two family squark masses are heavier than 3 TeV 
while $\tilde t_{2}$ and $\tilde b_{1,2}$ are about 2.9 TeV. The first two generation slepton masses
and $\tilde \tau_2$ are heavier 
than 1 TeV. The LSP neutralino and light stau masses are degenerate and about 773 GeV.
Here, we also notice that apart from representing
the neutralino-stau coannihilation scenario, this point also satisfies the $A$-resonance condition 
$|2 m_{\tilde \chi_{1}^{1}}-m_{A_{1}}|/m_{A_{1}} \lesssim$ 0.3. But it still has relatively large
dark matter relic density 0.23569, and small neutralino-proton spin independent and spin dependent cross sections.
Moreover, Point 2 and Point 3 share the similar
properties but having relatively heavier spectra.


\section{Discussions and Conclusion}
\label{summary}

We briefly reviewed the super-natural SUSY and addressed its subtle issues.
we pointed out that the NMSSM is a perfect framework for
super-natural SUSY since unlike the MSSM
it can be scale invariant and then has no mass parameter in its Lagrangian
before SUSY and gauge symmetry breakings. To generate the SUSY breaking soft mass to singlet,
we studied the moduli and dilaton dominant supersymmetry breaking scenarios in M-theory on $S^1/Z_2$.
In these scenarios, SUSY is broken by one and only one $F$-term of moduli or dilaton superfield,
and the SUSY breaking soft terms can be determined via the K\"ahler potential and
superpotential from Calabi-Yau compactification of M-theory on $S^1/Z_2$. Thus, according to
the super-natural SUSY, the SUSY EW fine-tuning measure is predicted to be of unity order.
In the moduli dominant SUSY breaking scenario, we found that the right-handed sleptons
are relatively light around 1 TeV, and stau can be even as light as 580 GeV and degenerate
with the LSP neutralino. Moreover, charginos are $\gtrsim$ 1 TeV,
the light stop masses are around 2 TeV or larger, the first two-generation
squark masses are about 3 TeV or larger, and gluinos are heavier than squarks as well.
In the dilaton dominant SUSY breaking scenario, the above qualitative picture is preserved
but the particle spectra are heavier as compared to moduli dominant SUSY breaking scenario.
In addition to it,  we have Higgs $H_{2}/A_{1}$-resonance solutions.
In both scenarios, the minimum value of DM relic density is about 0.2. To realize the correct
DM relic density, we can employ the dilution effect from supercritical string cosmology or introduce the axino
as the lightest supersymmetric particle.

\begin{table}[b]\hspace{-1.0cm}
\centering
\begin{tabular}{|c|cc|}
\hline
\hline
                 & Point 1 & Point 2 \\

\hline
$x$            &   0.67683       & 0.92663\\
$M_{3/2}$      &   3396          & 4329.3\\
$\tan\beta$    &   26.354        & 27.518\\
$\lambda$      &   $9.9924 \times 10^{-3}$ & $1.5796 \times 10^{-2}$ \\
\hline
$m_{0}$        &  625.13        & 1021.7\\      
$M_{1/2}$      &  1370.7        &  2082.2    \\
$A_0$          &  -1875.4       & -3065   \\
\hline
\hline
$m_{H_{1}}$            & 123  & 124.7  \\
$m_{H_{2}}$            & 1767 & 2731  \\
$m_{H_{3}}$            & 1840 &  2979   \\
$m_{A_{1}}$            & 1840 & 2731     \\
$m_{A_{2}}$            & 2542 & 4201 \\
$m_{H^{\pm}}$          & 1842 & 2732  \\
\hline
$m_{\tilde{\chi}^0_{1,2}}$
                 & 596, 1119     & 923, 1716       \\

$m_{\tilde{\chi}^0_{3,4,5}}$
                 & 1891, 1895, 2295    & 2810, 2813, 3837       \\

$m_{\tilde{\chi}^{\pm}_{1,2}}$
                 & 1119, 1895    & 1716, 2813       \\
\hline
$m_{\tilde{g}}$  & 2957 & 4369  \\
\hline $m_{ \tilde{u}_{L,R}}$
                 & 2722, 2618     & 4025, 3865          \\
$m_{\tilde{t}_{1,2}}$
                 & 1923, 2377    & 2809, 3466            \\
\hline
$m_{ \tilde{d}_{L,R}}$
                 & 2723, 2605     & 4026, 3845           \\
$m_{\tilde{b}_{1,2}}$
                & 2344, 2344     & 3443, 3443            \\
$m_{\tilde{\nu}_{1,2}}$
                 &1084  &   1680\\
$m_{\tilde{\nu}_{3}}$
                 & 1019  &   1567\\
\hline
$m_{ \tilde{e}_{L,R}}$
                & 1086, 801    & 1682, 1271          \\
$m_{\tilde{\tau}_{1,2}}$
                & 596, 1028    & 928, 1573            \\
\hline
$\sigma_{SI}({\rm pb})$
                & $9.88\times 10^{-12}$ & $ 3.91\times 10^{-12}$ \\

$\sigma_{SD}({\rm pb})$
                & $3.45\times 10^{-9}$ & $ 6.98\times 10^{-10}$ \\
$\Omega_{CDM}h^{2}$&  0.2085 & 0.6458  \\
\hline
\hline
\end{tabular}
\caption{Sparticle and Higgs masses are in GeV units and $\mu>0$.
All of these points satisfy  the sparticle mass, B-physics constraints described in Section~\ref{sec:scan}.
Points 1  and 2 display neutralino-stau coannihilation scenario when $m_{H_{3}} \approx m_{A_{1}}  \approx m_{H^{\pm}}$
and $m_{H_{2}} \approx m_{A_{1}}  \approx m_{H^{\pm}}$ respectively.
}
\label{table1}
\end{table}
\begin{table}[b]\hspace{-1.0cm}
\centering
\begin{tabular}{|c|ccc|}
\hline
\hline
                 & Point 1 & Point 2 & point 3\\

\hline
$x$            &   0.12148       & 0.11913  & 0.10126\\
$M_{3/2}$      &   1136.6        & 2145.5   & 2551.7\\
$\tan\beta$    &   40.695        & 40.554   & 40.491\\
$\lambda$      &   $9.0562 \times 10^{-3}$ & $1.2251 \times 10^{-2}$ & $2.4202 \times 10^{-2}$\\
\hline
$m_{0}$        &  998.06        & 1889.1 & 2292.7\\
$M_{1/2}$      &  1755.4        & 3320.6 & 4013.3    \\
$A_0$          &  -1738.8       & -3290.4 &  -3986.7 \\
\hline
\hline
$m_{H_{1}}$            & 123  & 125   & 125\\
$m_{H_{2}}$            & 993  & 2299  & 2924\\
$m_{H_{3}}$            & 1811 & 3252  & 3874 \\
$m_{A_{1}}$            & 1811 & 3252  & 3874 \\
$m_{A_{2}}$            & 1990 & 4013  & 4938\\
$m_{H^{\pm}}$          & 1812 & 3553  & 3874\\
\hline
$m_{\tilde{\chi}^0_{1,2}}$
                 & 773, 1442      &1502, 2760  &1830, 3346      \\

$m_{\tilde{\chi}^0_{3,4,5}}$
                 & 1516, 2184, 2188     &3263, 3864, 3867  &4083, 4579, 4581        \\

$m_{\tilde{\chi}^{\pm}_{1,2}}$
                 & 1442, 2188     &2760, 3867  &3346, 4582      \\
\hline
$m_{\tilde{g}}$  &3732  & 6770  & 8085\\
\hline $m_{ \tilde{u}_{L,R}}$
                 & 3472, 3339  & 6277, 6021   &7491, 7180         \\
$m_{\tilde{t}_{1,2}}$
                 & 2556, 2998     &4625, 5403    & 5515, 6446        \\
\hline
$m_{ \tilde{d}_{L,R}}$
                 & 3473, 3322      &6277, 5987   & 7491, 7139         \\
$m_{\tilde{b}_{1,2}}$
                & 2950, 2950      & 5369, 5369    &6416, 6416         \\
$m_{\tilde{\nu}_{1,2}}$
                 & 1506  & 2827     &  3415       \\
$m_{\tilde{\nu}_{3}}$
                 &1372  & 2580  & 3117\\
\hline
$m_{ \tilde{e}_{L,R}}$
                & 1508, 1184     & 2828, 2237    &3416, 2710        \\
$m_{\tilde{\tau}_{1,2}}$
                & 774, 1381     & 1503, 2584    & 1834, 3121         \\
\hline
$\sigma_{SI}({\rm pb})$
                & $8.96\times 10^{-12}$ & $ 2.62\times 10^{-12}$ & $ 1.92\times 10^{-12}$\\

$\sigma_{SD}({\rm pb})$
                & $2.95\times 10^{-9}$ & $ 3.40\times 10^{-10}$ & $ 1.79\times 10^{-10}$\\
$\Omega_{CDM}h^{2}$&  0.23569 & 0.74549 & 0.99767  \\
\hline
\hline
\end{tabular}
\caption{Sparticle and Higgs masses are in GeV units and $\mu>0$.
All of these points satisfy  the sparticle mass, B-physics constraints described in Section~\ref{sec:scan}.
Point 1 displays neutralino-stau coannihilation and Point 2 and Point 3 represents $m_{A_{1}}$-resonance solutions though stau is also very near to neutralino in mass. For these points $m_{H_{3}} \approx m_{A_{1}}  \approx m_{H^{\pm}}$. 
}
\label{table2}
\end{table}

\section*{Acknowledgments}

This research was supported in part by the 
Natural Science Foundation of China under grant numbers 11135003, 11275246, and 11475238 (TL).


\end{document}